\documentclass[a4paper,12pt,nofootinbib,notitlepage,preprintnumbers]{revtex4-1}

\usepackage{amsmath}
\usepackage{amssymb}
\usepackage{xcolor}
\usepackage{bbm}
\usepackage{braket}
\usepackage{xspace}
\usepackage{mathtools}
\usepackage{graphicx}
\usepackage{enumerate}
\usepackage{hyperref}
\usepackage{tabularx}
\usepackage{multirow}
\usepackage{enumitem}
\usepackage{etoolbox}

\renewcommand{\a}{\mathbf{a}}
\renewcommand{\b}{\mathbf{b}}
\renewcommand{\k}{\mathbf{k}}
\newcommand{\p}{\mathbf{p}}

\renewcommand{\r}{\mathbf{r}}

\renewcommand{\P}{\mathbf{P}}

\newcommand{\0}{\mathbf{0}}
\newcommand{\2}{\mathbf{2}}
\newcommand{\3}{\mathbf{3}}
\newcommand{\n}{\mathbf{n}}
\newcommand{\m}{\mathbf{m}}

\newcommand{\Cbb}{\mathbbm{C}}
\newcommand{\Nbb}{\mathbbm{N}}
\newcommand{\Zbb}{\mathbbm{Z}}

\newcommand{\Ac}{\mathcal{A}}
\newcommand{\Bc}{\mathcal{B}}
\newcommand{\Cc}{\mathcal{C}}
\newcommand{\Gc}{\mathcal{G}}
\newcommand{\Hc}{\mathcal{H}}
\newcommand{\Kc}{\mathcal{K}}
\newcommand{\Lc}{\mathcal{L}}
\newcommand{\Mc}{\mathcal{M}}
\newcommand{\Nc}{\mathcal{N}}
\newcommand{\Oc}{O}
\newcommand{\Pc}{\mathcal{P}}
\newcommand{\Rc}{\mathcal{R}}
\newcommand{\Tc}{\mathcal{T}}

\DeclareMathOperator{\im}{Im}

\newcommand{\bs}[1]{\boldsymbol{ #1 }}
\newcommand{\wt}[1]{\widetilde{ #1 }}

\newcommand{\bh}[1]{\mathbf{\hat{ #1 }}}
\newcommand{\overbar}[1]{\,\,\overline{\!\!{#1}}}

\newcommand{\cf}{cf.\xspace}
\newcommand{\eg}{e.g.\xspace}
\newcommand{\ie}{i.e.\xspace}

\newcommand{\nn}{\nonumber}
\newcommand{\diff}{\textrm{d}}
\newcommand{\ts}{\textsuperscript}

\newcolumntype{Y}{>{\centering\arraybackslash}X}
\newcolumntype{s}{>{\centering\arraybackslash\hsize=.2\hsize}X}

\newcommand\litem[1]{\item{$\boldsymbol{#1}$}}
\newcommand*\mathbshead[2]{\texorpdfstring{$\boldsymbol{#1}$}{#2}}

\makeatletter
\patchcmd{\subsubsection}{\itshape}{\itshape\bfseries}{}{}
\makeatother

\hypersetup{
pdftitle     = {Partial-wave projection of the one-particle exchange in three-body scattering amplitudes},
pdfsubject   = {Scattering Theory},
pdfkeywords  = {Scattering, Partial Waves, QCD, Hadron, Physics, Lattice},
pdfauthor    = {Andrew W. Jackura},
pdfnewwindow = {true},
colorlinks   = {true},
linkcolor    = {blue},
citecolor    = {blue},
filecolor    = {blue},
urlcolor     = {blue}
} 

\newcommand{\wm}{Department of Physics, 
William \& Mary, 
Williamsburg, VA 23187, USA}
\newcommand{\ucb}{Department of Physics, 
University of California, 
Berkeley, CA 94720, USA}   
\newcommand{\lbnl}{Nuclear Science Division, 
Lawrence Berkeley National Laboratory, Berkeley, 
CA 94720, USA}

\begin{document}

\title{Partial-wave projection of the one-particle exchange in three-body scattering amplitudes}


\author{Andrew W. Jackura}
\email[e-mail: ]{awjackura@wm.edu}
\affiliation{\wm}

\author{Ra\'ul A. Brice\~no}
\email[e-mail: ]{rbriceno@berkeley.edu}
\affiliation{\ucb}
\affiliation{\lbnl}

\begin{abstract}
As the study of three-hadron physics from lattice QCD matures, it is necessary to develop proper analysis tools in order to reliably study a variety of phenomena, including resonance spectroscopy and nuclear structure. Reconstructing the three-particle scattering amplitude requires solving integral equations, which can be written in terms of data-constrained dynamical functions and physical on-shell quantities. The driving term in these equations is the so-called one-particle exchange, which leads to a kinematic divergence for particles on-mass-shell. A vital component in defining three-particle amplitudes with definite parity and total angular momentum, which are used in spectroscopic studies, is to project the one-particle exchange into definite partial waves. We present a general procedure to construct exact analytic partial wave projections of the one-particle exchange contribution for any system composed of three spinless hadrons. Our result allows one full control over the analytic structure of the projection, which we explore for some low-lying partial waves with applications to three pions.
\end{abstract}

\date{\today}

\maketitle

\section{Introduction}
\label{sec:intro}

Applications of reaction theory to three-body systems have seen a resurgence due to modern theoretical hadronic spectroscopy. The success of two-hadron resonance studies from Quantum Chromodynamics (QCD) using numerical lattice QCD~\cite{Dudek:2010ew,Beane:2011sc,Pelissier:2012pi,Dudek:2012xn,Liu:2012zya,Beane:2013br,Wilson:2014cna,Dudek:2014qha,Orginos:2015aya,Berkowitz:2015eaa,Lang:2015hza,Wilson:2015dqa,Dudek:2016cru,Briceno:2016mjc,Moir:2016srx,Bulava:2016mks,Hu:2016shf,Alexandrou:2017mpi,Bali:2017pdv,Wagman:2017tmp,Andersen:2017una,Briceno:2017qmb,Woss:2018irj,Brett:2018jqw,ExtendedTwistedMass:2019omo,Mai:2019pqr,Woss:2019hse,Wilson:2019wfr,Cheung:2020mql,Rendon:2020rtw,Woss:2020ayi,Horz:2020zvv} in conjunction with non-perturbative mappings between finite-volume spectra and reaction amplitudes has allowed the community to pursue implementing such an analysis strategy for excited hadrons which have coupling to three-hadron decay modes.

The framework to compute non-perturbative reaction amplitudes from QCD relies on a methodology first presented by L\"uscher~\cite{Luscher:1985dn,Luscher:1986n2,Luscher:1990ux} for two-particle systems~\cite{Rummukainen:1995vs, Kim:2005gf,He:2005ey,Davoudi:2011md,Hansen:2012tf,Briceno:2012yi,Briceno:2013lba, Briceno:2014oea,Romero-Lopez:2018zyy}, with extensions to three-body systems developed in the last decade~\cite{Hansen:2014eka,Hansen:2015zga,Hansen:2016ync,Briceno:2017tce,Briceno:2018mlh,Briceno:2018aml,Briceno:2019muc,Blanton:2019igq,Hansen:2020zhy,Blanton:2020gha,Polejaeva:2012ut,Guo:2017crd,Guo:2017ism,Hammer:2017uqm,Hammer:2017kms,Meng:2017jgx,Pang:2019dfe,Muller:2021uur,Briceno:2017tce,Blanton:2020gmf,Blanton:2021mih,Hansen:2020zhy,Blanton:2020gha,Blanton:2020gmf,Blanton:2021mih,Jackura:2022gib,Mai:2017vot,Jackura:2018xnx,Mikhasenko:2019vhk,Dawid:2020uhn}. The procedure for three-hadrons is given as follows. Finite-volume correlation functions of operators with non-zero overlap to the desired quantum numbers are computed via numerical Monte Carlo methods and the subsequent spectrum is determined by novel techniques within lattice QCD. The finite-volume energy spectrum is then used in conjunction with formalisms known as \emph{quantization conditions} which relate short-distance dynamical objects known as $K$ matrices to the spectrum through geometric functions characterizing the distortions due to the periodic, finite-volume. Practically, one uses this avenue to constrain the $K$ matrices which seed into a set of integral equations which describe the on-shell scattering of the three hadrons. Examples of this computational procedure are given in~\cite{Briceno:2018mlh,Blanton:2019vdk,Romero-Lopez:2019qrt,Fischer:2020jzp,Jackura:2020bsk,Hansen:2020otl}. 

A major challenge in the study of three-particle reactions via lattice QCD is the last stage of the analysis, where physical amplitudes are reconstructed from the data-constrained $K$ matrices. For spectroscopy, one usually desires the resulting scattering amplitudes to be of definite spin-parity $J^P$ so that one may search for the spectral content by means of analytic continuation. Although there has been substantial progress on this end~\cite{Briceno:2018mlh,Blanton:2019vdk,Romero-Lopez:2019qrt,Fischer:2020jzp,Jackura:2020bsk,Hansen:2020otl,Dawid:2023jrj,Dawid:2023kxu}, most studies have focused on the restricted scenario where all the particles are identical spinless bosons in which all angular momenta are projected to $S$ wave.
\footnote{An exception is the exploratory study of an isovector $a_1$ meson~\cite{Mai:2021nul,Sadasivan:2021emk}, which numerically projected the scattering equations into $J^P = 1^+$ and neglected all other partial wave channels except $a_1 \to \rho\pi(^3S_1)$.}

In this work, we focus on lifting this technical restriction by presenting the operations needed to project the $\3\to\3$
\footnote{We use the notation $\n\to\m$ to indicate a reaction involving $\n$ incoming and $\m$ outgoing stable hadrons.}
scattering amplitude into any definite $J^P$ partial wave. We consider the partial wave expansion of the scattering amplitude of three arbitrary spinless particles, that is the particles can be identical or distinguishable. The exact details of the relativistic $\3\to\3$ scattering amplitude can be found in Sec.~\ref{sec:amplitudes}, as well as the introduction of relevant kinematic variables. In Sec.~\ref{sec:partial}, we review key concepts used for partial-wave projecting the scattering amplitude. As is emphasized there, the procedure followed is to define amplitudes within the helicity basis that are then projected to definite $J^P$.

At the center of our analysis is the one-particle exchange (OPE) process, a known kinematic function central to the integral equations~\cite{Fleming:1964zz,Holman:1965mxx,Kamal:1965vto,Jackura:2022gib}. The OPE has a complicated angular dependence which arises when two of the particles couple to some definite spin before recoiling against the third, making it the most challenging amplitude to project to definite partial waves. Schematically, the exchange propagator of the OPE, denoted by $\Gc$, takes the form
\begin{align}
	\Gc = \frac{\Hc}{u - m_e^2} \, , \nn 
\end{align}
where $\Hc$ is a dense matrix in the angular momentum of the incoming and outgoing pairs which we call the \emph{spin-helicity matrix}, and $u$ is the momentum-squared of the exchange particle which has mass $m_e$. The functions $\Hc$ and $u$ depend on the kinematics of the exchanged spectator particles, including the scattering angle. The main goal of this work is to provide a generic procedure to obtain an analytic representation of the partial wave projection of $\Gc$. Since our focus is primarily for lattice QCD applications, although this procedure can also be used in phenomenological studies, we use the definition of $\Hc$ a presented in Ref.~\cite{Hansen:2014eka,Hansen:2015zga,Jackura:2022gib}.

Details of the analytic partial wave projection of the OPE are given in Sec.~\ref{sec:ope}, which makes use the procedure outlined in Sec.~\ref{sec:partial} to derive a generic result for the partial wave OPE for any target $J^P$ quantum number. Our result can be expressed in terms of entirely known functions, taking the form
\begin{align}
    \label{eq:main_eq}
	\Gc^{J^P} = \Kc_{\Gc}^{J^P} + \Tc^{J^P} \, Q_0(\zeta_{pk}) \, , 
  \end{align}
where $\Gc^{J^P}$ is the exchange propagator projected to definite spin-parity $J^P$, $\Kc_{\Gc}^{J^P}$ and $\Tc^{J^P}$ are functions of external kinematics and include Clebsch-Gordan coefficients which couple the system to $J^P$. The functions $\Kc_{\Gc}^{J^P}$ and $\Tc^{J^P}$ are matrices in space of partial waves which contribute to a particular $J^P$, and are completely determined by the spin-helicity matrix $\Hc$, as shown in Sec.~\ref{sec:ope}. The $\Gc^{J^P}$ amplitude contains a branch cut in the complex energy plane which is due to on-shell particle exchange. This non-analytic behavior of the OPE is encoded entirely in $Q_0$, the zero-degree Legendre functions of the 2\ts{nd} kind, which depends on external kinematic variables through the function $\zeta_{pk}$, which is defined in the main body of the text. Our result allows one to control the entire analytic behavior of the amplitude which is vital in the analytic continuation of three-body amplitudes to complex energy planes~\cite{Dawid:2023jrj,Dawid:2023kxu}.

In Sec.~\ref{sec:cases}, we use our main equation~\eqref{eq:main_eq} to provide explicit expressions for the OPE amplitude for key low-lying partial waves. Applications of these results are given in Sec.~\ref{sec:application}, where we show numerical results for relevant channels in $3\pi$ systems to illustrate some of the analytic properties of these functions as discussed in the main text. Our procedure is summarized in Sec.~\ref{sec:summary}. To aid the reader, we provide three technical Appendices, App.~\ref{sec:app.angular_momentum}, \ref{sec:app.C_integral_eval}, and \ref{sec:app.generic_frames}, that include details of various special functions that are used throughout this work, a derivation of a key integral used in the analytic partial wave projection, and an alternative version of our approach using arbitrary reference frames. For the reader who wishes to use our explicit partial wave projected OPE amplitudes directly in their analyses, a fourth Appendix, App.~\ref{sec:app.formulary}, collects the cases presented in Sec.~\ref{sec:cases} along with brief explanations of the required kinematic variables.

\section{Amplitudes \& Kinematics}
\label{sec:amplitudes}

In the following, we consider the scattering of three spinless particles. In this work, we do not restrict the particles to be degenerate or identical, however, we do not consider any additional internal symmetries. \eg hadronic flavor quantum numbers.
\footnote{It is straightforward to include restrictions due to additional symmetries, \eg by including the appropriate SU(2) Clebsch-Gordan coefficients for three hadrons with isospin symmetry, \cf Refs.~\cite{Blanton:2020gmf,Blanton:2021mih,Hansen:2020zhy}.} 
Since our focus is ultimately on the on-shell exchange mechanism, we find this generalization benefits future applications as we provide a generic result to accommodate not only cases such as elastic $\pi\pi\pi$ scattering, but also those such as $K\bar{K}\pi \to K\bar{K}\pi$ where $K\bar{K} \to \eta\pi$ allows for $\eta$ meson exchanges between $K\bar{K}$ pairs.

Therefore, we consider a three-body reaction of the form
\begin{align}
	\varphi_k(k) + \varphi_a(a) + \varphi_{a'}(a') \to \varphi_p(p) + \varphi_b(b) + \varphi_{b'}(b') \, , \nn 
\end{align}
where $\varphi_k(k)$ represents a single spinless particle carrying a four-momentum $k = (\omega_k,\k)$ with its energy $\omega_k$ fixed by its mass $m_k$ and momentum $\k$ through the usual relativistic on-shell dispersion relation $\omega_k^2 = m_k^2 + \k^2$. Similar definitions hold for the other particles. Here we adopt the notation that the mass of the particle will be labeled by its momentum. We normalize the single particle state by the usual Lorentz invariant measure 
\begin{align}
	\braket{\k'|\k} = (2\pi)^3 \, 2\omega_{k} \, \delta^{(3)}(\k'-\k) \, , \nn 
\end{align}
where $\delta^{(3)}$ is the three-dimensional Dirac delta distribution. The initial system carries a total four-momentum $P = (E,\P)$, where $E$ is the total energy and $\P$ is the total momentum, which in terms of the constituent momenta is $P = k + a + a'$. Similarly, $P' = p + b + b' = (E',\P')$ for the final state four-momentum. A three-particle state is constructed by the usual tensor product of single-particle states, which we denote as $\ket{P,\k,\a}$. Here we trade the momentum $\a'$ for the total momentum as it is conserved in reactions and $\a' = \P - \k - \a$.

%
\begin{figure}[t]
	\centering
	\includegraphics[width=0.65\textwidth]{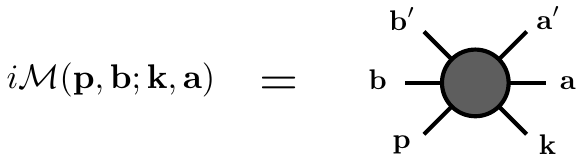}
	\caption{The fully connected $\3\to\3$ amplitude $i\Mc$ with momentum assignments. All external legs represent incoming and outgoing on-shell particles constrained by total momentum conservation.}
	\label{fig:amplitude}
\end{figure}
The $\3\to\3$ scattering amplitude $\Mc$, depicted in Fig.~\ref{fig:amplitude}, is defined as the fully connected $S$ matrix element
\begin{align}
    \label{eq:M3_def}
	\braket{P',\p,\b ;\mathrm{out} | P,\k,\a ;\mathrm{in}}_{\mathrm{conn.}} = (2\pi)^{4}\delta^{(4)}(P'-P) \, i\Mc(\p,\b;\k,\a) \, ,
\end{align}
where ``conn.'' indicates only the fully connected contribution is to be taken, and ``in/out'' refer to the asymptotically far past/future. We have also factored out a Dirac delta from the amplitude which ensures total momentum is conserved, $P = P'$. The amplitude depends on the total three-body center-of-momentum (CM) frame energy $E = \sqrt{s}$, where $s \equiv P^2 = P^{\mu}P_{\mu} = E^2 - \P^2$ is the Mandelstam invariant. The physical scattering threshold is given by $E^{(\mathrm{thr.})} = \max(m_k+m_a+m_{a'}, m_p + m_b + m_{b'})$. In this work, we suppress the dependence of $s$ for all amplitudes to simplify the notation. The amplitude depends on seven more kinematic variables which are formed from the set of initial and final state momenta.

In order to construct useful kinematic variables, it is convenient to consider the kinematic configuration of the three-body system as one consisting of two particles in a \emph{pair} with an associated \emph{spectator} being the third particle. In most of this work, we choose to label the initial state spectator with momentum $k$, while the associated pair is composed of the particles with momenta $a$ and $a'$. Likewise, for the final state, the spectator has momentum $p$ and the pair consists of the particles with momenta $b$ and $b'$. 

Each pair has a four-momentum given by $P_k = (E_k,\P_k) \equiv P - k$ and $P_p = (E_p,\P_p) \equiv P - p$ for the initial and final state, respectively, where the subscripts $k$ and $p$ indicate which spectator is associated with the pair. The invariant mass-squared of the pairs is given by
\begin{align}
	\label{eq:pair_masses}
	\sigma_{k} \equiv P_k^2 = (P-k)^2 \, , \qquad \sigma_{p} \equiv P_p^2 = (P-p)^2 \, .
\end{align}
Focusing first on the initial state, for a fixed $s$ the physical region of the pair invariant mass is limited to $\sigma_{k}^{(\mathrm{thr.})} \le \sigma_k \le (\sqrt{s}-m_k)^2$, where $\sigma_{k}^{(\mathrm{thr.})}$ is the physical scattering threshold for that pair, $\sigma_{k}^{(\mathrm{thr.})} = (m_a + m_{a'})^2$. Momentum conservation constrains the pair invariant masses through the usual Mandelstam condition, 
\begin{align}
	\label{eq:mandelstam_cond}
	\sigma_k + \sigma_a + \sigma_{a'} = s + m_k^2 + m_a^2 + m_{a'}^2 \, ,
\end{align}
where $\sigma_a$ and $\sigma_{a'}$ are the pair invariant masses considering $a$ and $a'$ as spectators, respectively. The physical scattering region of the three particles is therefore bounded by the condition $\Phi(k,a) \ge 0$, where $\Phi(k,a)$ is the Kibble boundary function defined as~\cite{Kibble:1960zz,byckling1973particle,Collins:1971ff}
\begin{align}
	\Phi(k,a)  = \sigma_k \sigma_a \sigma_{a'} & - \sigma_k(s - m_a^2)(m_k^2 - m_{a'}^2) - \sigma_a(s - m_k^2)(m_a^2 - m_{a'}^2) \, \nn \\[5pt]
	& - (s m_{a'}^2 - m_a^2 m_k^2)(s + m_{a'}^2 - m_a^2 - m_k^2 ) \, .
\end{align}
Similar restrictions hold for the final state particles, with expressions given by the substitution $\{k,a,a'\} \to \{p,b,b'\}$ in the above conditions.

In Sec.~\ref{sec:ref_frames}, we specify three reference frames which we use to define additional kinematic variables used in the partial wave projection.

\subsection{Reference Frames}
\label{sec:ref_frames}

\begin{figure*}[t!]
	\centering
	\includegraphics[width=0.98\textwidth]{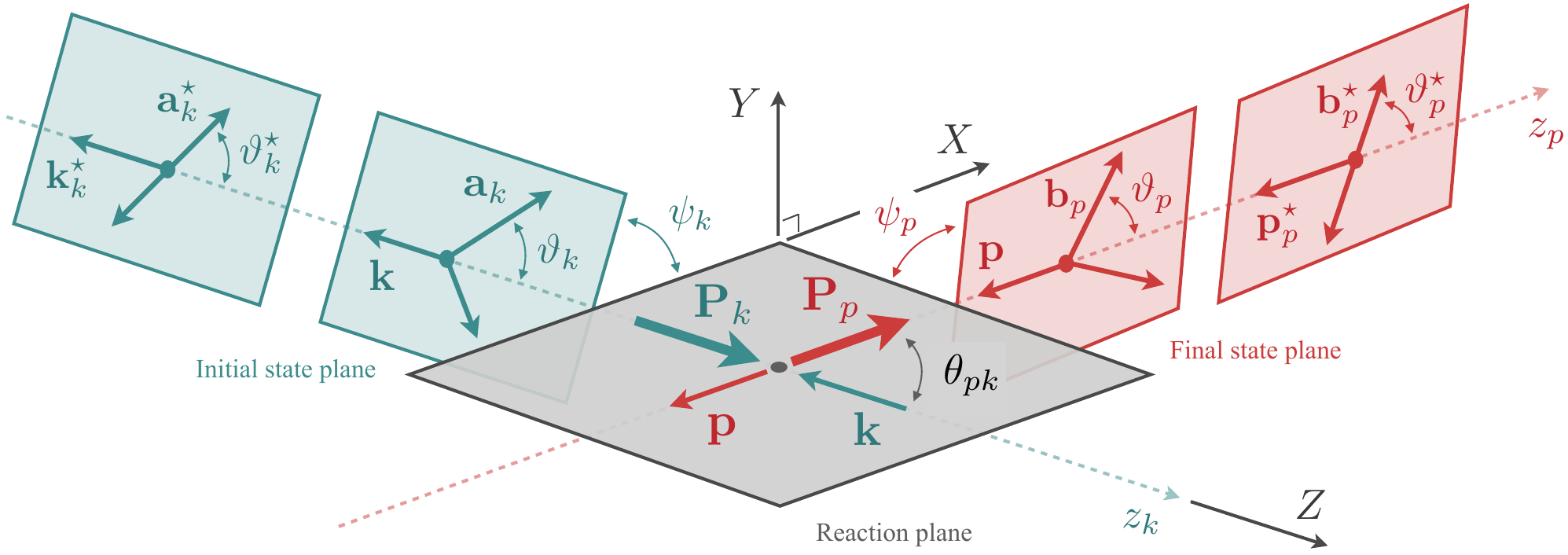}
	\caption{Reference frames for the $\3\to\3$ amplitude as described in the text. Shown in blue is the initial three-body state plane, in red the final three-body state, and in gray the reaction plane. Kinematics in the reaction plane are shown in the total CM frame ($\P = \0)$, while the initial state planes are shown for both the initial pair CM frame ($\P_{k,k}^{\star} = \0$) and the total CM frame. The final state planes are shown in the total CM frame and the final pair CM frame ($\P_{p,p}^{\star} = \0$).}
	\label{fig:reference_frames}
\end{figure*}

Three reference frames are required in our analysis of the partial wave projection of the $\3\to\3$ amplitude. Here we define the essential characteristics of these frames, and will refer to these in our constructions of partial waves in Sec.~\ref{sec:partial} and give additional kinematic relations when we discuss the application to the exchange propagator in Sec.~\ref{sec:ope}. These reference frames are illustrated in Fig.~\ref{fig:reference_frames}, and are designated the ``initial pair CM frame'', the ``final pair CM frame'', and the ``total CM frame''. We define these frames as follows:
%
\begin{enumerate}
	\item \emph{Initial pair CM frame} --  The initial pair CM frame is defined by $\P_k = \P - \k = \0$. It is common to introduce a notation to indicate a given kinematic variable is evaluated in some specific reference frame. Commonly in the literature one uses $\star$ superscript to indicate such a situation for the CM frame. In our case, however, we need to be careful as there are three CM frames of interest. Therefore, we adopt a notation for the initial pair rest frame that a $\star$ superscript \emph{along with} a $k$ subscript indicating that the kinematic variable is evaluated in this frame. While this results in slightly cumbersome notation, we feel this will alleviate future confusion for implementing the results of this work. As an example, the defining relation for this frame can be written as $\P_{k,k}^{\star} = \P_k^{\star} - \k_k^{\star} = \0$, where $\P_{k,k}^{\star}$ indicates the initial state pair momentum is evaluated in its rest frame. 
    \footnote{An example where this notation is vital is for $\P_{p,k}^{\star}$, which is the final state pair momentum evaluated in the initial pair rest frame. Such evaluations become necessary as detailed in Sec.~\ref{sec:ope}.}
	 
	In this frame, the pair has back-to-back momentum $\a_k^{\star} = -\a_k'^{\star}$, with its magnitude fixed by the pair invariant mass
    \footnote{The difference between the four-momentum $a = (\omega_a,\a)$ and the magnitude of its three-momentum $a = \lvert \a\rvert$ is clear from context.}
    %
	\begin{align}
		a_k^{\star} \equiv \lvert \a_k^{\star} \rvert = \frac{1}{2\sqrt{\sigma_k}} \, \lambda^{1/2}(\sigma_k,m_a^2,m_{a'}^2) \, ,
	\end{align}
	where $\lambda(x,y,z) = x^2 + y^2 + z^2 - 2 ( x y + y z + z x )$ is the K\"all\'en triangle function. Note that $\lambda(x,y,z)$ is symmetric under interchange of the variables $x,y,z$. Note also that in the case where $m_{a} = m_{a'} = m$, then $a_k^{\star}$ reduces to $a_k^{\star} = \sqrt{\sigma_k/4 - m^2}$ . 
	
	As illustrated in Fig.~\ref{fig:reference_frames}, we define a coordinate system with a $z_k$-axis for the initial state defined to be anti-parallel to the spectator momentum, \ie $\bh{z}_k \equiv -\bh{k}$. 
    \footnote{We use the notation $\bh{r}$, the unit vector of $\r$, to indicate the polar and azimuthal angles, $(\theta_r,\varphi_r)$. Note that we use the standard convention for the domain of the polar and azimuthal angles, $\theta_r \in [0,\pi]$, and $\varphi_r \in [0,2\pi)$. }
	Thus, the first particle in the pair (taken to by $\varphi_{a}$) has its momentum oriented at a polar angle $\vartheta_k^{\star}$ with respect to this $z_k$-axis. Furthermore, the three momenta form a plane (the initial state plane) defined by the normal vector $\k \times \a_k$, oriented with respect to the \emph{reaction plane} (with a coordinate system $XYZ$ which is defined later) by an azimuthal angle $\psi_k^{\star}$. This angle is preserved upon Lorentz boosts along $z_k$, \ie $\psi_k = \psi_k^{\star}$ from the total CM frame to the initial pair CM frame. The boost velocity from the initial pair CM frame to the total CM frame is given by $\bs{\beta}_k = \mathbf{P}_k / E_k$.
	%
    %
	%
	\item \emph{Final pair CM frame} -- The final pair CM frame, defined by $\P_{p,p}^{\star} = \0$, is constructed analogously to the initial pair CM frame. The notation of a $\star$ superscript \emph{with} a $p$ subscript indicates kinematic variables are in this frame. Another body-fixed coordinate system is assigned to this frame, with its $z_p$-axis is defined by $\bh{z}_p \equiv -\bh{p}$ and the ``final state plane'' defined with a normal vector $\p \times \b_p$, which is depicted in Fig.~\ref{fig:reference_frames}. The pairs polar and azimuthal angles are $\vartheta_p^{\star}$ and $\psi_p^{\star}$, respectively. The azimuthal angle is again invariant under boost along $\bh{z}_p$, $\psi_p^{\star} = \psi_p$. The final pair momenta are defined back-to-back, $\b_p^{\star} = -\b_p'^{\star}$, with a magnitude fixed by $\sigma_p$
    \begin{align}
        b_p^{\star} \equiv \lvert \b_p^{\star} \rvert = \frac{1}{2\sqrt{\sigma_p}} \, \lambda^{1/2}(\sigma_p,m_b^2,m_{b'}^2) \, .
    \end{align}
    %
	%
    %
	%
	\item \emph{Total CM frame} -- The final reference frame in our analysis is the total CM frame, defined by $\P = \0$. Unlike the initial and pair CM frames, we do not include a special notation to indicate a kinematic variable is evaluated in the total CM frame. This frame proves convenient to define the \emph{reaction plane}, which connects the initial three-particle state to the final state. Both the initial and final state momenta are equally evaluated in this frame. Specifically, the magnitudes of the initial and final spectator momenta are fixed by their pair invariant masses,
	\begin{align} 
	\label{eq:pk_cm}
		p \equiv \lvert \p \rvert = \frac{1}{2\sqrt{s}} \, \lambda^{1/2}(s,\sigma_{p},m_p^2), \qquad k \equiv \lvert \k \rvert = \frac{1}{2\sqrt{s}} \, \lambda^{1/2}(s,\sigma_{k},m_k^2) \, .
	\end{align}
	These relations follow from Eq.~\eqref{eq:pair_masses}, where the inverse relations are readily given
	\begin{align}
		\sigma_k = s + m_k^2 - 2\sqrt{s}\,\omega_k \, , \quad \sigma_p = s + m_p^2 - 2\sqrt{s}\,\omega_p \, ,
	\end{align}
	where recall that $\omega_k = \sqrt{m^2 + k^2}$ and $\omega_p = \sqrt{m_p^2 + p^2}$. The angular degrees of freedom are not fixed. Instead of specifying the angles of the spectators, it proves more convenient to consider the angles of the pair momenta $P_k$ and $P_p$. To define the momentum orientations, we introduce a \emph{space-fixed} coordinate system denoted by $XYZ$. This coordinate system allows us to define our reaction plane, and allows us to think about the pair-spectator scattering system as a \emph{quasi-two-body} reaction. This quasi-two-body reaction is depicted in Fig.~\ref{fig:reaction_plane}, which for some fixed invariant masses $\sqrt{\sigma_k}$ and $\sqrt{\sigma_p}$ is specified by the total CM frame energy $\sqrt{s}$ and scattering angle between the spectators.
    \begin{figure}[t]
        \centering
        \includegraphics[width=0.55\textwidth]{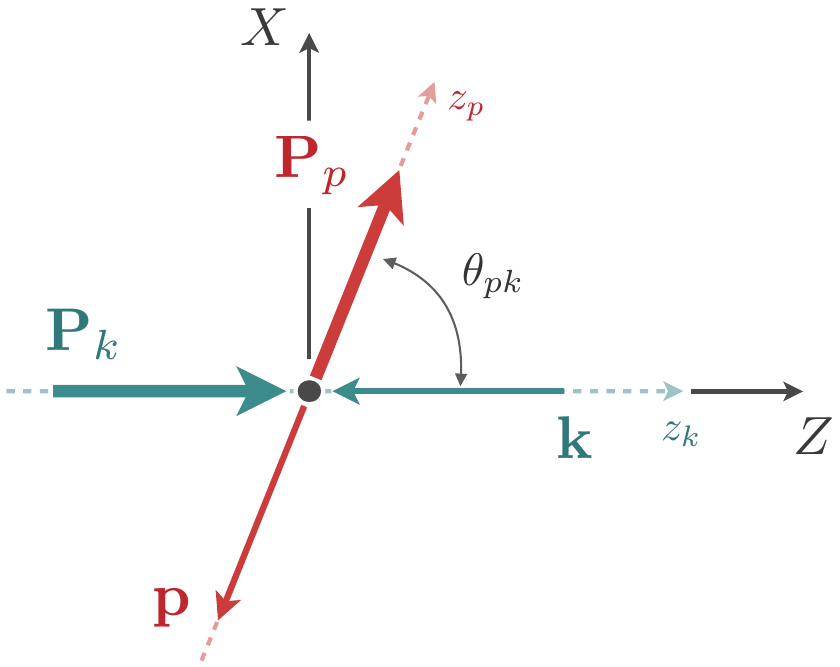}
        \caption{Orientations of the initial and final state pair momenta defined with respect to the external space-fixed coordinate system (denoted by $XYZ$). The angle between the initial and final momentum $\P_k$ and $\P_p$, respectively, is the effective CM frame scattering angle $\theta_{pk}$.}
        \label{fig:reaction_plane}
    \end{figure}
  
    Without loss of generality,
    \footnote{In App.~\ref{sec:app.generic_frames} we lift this choice of coordinates and illustrate the partial wave expansion with respect to a generic externally fixed coordinate system. Although important in future analyses, as discussed in App.~\ref{sec:app.generic_frames}, we find that working in a generic coordinate system is not vital to reach our results in this work. Therefore, we invite the interested reader to view App.~\ref{sec:app.generic_frames}.}
    we define the initial pair momentum $P_k$ to be aligned with the $+Z$-axis of some space-fixed coordinate system (with axes $XYZ$), that is $\bh{Z} \equiv \bh{P}_k$. Since $\P = \0$, the initial spectator momentum is then aligned with the $-Z$-axis. We then choose the final pair momentum $\P_p$ to lie in the $XZ$-plane, \ie the quasi-two-body reaction lies in the reaction plane. This plane is defined with the $+Y$-axis proportional to $\k \times \p = \P_k \times \P_p$.  We denote the total CM frame \emph{scattering angle} by $\theta_{pk}$, which is defined in the usual way
	\begin{align}
		\label{eq:cos_theta_pk}
		\cos\theta_{pk} & \equiv  \bh{P}_p \cdot \bh{P}_k  = \bh{p} \cdot \bh{k} \, .
	\end{align}
    Notice that $\theta_{pk}$ is simply the angle of $\P_p$ with respect to the $Z$-axis, $\cos\theta_{pk} = \P_p \cdot \bh{Z}$, with respect to our space-fixed coordinate system. Since we use the standard convention that $\theta_{pk} \in [0,\pi]$, this means that we can specify $\theta_{pk}$ completely with just $\cos\theta_{pk}$.
    
    Additionally, as mentioned in the previous reference frame definitions, the reaction plane serves as a convenient reference for the azimuthal angles of the initial and final three-body planes, $\psi_k$ and $\psi_p$, respectively.
\end{enumerate}
%

To conclude this section, we summarize the eight necessary kinematic variables relevant to project the $\3\to\3$ system to definite partial waves. For energy variables, we choose the total CM energy $s$, as well as the initial and final pair invariant mass-squares $\sigma_k$ and $\sigma_p$, respectively. An alternative to $\sigma_k$ and $\sigma_p$ is the magnitudes of the associated spectator momenta $k$ and $p$. Through Eq.~\eqref{eq:pk_cm} at a fixed $s$, these are completely interchangeable. We freely use either the set $\sigma_k,\sigma_p$ or $k,p$ where convenient, either for ease of notation or exploiting some physical relation. The final five variables orient our system, four of which are the initial and final pair polar and azimuthal angles defined in their respective rest frames, $\bh{a}_k^{\star}$ and $\bh{b}_p^{\star}$, respectively. The last variable is the total CM frame scattering angle $\theta_{pk}$. In the following section, we construct partial wave $\3\to\3$ amplitudes by integrating over the angular degrees of freedom with appropriate angular momentum weight functions.

\section{Partial Wave Projection}
\label{sec:partial}

Our first task is to define the generic partial wave projection for $\3\to\3$ scattering amplitudes. The scheme we follow is similar to that of Ref.~\cite{JPAC:2019ufm}, where we first couple the three-particle system to a definite total angular momentum $J$ through the helicity framework. Then, we re-couple the helicity partial wave to ones of definite parity using spin-orbit or $LS$ coefficients. The reason for going through this two-step process is that helicity transforms simply under Lorentz transformations compared to spin-projections against some space-fixed $z$-axis. Doing so makes the projection of the exchange amplitude simpler, as the angles in the total CM frame are simply related to those in either pair rest frame.

\subsection{Helicity Projection}
\label{sec:helicity}

\begin{figure}[t]
	\centering
	\includegraphics[width=0.6\textwidth]{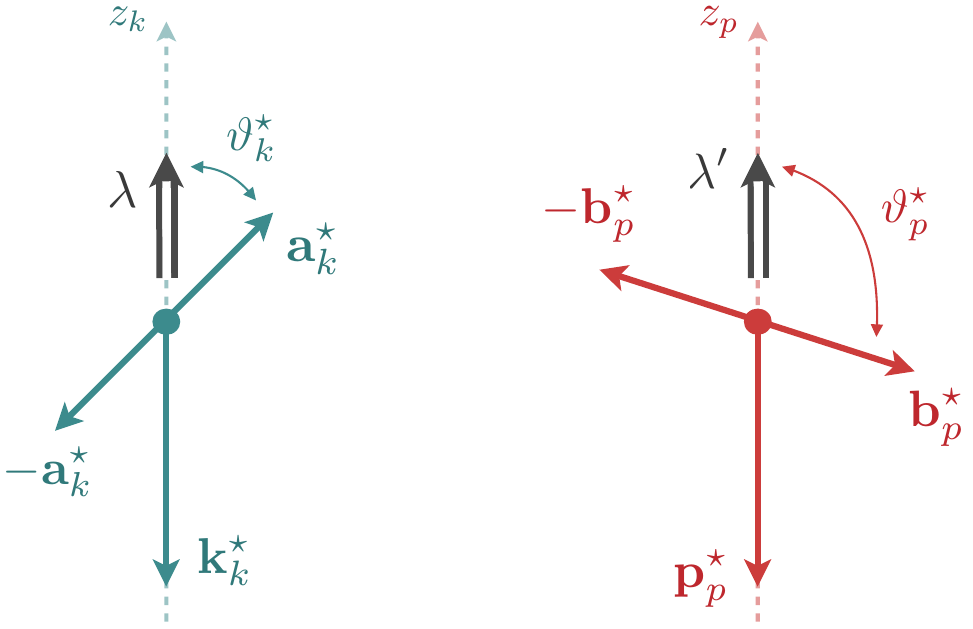}
	\caption{Kinematic configuration and spin-projection definitions for the initial (blue) and final (red) state pairs in their associated rest frame. Each pair has its spin-projection along the opposite direction of their associated spectators, giving $\lambda$ and $\lambda'$ a helicity interpretation.}
	\label{fig:kin_pair_cm_frame}
\end{figure}

The starting point is to project the amplitude in \emph{helicity partial waves}, that is partial waves of definite total angular momentum where the pairs have their spin projected quantized along their momentum direction. Following the decomposition used in Refs.~\cite{Hansen:2014eka,Hansen:2015zga,Jackura:2022gib}, we proceed by first partial wave projecting the pair into a definite angular momentum state
\begin{align}
	\ket{P,\k,\a} \propto \sqrt{4\pi} \, \sum_{\ell,\lambda} \ket{P,\k,\ell \lambda} \, Y_{\ell \lambda}^{*}(\bh{a}_k^{\star}) \, , \nn 
\end{align}
where $\ell$ is the angular momentum of the pair, $\lambda$ is its projection along the $z_k$-axis defined by the opposite sense of the spectator momentum $k$ (see Fig.~\ref{fig:kin_pair_cm_frame}.), and $Y_{\ell \lambda}(\bh{a}_k^{\star} ) = Y_{\ell \lambda}(\chi_k^{\star},\psi_{k}^{\star}) \equiv \braket{\bh{a}_k^{\star} | \ell \lambda}$ are the usual spherical harmonics. Since the spin-quantization axis is along the direction of the pair, we interpret $\lambda$ as the pair helicity.
\footnote{In the recent three-particle finite-volume frameworks for lattice QCD analyses, the pair angular momentum projection $m$ usually has a quantization axis taken to be some fixed $z$-axis of a volume. If one starts with this definition, then converting to a helicity quantization with $\lambda$ amounts to a unitary rotation of the pair state $\ket{\P_k,\ell \lambda}$, 
\begin{align}
	\ket{\P_k,\ell \lambda} = \sum_{m} D_{m\lambda}^{(\ell)}(\bh{\P}_k) \ket{\P_{k},\ell m} \, , \nn
\end{align}
where $D_{m\lambda}^{(\ell)}$ are the Wigner $D$ matrix elements which are discussed in App.~\ref{sec:app.angular_momentum}.}

The normalization of the state is not relevant for our discussion, as we freely absorb this factor into the definition of the amplitude. As we work with scattering states of three scalars, only $\ell \in \Nbb_0$ is allowed, with $\lambda \in \Zbb$ which spans $-\ell \le \lambda \le \ell$ for a given $\ell$. For the scattering amplitude, we arrive at the expansion
\begin{align}
	\label{eq:first_pw}
	\Mc(\p,\b;\k,\a) = 4\pi \sum_{\ell',\lambda'}\sum_{\ell,\lambda} Y_{\ell'\lambda'}(\bh{b}_{p}^{\star}) \, \Mc_{\ell'\lambda' , \ell \lambda}(\p,\k) \, Y_{\ell\lambda}^{*}(\bh{a}_{k}^{\star}) \, ,
\end{align}
where the factor of $4\pi$ is convention. Given the full amplitude, the projection is found by using the orthonormality of the spherical harmonics
\footnote{In this work we make frequent use of identities of mathematical special functions. For convenience, we collected a set of useful properties and appropriate references in App.~\ref{sec:app.angular_momentum}.}
%
\begin{align}
	\label{eq:helicity_proj}
	\Mc_{\ell'\lambda',\ell \lambda}(\p,\k) = \frac{1}{4\pi} \int\!\diff \bh{b}_p^{\star} \int\!\diff \bh{a}_k^{\star} \, Y_{\ell'\lambda'}^{*}(\bh{b}_p^{\star} ) \, \Mc(\p,\b;\k,\a) \, Y_{\ell\lambda}(\bh{a}_k^{\star} ) \, ,
\end{align}
where the integration measure is $\diff\bh{a}_k^{\star} \equiv \diff\psi_k^{\star} \,  \diff\cos\vartheta_k^{\star}$ with the integration domain being over $\vartheta_k^{\star} \in [0,\pi]$ and $\psi_k^{\star} \in [0,2\pi]$.

It is useful to consider the amplitude $\Mc_{\ell'\lambda' , \ell \lambda}$ as one describing the reaction of a spinless particle of mass $m_k$ and a \emph{quasi-particle} of mass $\sqrt{\sigma_k}$, spin $\ell$, and helicity $\lambda$, which transitions to a spinless particle of mass $m_p$ recoiling against another \emph{quasi-particle} of mass $\sqrt{\sigma_p}$, spin $\ell'$, and helicity $\lambda'$. We represent the quasi-two-body reaction as
\begin{align}
	\xi_k^{(\ell)}(P_k,\lambda) + \varphi_k(k) \to \xi_p^{(\ell')}(P_p,\lambda') + \varphi_p(p) \, , \nn 
\end{align}
where $\xi_k^{(\ell)}$ represents the quasi-particle of spin $\ell$. Note that this effective $\2\to\2$ processes only knows about particles $\varphi_{a}$, $\varphi_{a'}$ and $\varphi_{b}$, $\varphi_{b'}$ through formation and decay, thereby only restricting the threshold of the invariant mass. Thus the details of the kinematic configurations for these particles are not relevant in the rest of this construction. However, since the amplitude depends on the pair invariant masses, it contains an angular momentum barrier suppression as the energies of the pairs approach their threshold. For example, as the initial state pair invariant mass-squared $\sigma_k \to \sigma_{k}^{\mathrm{(thr.)}}$, then
\begin{align}
	\Mc_{\ell'\lambda',\ell \lambda}(\p,\k) \sim \left(a_k^{\star} \right)^{\ell} \, , \nn 
\end{align}
with a similar behavior for the final state, $\Mc_{\ell'\lambda',\ell \lambda} \sim \left(b_p^{\star} \right)^{\ell'}$ as $\sigma_p \to \sigma_{p}^{\mathrm{(thr.)}}$.

Once we have the effective helicity amplitude $\Mc_{\ell'\lambda',\ell \lambda}$, we can now couple the initial and final state to those of definite total angular momentum $J$ and projection $m_J$ defined with respect to the space-fixed $Z$-axis. The quasi-two-body state has the helicity partial wave expansion
\begin{align}
	\ket{P,\k, \ell\lambda} \propto \sum_{J,m_J}  \sqrt{2J+1} \, \ket{P,k,Jm_J,\ell \lambda} \,  D_{m_J\lambda}^{(J)}(\bh{P}_k) \, , \nn 
\end{align}
where $D_{m_J\lambda}^{(J)}$ are Wigner $D$ matrix elements (\cf App.~\ref{sec:app.angular_momentum}.). Again, we do not specify a normalization as we absorb this kinematic factor into the definition of the amplitude. Since we chose the pair momentum $P_k$ to have its momentum along the $+z_k$-axis, the angles we consider are those of this momentum, and not the spectator. We choose the phase convention of the Wigner D matrix elements such that
\begin{align}
	D_{m_J\lambda}^{(J)}(\bh{P}_k) & = D_{m_J\lambda}^{(J)}(\phi_{k},\theta_k,0) \, , \nn \\[5pt]
	& = e^{-im_J\phi_k} \, d_{m_J \lambda}^{(J)}(\theta_k) \, ,
\end{align}
where $\theta_k$, and $\phi_k$ are the polar and azimuthal angles of the momentum $\P_k$, respectively, and $d_{m_J \lambda}^{(J)}(\theta_k)$ are the $d$ matrix elements which are real for physical $\theta_k$, \ie $-1\le \cos\theta_k\le +1$, see Eq.~\ref{eq:wigner_d_jacobi}. Applying this basis expansion on both the initial and final states of Eq.~\eqref{eq:helicity_proj} yields 
\begin{align}
	\label{eq:helicity_expand}
	\Mc_{\ell'\lambda', \ell\lambda}(\p,\k) = \sum_{J',m_{J'}} \sum_{J,m_J} \sqrt{2J'+1} \, D_{m_{J'}\lambda'}^{(J')\,*}(\bh{P}_p) \, \Mc_{\ell'\lambda' , \ell\lambda}^{J'm_{J'},J m_J}(p,k) \, \sqrt{2J+1} \, D_{m_{J}\lambda}^{(J)}(\bh{P}_k) \, .
\end{align}
The normalization of Eq.~\eqref{eq:helicity_expand} is chosen such that for spinless pairs, $\ell ' = \ell = 0$, then \eqref{eq:helicity_expand} simplifies and the resulting expression has same normalization as Eq.~\eqref{eq:first_pw}. 
\footnote{See Eq.~\eqref{eq:Ylm_to_Wigner} in  App.~\ref{sec:app.angular_momentum} for the relations between the spherical harmonics and Wigner $D$ matrix elements.}

Rotational invariance of the entire three-body system imposes that total angular momentum $J$ is conserved, and the amplitude is independent of the projection $m_J$, 
\begin{align}
	\Mc_{\ell'\lambda' , \ell\lambda}^{J'm_{J'},J m_J} = \delta_{J' J} \delta_{m_{J'}m_J} \, \Mc_{\ell'\lambda' , \ell\lambda}^{J} \, .
\end{align}
Since the helicity partial wave amplitude is block diagonal in each $J$ sector, we can reduce the sums in the expansion to
\begin{align}
	\label{eq:helicity_expand_simplify}
	\Mc_{\ell'\lambda', \ell\lambda}(\p,\k) & = \sum_{J = J_{\min}}^{\infty} (2J+1) \,  \Mc_{\ell'\lambda' , \ell\lambda}^{J}(p,k) \, \sum_{m_J = -J}^{J} D_{m_{J}\lambda'}^{(J)\,*}(\bh{P}_p) \, D_{m_{J}\lambda}^{(J)}(\bh{P}_k) \, ,
\end{align}
where $J_{\min} = \max(\lvert \lambda' \rvert, \lvert\lambda\rvert)$ is the minimum $J$ value for the sum. The helicity partial wave amplitudes $\Mc_{\ell'\lambda' , \ell\lambda}^{J}$ depend on the three kinematic variables, total CM energy $\sqrt{s}$, and the initial and final spectator momenta $k$ and $p$, respectively (or alternatively the pair invariant masses $\sqrt{\sigma_k}$ and $\sqrt{\sigma_p}$).

Recall from Sec.~\ref{sec:ref_frames} that with respect to our chosen coordinate system, $\bh{P}_k = \bh{Z}$. Thus, for all $J$ the initial Wigner $D$ matrix element simplifies to
\begin{align}
    D_{m_{J}\lambda}^{(J)}(\bh{Z}) = \delta_{m_J \lambda} \, , \nn 
\end{align}
allowing us to trivially perform the sum to find
\begin{align}
    \label{eq:Wigner_D_addition}
    \sum_{m_J = -J}^{J} D_{m_{J}\lambda'}^{(J)\,*}(\bh{P}_p) \, D_{m_{J}\lambda}^{(J)}(\bh{P}_k) & = d_{\lambda\lambda'}^{(J)}(\theta_{pk}) \, ,
\end{align}
where we recall from Eq.~\eqref{eq:cos_theta_pk} that the CM frame scattering angle is defined by $\cos\theta_{pk} \equiv \bh{P}_p \cdot \bh{P}_k$, and since the pair momenta lie in the $XZ$-plane, there is no azimuthal angular dependence. Therefore, the helicity partial wave expansion is given by
\begin{align}
    \label{eq:helicity_pw_expand}
	\Mc_{\ell'\lambda', \ell\lambda}(\p,\k) = \sum_{J}(2J+1) \, \Mc_{\ell'\lambda' , \ell \lambda}^J(p,k) \, d_{\lambda\lambda'}^{(J)}(\theta_{pk})\, ,
\end{align}
with the projection given by
\begin{align}
    \label{eq:helicity_pw_proj}
	\Mc_{\ell'\lambda',\ell \lambda}^J(p,k) = \frac{1}{2} \int_{-1}^{+1} \!\diff\cos\theta_{pk} \, d_{\lambda\lambda'}^{(J)}(\theta_{pk}) \, \Mc_{\ell'\lambda',\ell \lambda}(\p,\k) \, .
\end{align}

The helicity partial wave amplitudes do not possess definite parity~\cite{Jacob:1959at}, and we must take appropriate linear combinations to recover definite parity amplitudes. In the following section, we construct definite parity amplitudes by connecting to the spin-orbit basis.
\footnote{One could of course define a partial wave projection directly into the spin-orbit basis without going through the helicity basis first. However, since our goal is partial wave projection the OPE contribution to the three-body amplitude, we find it more convenient to first project it into the helicity basis, and then form linear combinations of definite parity states. The reasoning is due to the complicated angular dependence of the OPE function, and the helicity basis allows us to easily define relations between the different reference frames which impact the OPE definition, which will be detailed in Sec.~\ref{sec:ope}.}

\subsection{Spin-Orbit Projection}
\label{sec:spin_orbit}

Spin-orbit amplitudes are those of definite spatial parity. These amplitudes are important to construct for the spectroscopy as hadrons appear as resonant states of amplitudes, and hadrons have definite spin-parity $J^P$. Given the helicity partial wave projections for the $\3\to\3$ amplitude in Sec.~\ref{sec:helicity}, we can easily construct amplitudes of definite parity by taking appropriate linear combinations. We use the fact that the amplitudes $\Mc_{\ell'\lambda',\ell\lambda}$ can be interpreted as a quasi-two-body amplitude where one particle has a helicity. We can therefore use standard techniques~\cite{Jacob:1959at} from the partial wave projection of a two-body helicity state to spin-orbit state to obtain 
\begin{align}
	\ket{P,k,^{2S+1}\!L_J, m_J,\ell} \propto \sum_{\lambda} \ket{P,k,J m_J,\ell \lambda} \, \Pc_{\lambda}^{(\ell)}(^{2S+1}L_J) \, , \nn 
\end{align}
which when applied to our helicity amplitude yields
\begin{align}
    \label{eq:LS_amp_recouple}
	\Mc_{L'S',LS}^{(\ell'\ell),J}(p,k) = \sum_{\lambda',\lambda} \Pc_{\lambda'}^{(\ell')}( ^{2S'+1}L'_J ) \, \Mc_{\ell'\lambda',\ell\lambda}^{J}(p,k) \, \Pc_{\lambda}^{(\ell)}( ^{2S+1}L_{J} ) \, .
\end{align}
Here $S$ is the total intrinsic spin of the pair-spectator system, $L$ is the orbital angular momentum between an initial state pair and its spectator in their total CM frame, and $S'$ and $L'$ are similarly defined for the final state. The total angular momentum $J$ of the three-body system therefore has values $\lvert L-S \rvert \le J \le L+S$ and $\lvert L'-S' \rvert \le J \le L'+S'$. The parity of the three-particle state with a total angular momentum $J$ is 
\begin{align}
	P = \eta (-1)^{S + L} = \eta (-1)^{S' + L'} \, , \nn 
\end{align}
where $\eta$ is the product of the intrinsic parities of the three particles, \eg for three pseudoscalar pions the product of intrinsic parities is $\eta = (-1)^3 = -1$. Since the strong interaction conserves parity, only transitions where $S+L$ and $S'+L'$ are both even or both odd are allowed.

To couple the helicity basis to the spin-orbit basis, we have introduced the \emph{spin-orbit coupling coefficients} $\Pc_{\lambda}^{(\ell)}$, which are defined in terms of Clebsch-Gordan coefficients as
\begin{align}
    \label{eq:spin_orbit_P}
	\Pc_{\lambda}^{(\ell)}( ^{2S+1}L_J) = \sqrt{\frac{2L+1}{2J+1}} \, \braket{J\lambda | L 0 , S \lambda} \, \delta_{\ell S} \, .
\end{align}
The Kronecker delta enforces that the total spin is equal to that of the pair, $S = \ell$, as expected for our three spinless particles.
\footnote{In anticipation of extensions for external particles with spin we define the spin-orbit coupling coefficients with the redundant $\ell = S$, which in the case for particles with spin the Kronecker delta will be replaced with an additional Clebsch-Gordan coefficient which couples the pair and spectator spins to total $S$.}
From the completeness relation of the Clebsch-Gordan coefficients, one immediately sees that the spin-orbit couplings satisfy
\begin{align}
	\sum_{\lambda} \Pc_{\lambda}^{(\ell)}( ^{2S+1}L'_J)\Pc_{\lambda}^{(\ell)}( ^{2S+1}L_J) = \delta_{L'L} \, . 
\end{align}
The spin-orbit amplitudes describe the transition $^{2S+1}L_J\, \to \, ^{2S'+1}L'_{J}$ for the quasi-two-body reaction $\xi_k^{(\ell)} + \varphi_k \to \xi_p^{(\ell')} + \varphi_p$. Therefore, for fixed $\sigma_k$ and $\sigma_p$, the amplitudes have the usual threshold behavior from orbital angular momentum barrier suppression,
\begin{align}
\label{eq:MLpSpLS_thr}
	\Mc_{L'S',LS}^{(\ell'\ell),J}(p,k) \sim p^{L'}k^{L} \, , \nn 
\end{align}
as $p,k \to 0$ for fixed $\sigma_k,\sigma_p$.

As an example of the form of spin-orbit coupling coefficients, let us consider a system of three pions. For a pair of pions in relative $S$ wave, then $\ell = 0$ and the only allowed $L$ is $L = J$. So, the spin-orbit coefficient is simply
\begin{align}
	\Pc_{\lambda}^{(0)}(^{1}J_J) = \delta_{\lambda 0 } \, .
\end{align}
If the pair of pions is in an angular momentum state $\ell = 1$, \ie the resonant $P$ wave channel, then the pair-spectator system is then a triple state with $S = 1$. For some target total angular momentum $J$, the allowed orbital angular momenta are $L = J-1, J, J+1$. Therefore, the spin-orbit coefficients can be simplified to the form
\begin{align}
	\Pc_{\lambda}^{(1)}(^{3}L_J) = 
	\begin{cases}		
		\, \sqrt{\dfrac{J+1}{2(2J+1)}} \, \lvert \lambda \rvert + \sqrt{\dfrac{J}{2J+1}} \, \delta_{\lambda,0} \, ,  & L = J-1 \, , \\[20pt]
		\, -\dfrac{\lambda}{\sqrt{2}}  \, ,  & L = J \, , \\[20pt]
		\, \sqrt{\dfrac{J}{2(2J+1)}} \, \lvert \lambda \rvert - \sqrt{\dfrac{J+1}{2J+1}} \, \delta_{\lambda,0} \, , & L = J+1 \, .
	\end{cases}
\end{align}
Since the pion is a pseudoscalar, the product of the intrinsic parities is $\eta = -1$, and therefore the total parity of the system is $P = (-1)^{L}$. For a target $J^P = 1^+$ and the initial and final pairs both being vectors $\ell = \ell' = 1$, then only $S$ and $D$ waves contribute giving an two-dimensional amplitude with $^3S_1$ and $^3D_1$.

\section{One Particle Exchange Amplitude}
\label{sec:ope}

We construct analytic representations of the $\3\to\3$ scattering amplitude by enforcing $S$ matrix unitarity on Eq.~\eqref{eq:M3_def}. One can show that a driving kinematic singularity of the amplitude is due to the exchange of an on-shell particle with mass $m_e$ and momentum $P - k - p$ between two-body sub-processes~\cite{Jackura:2022gib}. The imaginary part of the $\3\to\3$ amplitude at this kinematic point, specifically for the $\k$ and $\p$ spectators, is
\begin{align}
	\im \, \Mc(\p,\b;\k,\a) \supset \Mc_{2}^{*}(\sigma_p;\bh{b}_p^{\star},\bh{k}_p^{\star}) \, \pi \, \delta(u_{pk} - m_e^2) \, \Mc_{2}(\sigma_k;\bh{p}_k^{\star},\bh{a}_k^{\star}) \, \nn ,
\end{align}
where the angles $\bh{k}_p^{\star}$ and $\bh{p}_k^{\star}$ correspond to the orientations of the spectator in the rest frame of the \emph{opposite} pair indicated, \eg $\bh{k}_p^{\star}$ is the unit vector of the initial spectator defined in the final pair rest frame defined by its spectator $\p$.
\footnote{Since the OPE involves pair-spectator systems in both its external and intermediate states, the thresholds for the pair invariant masses extend to the cases $\sigma_k^{(\mathrm{thr.})} = \max(m_a + m_{a'}, m_p + m_e)$ and $\sigma_p^{(\mathrm{thr.})} = \max(m_{b} + m_{b'}, m_k + m_e)$ for the initial and final pair, respectively.
}
We also defined
\begin{align}
    u_{pk} \equiv (P-k-p)^2 \, , \nn 
\end{align}
as the momentum-squared of the exchanged particle. We focus only on the $k$ and $p$ spectators here, but note that other spectator combinations will result in similar contributions to the imaginary part.

\begin{figure}[t]
	\centering
	\includegraphics[width=0.65\textwidth]{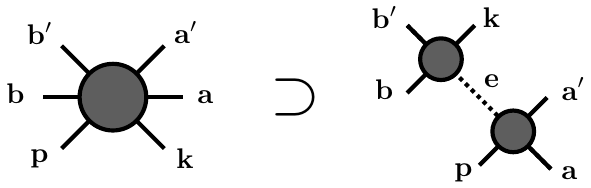}
	\caption{The OPE contribution to the on-shell $\3\to\3$ amplitude $i\Mc$ with momentum assignments. The dashed line indicates that we have removed short-distance contributions of the exchange propagator. The $\2\to\2$ subprocesses $i\overbar{\Mc}_2$ are denoted by the grey-filled circles on either side of the exchange propagator.}
	\label{fig:ope}
\end{figure}

The aforementioned pole singularity of the $\3\to\3$ amplitude is encoded in the OPE. Defining the momenta of the initial and final spectators respectively to be $\k$ and $\p$, the OPE, depicted diagrammatically in Figure~\ref{fig:ope}, can in general be written as
\footnote{Equation~\eqref{eq:ope_def} can be argued by constructing on-shell representations through either $S$ matrix unitarity~\cite{Mai:2017vot,Jackura:2018xnx,Jackura:2022gib} or summing Feynman graphs to all-orders within some generalized effective field theory and projecting intermediate states on their mass-shell~\cite{Hansen:2014eka,Hansen:2015zga}. As with all on-shell representations, the OPE is defined up to some real part in the physical region which is absorbed into the global $K$ matrix which desribes short-distance three-body dynamics. For example, in the resulting integral equations of the aforementioned references, one usually includes a cutoff function to render the momentum integrals UV finite. Since our focus here is on the partial wave projection of the function, we omit the cutoff function for convenience.}
%
\begin{align}
	\label{eq:ope_def}
	i\Mc(\p,\b;\k,\a) & \supset i\overbar{\Mc}_{2}(\sigma_p;\bh{b}_p^{\star},\bh{k}_p^{\star}) \, \frac{i}{u_{pk} - m_e^2 + i\epsilon} \, i\overbar{\Mc}_{2}(\sigma_k;\bh{p}_k^{\star},\bh{a}_k^{\star})  \equiv i\Mc_{\mathrm{OPE}}(\p,\b;\k,\a) \, .
\end{align}
On either side of the exchange propagator is a modified $\2\to\2$ amplitude, $\overbar{\Mc}_2$. The modification chosen, which is not unique, assures that $\overbar{\Mc}_2$ agrees with ${\Mc}_2$ in the limit that the exchanged particle goes on shell, while assuring that no unphysical kinematic singularities are introduced. Explicitly, these amplitudes are defined through the following angular momentum expansion,
\begin{subequations}
	\begin{align}
		\label{eq:Mpartial1}
		\overbar{\Mc}_{2}(\sigma_k;\bh{p}_k^{\star},\bh{a}_k^{\star}) = 4\pi \sum_{\ell', \lambda'} \sum_{\ell, \lambda} \left( \frac{p_k^{\star}}{q_k^{\star}} \right)^{\ell'}\,Y_{\ell' \lambda'}(\bh{p}_k^{\star}) \, \Mc_{2, \ell'\lambda';\ell \lambda}(\sigma_k) \, Y_{\ell \lambda}^{*}(\bh{a}_k^{\star}) \, , \\[5pt]
		\label{eq:Mpartial2}
		\overbar{\Mc}_{2}(\sigma_p;\bh{b}_k^{\star}, \bh{k}_p^{\star}) = 4\pi \sum_{\ell', \lambda'} \sum_{\ell, \lambda} Y_{\ell' \lambda'}(\bh{b}_p^{\star}) \, \Mc_{2,\ell'\lambda';\ell \lambda}(\sigma_p) \, \left( \frac{k_p^{\star}}{q_p^{\star}} \right)^{\ell} \, Y_{\ell \lambda}^{*}(\bh{k}_p^{\star}) \, .
	\end{align}
\end{subequations}
Angular momentum barrier factors are included to suppress the kinematic divergence induced by the spherical harmonics as $p_k^{\star}$ and $k_p^{\star}$ go to zero in their respective amplitudes. These momenta are defined in the CM frame of the pair of the \emph{opposite} spectator, specifically one can show
\begin{align}
	p_{k}^{\star} = \frac{1}{2 \sqrt{ \sigma_{k}}} \, \lambda^{1/2} ( \sigma_{k},  m_p^2, u_{pk}), \qquad k_{p}^{\star} = \frac{1}{2\sqrt{ \sigma_{p}}}  \, \lambda^{1/2} (\sigma_{p}, m_k^2, u_{pk}) \, .
\end{align}
The barrier factors are chosen to be unity at the on-shell point $u_{pk} = m_e^2$, where we define the momenta
\begin{align}
    \label{eq:pair_q}
	q_{k}^{\star} & \equiv p_{k}^{\star} \, \Big\rvert_{u_{pk} = m_e^2} = \frac{1}{2 \sqrt{ \sigma_{k}}} \, \lambda^{1/2} ( \sigma_{k}, m_p^2, m_e^2 ) \, ,  \nn \\[5pt] 
	q_{p}^{\star} & \equiv k_{p}^{\star} \, \Big\rvert_{u_{pk} = m_e^2} = \frac{1}{2\sqrt{ \sigma_{p}}}  \, \lambda^{1/2} (\sigma_{p}, m_k^2, m_e^2 ) \, ,
\end{align}
Finally note that rotational invariance of the two-body subsystems diagonalize their respective $\2\to\2$ partial wave amplitude $\Mc_{2,\ell'\lambda';\ell \lambda}(\sigma) = \delta_{\ell'\ell} \, \delta_{\lambda'\lambda} \, \Mc_{2,\ell}(\sigma)$.

\subsection{Exchange Propagator}
\label{sec:exchange}

Given the OPE amplitude defined in Eq.~\eqref{eq:ope_def}, we manipulate it to be amenable for an analytic partial wave projection to total angular momentum $J$. This means isolating the dependence on the total scattering angle $\theta_{pk}$. Using the on-shell representation defined in Eq.~\eqref{eq:ope_def} with Eqs.~\eqref{eq:Mpartial1} and \eqref{eq:Mpartial2}, we write the OPE amplitude as
\begin{align}
	i\Mc_{\mathrm{OPE}} = 4\pi \sum_{\ell',\lambda'}\sum_{\ell,\lambda} Y_{\ell'\lambda'}(\bh{b}_p^{\star}) \, i\Mc_{\ell'}(\sigma_p) \, i\Gc_{\ell'\lambda', \ell \lambda}(\p,\k) \, i\Mc_{\ell}(\sigma_k) \, Y_{\ell \lambda}^{*}(\bh{a}_k^{\star}) \, ,
\end{align}
which in effect performs the first partial wave expansion on the initial and final state pairs as given in Eq.~\eqref{eq:first_pw}. Here we define the kinematic exchange propagator $\Gc$ as
\begin{align}
	\label{eq:ope}
    \Gc_{\ell'\lambda',\ell\lambda}(\p,\k) \equiv \frac{\Hc_{\lambda'\lambda}^{(\ell'\ell)}(\p,\k)}{u_{pk} - m_e^2 + i\epsilon}  \,  \, ,
\end{align}
where the $\epsilon \to 0^+$ limit is understood, and $\Hc$ is the spin-dependent numerator
\footnote{We emphasize that we use the regular spherical harmonics as opposed to the \emph{real} harmonics originally used in the original derivation using the finite-volume framework~\cite{Hansen:2014eka,Hansen:2015zga}, which are simply unitary transformations of the regular spherical harmonics, $\mathbf{Y}_{\ell} = \mathbf{U}_{\ell} \cdot \mathbf{S}_{\ell}$ with $\mathbf{S}_{\ell}$ denoting the real spherical harmonics of degree $\ell$.}
which we define as the \emph{spin-helicity matrix},
\begin{align}
	\label{eq:spin_helicity}
	\Hc_{\lambda'\lambda}^{(\ell'\ell)}(\p,\k) \equiv \left(\frac{k_p^{\star}}{q_p^{\star}}\right)^{\ell'} 4\pi \, Y_{\ell'\lambda'}^{*}(\bh{\k}_p^{\star}) Y_{\ell\lambda}(\bh{\p}_k^{\star}) \left(\frac{p_k^{\star}}{q_k^{\star}}\right)^{\ell} \, .
\end{align}
From the properties of the spherical harmonics, the spin-helicity matrix, obeys the reflection property
\begin{align}
	\label{eq:spin_helicity_reflection}
	\Hc_{\lambda'\lambda}^{(\ell' \ell)}(\p,\k) & = (-1)^{\lambda'+\lambda} \, \Hc_{-\lambda' -\lambda}^{(\ell' \ell) \, *}(\p,\k)\, .
\end{align}

In order to analytically perform the partial wave projection, we manipulate the exchange propagator~\eqref{eq:ope} into a form to make explicit the dependence of the angular variable $\theta_{pk}$. We therefore need to express Eqs.~\eqref{eq:ope} and \eqref{eq:spin_helicity} with respect to our reaction plane defined in the space-fixed coordinate system illustrated in Fig.~\ref{fig:reaction_plane}. For convenience we define $z_{pk}$ as the cosine of the scattering angle,
\begin{align}
	z_{pk} \equiv \cos\theta_{pk} \, , \nn 
\end{align}
and work with $z_{pk}$. Upon inspection of the propagator of Eq.~\eqref{eq:ope}, we find that the $z_{pk}$-dependence will reside in the pole term through $u_{pk} = (P-k-p)^2$, and through the arguments of the spherical harmonics which are related to the scattering angle by Lorentz transformations. The dependence on $z_{pk}$ leads to singular behavior in both when the propagator goes on the mass shell and through kinematic factors associated with the spin of the pairs. In the following, we derive a generic form for the OPE which identifies the angular dependence including the isolation of the singular behavior of the function on $z_{pk}$.

The OPE is a $u$-channel process in the effective $\xi_k^{(\ell)}(P_k,\lambda) + \varphi_k(k) \to \xi_p^{(\ell')}(P_p,\lambda') + \varphi_p(p)$ reaction. The invariant momentum transfer is related to the cosine of the scattering angle in the usual way, 
\begin{align}
	u_{pk} & = (P-k-p)^2 \, , \nn \\[5pt]
    & = \sigma_k+ m_p^2 - 2E_k\omega_p - 2pk \, z_{pk} \, , \nn \\[5pt]
	& \equiv u_{pk}^{(0)} - 2pk ( 1 + z_{pk}) \, ,
\end{align}
where $E_k = \sqrt{s} - \omega_k$, and $u_{pk}^{(0)} \equiv \sigma_k+ m_p^2 - 2E_k\omega_p + 2pk$ is the backward limit ($z_{pk} = -1$) of $u_{pk}$. The value of $z_{pk}$ at the on-shell point, $u_{pk} = m_e^2$, is given a special notation,
\begin{align}
	\label{eq:ope_arg}
	\zeta_{pk} & \equiv z_{pk} \, \Big\rvert_{u_{pk} = m_e^2} \, , \nn \\[5pt]
    & = -1 - \frac{m_e^2 - u^{(0)}_{pk}}{2pk} \, \nn \\[6pt]
	& =  \frac{2s(\sigma_{k} + m_p^2 - m_e^2) - (s + \sigma_{k} - m_k^2)(s + m_p^2 - \sigma_{p})}{\lambda^{1/2}(s,\sigma_{k},m_k^2) \lambda^{1/2}(s,\sigma_{p},m_p^2)} \, ,
\end{align}
where we have used, along with $k$ and $p$ defined in Eq.~\eqref{eq:pk_cm}, the relations
\begin{align}
	E_k = \sqrt{s} - \omega_{k} = \frac{1}{2\sqrt{s} } (s + \sigma_{k} - m_k^2), \qquad \omega_{p} = \frac{1}{2\sqrt{s} } (s + m_p^2 - \sigma_{p} ),
\end{align}
which follow from the definition of $s$ in the total CM frame. Note that we have not explicitly written the $+i\epsilon$ shift which avoids the pole. However, one can include this shift by either substituting $m_e^2 \to m_e^2 - i\epsilon$ or $\zeta_{pk} \to \zeta_{pk} + i\epsilon$. The OPE pole of Eq.~\eqref{eq:ope} in terms of $z_{pk}$ is then $u_{pk} - m_e^2 = 2pk( \zeta_{pk} - z_{pk} )$.

\begin{figure*}[t!]
	\centering
	\includegraphics[width=1.0\textwidth]{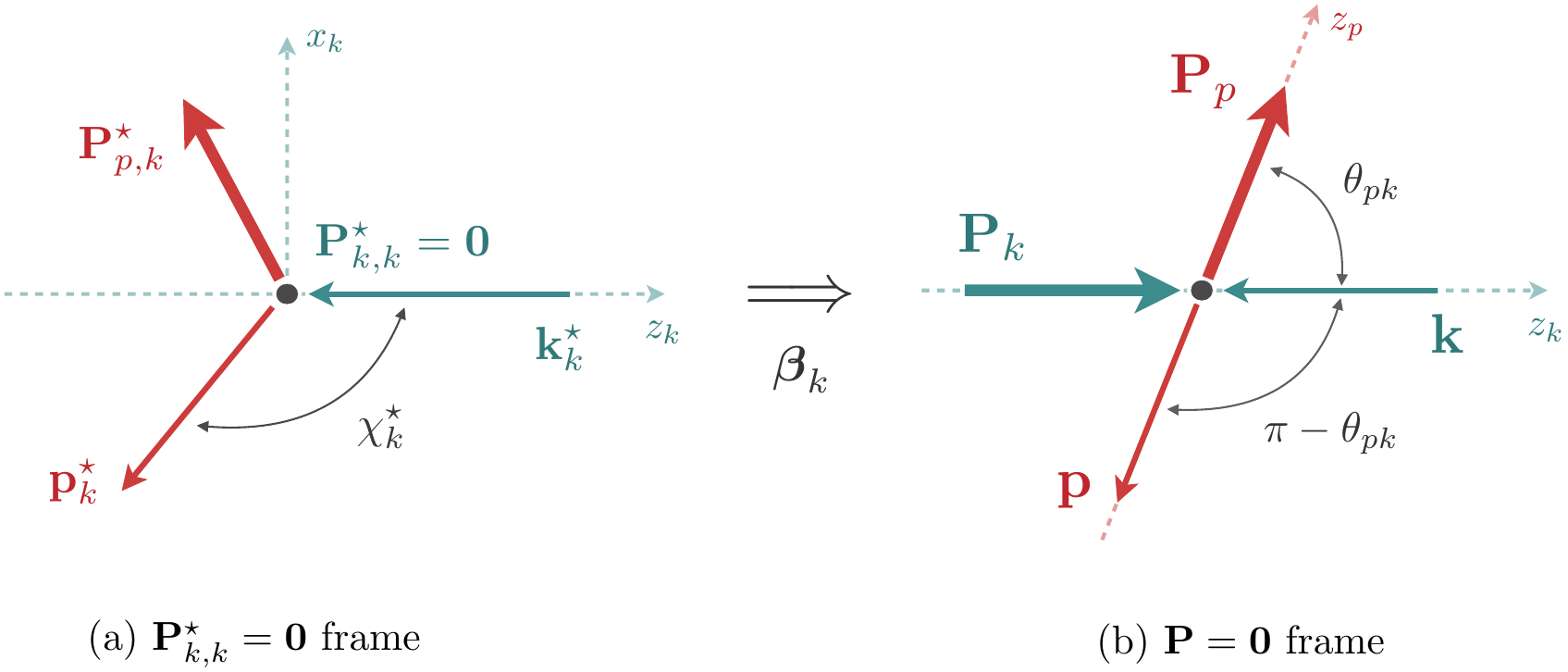}
	\caption{Kinematics of the OPE in (a) the initial state pair rest frame ($\P_{k,k}^{\star} = \0$) and (b) the total CM frame ($\P = \0$). The boost velocity from the $\P_{k,k}^{\star} = \0$ frame to the $\P = \0$ frame is $\bs{\beta}_k = \P_{k}/E_k = (\P - \k) / (\sqrt{s} - \omega_k)$.}
	\label{fig:ope_frames_initial_pair_rest}
\end{figure*}

\begin{figure*}[t!]
	\centering
	\includegraphics[width=1.0\textwidth]{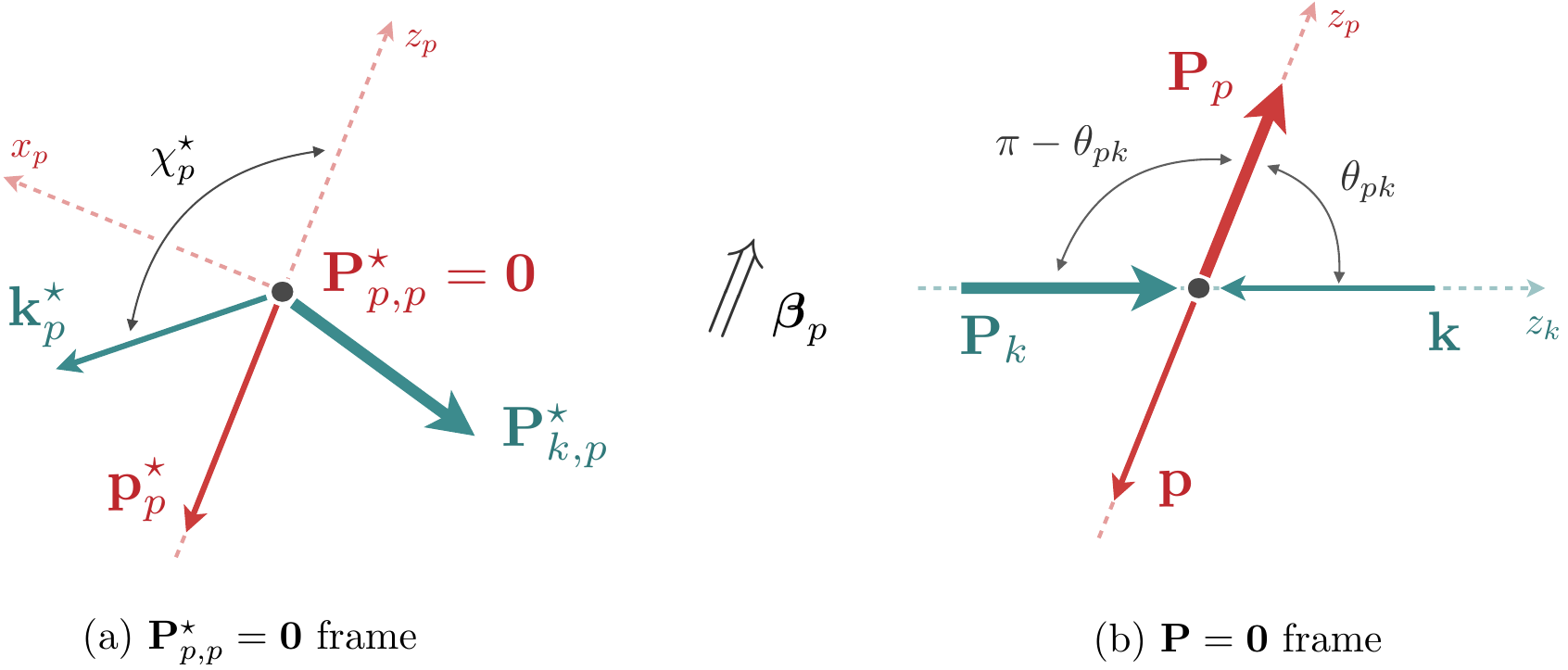}
	\caption{Similar to Fig.~\ref{fig:ope_frames_initial_pair_rest} but for the (a) final state pair frame ($\P_{p,p}^\star = \0$) where the boost to the (b) total CM frame is given by $\bs{\beta}_p = \P_p / E_p = (\P-\p) / (\sqrt{s} - \omega_p)$.}
	\label{fig:ope_frames_final_pair_rest}
\end{figure*}

For the $z_{pk}$-dependence in the spin-helicity matrix, Eq.~\eqref{eq:spin_helicity}, we make use of the Lorentz transformations between the total CM frame ($\P = \0$) where $\theta_{pk}$ is defined, and the pair CM frames where the orientations of $\k_p^{\star}$ and $\p_k^{\star}$ are defined. These Lorentz transformations are illustrated in Figs.~\ref{fig:ope_frames_initial_pair_rest} and \ref{fig:ope_frames_final_pair_rest} for the initial pair and final pair rest frames, respectively. Recall that in the reaction plane, \ie the $XZ$-plane, the pair momenta have zero azimuthal angle in the CM frame. 

Focusing first on the initial pair rest frame where $\P_{k,k}^{\star} = (\P-\k)_k^{\star} = \0$, as illustrated in Fig.~\ref{fig:ope_frames_initial_pair_rest} (a), we define $\chi_k^{\star}$ as the polar angle of $\p_k^{\star}$ in the initial pair rest frame, $\cos\chi_k^{\star} = \p_k^{\star}\cdot \bh{z}_k = - \p_k^{\star} \cdot \k_k^{\star}$. With respect to the external coordinate system, the azimuthal angle of $\p_k^\star$ is $\pi$ since the vector is oriented with respect to the negative $x_k$-axis. Therefore, the orientation of $\p_k^\star$ is given by the angles $(\chi_k^\star,\pi)$. In the final pair rest frame $\P_{p,p}^\star = (\P - \p)_p^\star = \0$ as shown in Fig.~\ref{fig:ope_frames_final_pair_rest} (a), the polar angle of $\k_p^{\star}$ is $\chi_p^\star$ which is defined as $\cos\chi_p^{\star} = \k_p^{\star}\cdot \bh{z}_p = - \k_p^{\star} \cdot \p_p^{\star}$. With respect to our coordinate system, the azimuthal angle is zero, thus the polar and azimuthal angles of $\k_p^\star$ in this frame are $(\chi_p^\star,0)$. Therefore, the angular dependence of the spin-helicity matrix is of the form
\begin{align}
    \label{eq:spin_helicity_pair_rest_frames}
    \Hc_{\lambda'\lambda}^{(\ell'\ell)}(\p,\k) & \propto  Y_{\ell'\lambda'}^{*}(\chi_p^\star,0) Y_{\ell\lambda}(\chi_k^\star,\pi) \, , \nn \\[5pt]
    & = (-1)^{\lambda} \, Y_{\ell'\lambda'}(\chi_p^\star,0) Y_{\ell\lambda}(\chi_k^\star,0) \, .
\end{align}
%

The relation between the $\chi_k^\star$ and $\chi_p^\star$ and $z_{pk}$ is found by Lorentz boosting between the frames in which these angles are defined and the total CM frame. The Lorentz boost along the $z_k$ axis from the initial pair rest frame to the total CM frame, \cf Fig.~\ref{fig:ope_frames_initial_pair_rest} (b), yields the relation
\begin{subequations}
\label{eq:k_boost}
    \begin{align}
        p_k^{\star} \cos\chi_k^{\star} & = \gamma_k \, \left[ \, p\cos(\pi - \theta_{pk}) - \omega_p \beta_k\, \right] \, , \\[5pt]
        p_k^{\star} \sin\chi_k^{\star} & = p\sin(\pi - \theta_{pk}) \, , 
    \end{align}
\end{subequations}
where $\beta_k = \lvert \P_k\rvert / E_k = k / (\sqrt{s} - \omega_k)$ is the magnitude of the boost velocity and  $\gamma_k = (1 - \beta_k^2)^{-1/2} = (\sqrt{s} - \omega_k )/ \sqrt{\sigma_k}$. Note that $\cos(\pi-\theta_{pk}) = -\cos\theta_{pk} = -z_{pk}$ and $\sin(\pi-\theta_{pk}) = \sin\theta_{pk} = \sqrt{1 - z_{pk}^2}$. A similar analysis for the initial state spectator momentum in the final state pair rest frame, $\k_p^{\star}$, yields the transformation (\cf Fig.~\ref{fig:ope_frames_final_pair_rest})
\begin{subequations}
    \label{eq:p_boost}
    \begin{align}
        k_p^{\star} \cos\chi_p^{\star} & = \gamma_p \, \left[ \, k\cos(\pi - \theta_{pk}) - \omega_k \beta_p\, \right] \, , \\[5pt]
        k_p^{\star} \sin\chi_p^{\star} & = k\sin(\pi - \theta_{pk}) \, ,
    \end{align}
\end{subequations}
where $\beta_p = \lvert \P_\p\rvert / E_p = p / (\sqrt{s} - \omega_p)$, $\gamma_p = (1 - \beta_p^2)^{-1/2} = (\sqrt{s} - \omega_p )/ \sqrt{\sigma_p}$. From Eqs.~\eqref{eq:k_boost} and \eqref{eq:p_boost}, we see that the spherical harmonics contain dependencies on $s$, $p$, $k$, as well as $z_{pk}$.

The analytic structure of the $z_{pk}$-dependence of $\Hc$ can be understood by its partial wave expansion. Since the helicity dependence of the OPE is entirely contained in $\Hc$, it admits a partial wave expansion similar to the full helicity amplitude as given in Eq~\eqref{eq:helicity_pw_expand}. We write its expansion as
\footnote{In a slight abuse of notation, we express $d_{\lambda\lambda'}^{(J)}(\theta_{pk})$ as a function of $z_{pk} = \cos\theta_{pk}$, $d_{\lambda\lambda'}^{(J)}(z_{pk})$, so that we can write everything as a function of $z_{pk}$.}
%
\begin{align}
	\label{eq:spin_helicity_pw}
	\Hc_{\lambda'\lambda}^{(\ell'\ell)}(\p,\k) = \sum_{J}(2J+1) \, \Hc_{\lambda'\lambda}^{(\ell'\ell), J}(p,k) \, d_{\lambda\lambda'}^{(J)}(z_{pk}) 
\end{align}
In the complex $z_{pk}$-plane, the spin-helicity function contains singularities associated with the Wigner $d$ functions as seen by Eq.~\eqref{eq:spin_helicity_pw}. By definition, \cf App.~\ref{sec:app.angular_momentum} and references therein, the $z_{pk}$ dependence in the Wigner $d$ matrix elements is of the form
\begin{align}
	d_{\lambda\lambda'}^{(J)}(z_{pk}) \propto \xi_{\lambda\lambda'}(z_{pk}) \, P_{\rho}^{(\mu,\nu)}(z_{pk}) \, ,
\end{align}
where $\xi_{\lambda\lambda'}$ is called the \emph{half-angle factor} and $P_{\rho}^{(\mu,\nu)}$ are the Jacobi polynomials with $\mu = \lvert \lambda - \lambda' \rvert$, $\nu = \lvert \lambda + \lambda' \rvert$, and $\rho = J - (\mu + \nu) / 2 = J - J_{\min}$. The Jacobi polynomials $P_{\rho}^{(\mu,\nu)}(z_{pk})$ are regular functions of $z_{pk}$ for all indices $\mu$, $\nu$, and $\rho$, while the half-angle factor, defined by
\begin{align}
\label{eq:xi_def}
	\xi_{\lambda\lambda'}(z_{pk}) = \left( \frac{1-z_{pk}}{2} \right)^{\frac{\lvert \lambda - \lambda' \rvert }{2} }\left( \frac{1+z_{pk}}{2} \right)^{ \frac{\lvert \lambda + \lambda' \rvert }{2} } \, ,
\end{align}
contains potentially square-root singularities depending on the system of helicities. In order to obtain an analytic representation for the OPE, we isolate the singular dependencies in $z_{pk}$ as they will impact the partial wave projection. Therefore, we conclude that the spin-helicity function for any $\ell',\ell$ has the generic structure,
\footnote{This behavior has been known from the scattering of two spinning particles, see for example Ref.~\cite{Martin:102663} and references therein.}
%
\begin{align}
	\label{eq:spin_hel_singular}
	\Hc_{\lambda'\lambda}^{(\ell'\ell)}(\p,\k) \equiv \xi_{\lambda\lambda'}(z_{pk}) \, \Ac_{\lambda'\lambda}^{(\ell'\ell)}(p,k,z_{pk}) \, ,
\end{align}
where $\Ac_{\lambda'\lambda}^{(\ell'\ell)}$ is defined to be a regular function in $z_{pk}$ for the physical kinematics. For example, consider the scattering with $\ell' = \ell = 1$, with the helicities $\lambda' = +1$, $\lambda = 1$. Then, as detailed in Sec.~\ref{sec:cases}, the spin-helicity function behaves like $\Hc_{+10}^{(11)} = \sin\chi_p^\star \cos\chi_k^\star$. This function is non-analytic in $z_{pk}$ since $\sin\chi_p^\star \propto \sin\theta_{pk} = \sqrt{1 - z_{pk}^2}$ from the Lorentz transformations Eq.~\eqref{eq:p_boost}. However, $\xi_{01}(z_{pk}) \propto \sqrt{1 - z_{pk}^2}$, thus the spin-helicity function factorizes into the non-analytic half-angle factor and a regular function in $z_{pk}$.

It is further useful to define the $\Ac$ coefficient at the on-shell point $z_{pk} = \zeta_{pk}$,
\begin{align}
	\label{eq:Acoef}
	\Ac_{\lambda'\lambda}^{(\ell'\ell)}(p,k) \equiv \Ac_{\lambda'\lambda}^{(\ell'\ell)}(p,k,z_{pk}) \, \Big\rvert_{z_{pk} = \zeta_{pk}} \, ,
\end{align}
which allows us to express the function as a term at the propagator pole and a term which is the difference of the pole and non-pole term.
\begin{align}
	\Ac_{\lambda'\lambda}^{(\ell'\ell)}(p,k,z_{pk}) & = \Ac_{\lambda'\lambda}^{(\ell'\ell)}(p,k) + \left[ \Ac_{\lambda'\lambda}^{(\ell'\ell)}(p,k,z_{pk}) - \Ac_{\lambda'\lambda}^{(\ell'\ell)}(p,k) \right] \, , 
\end{align}
Near the pole, the difference vanishes as $(\zeta_{pk} - z_{pk})$, thus it is convenient to define a new function which is regular near this pole,
\begin{align}
	\label{eq:Bcoef}
	\Bc_{\lambda'\lambda}^{(\ell'\ell)}(p,k,z_{pk}) \equiv \frac{1}{\zeta_{pk} - z_{pk}} \, \left[ \Ac_{\lambda'\lambda}^{(\ell'\ell)}(p,k,z_{pk}) - \Ac_{\lambda'\lambda}^{(\ell'\ell)}(p,k) \right]  \, .
\end{align}

The introduction of the $\Bc$ coefficient allows us to completely isolate the analytic behavior of the OPE in $z_{pk}$ into the generic form,
\begin{align}
	\label{eq:ope_manip}
	\Gc_{\ell'\lambda',\ell\lambda}(\p,\k) = \frac{1}{2pk} \xi_{\lambda'\lambda}(z_{pk}) \, \left[ \, \frac{\Ac_{\lambda'\lambda}^{(\ell'\ell)}(p,k)}{\zeta_{pk} - z_{pk}} + \Bc_{\lambda'\lambda}^{(\ell'\ell)}(p,k,z_{pk}) \, \right] \, .
\end{align}
As discussed in the beginning of this section, the singular behavior of $\Gc$ in the variable $z_{pk} = \cos\theta_{pk}$ comes from two locations. First there is the overall half-angle factor, hidden in the spin-helicity function, which exhibits kinematic singularities due to the spin of the initial and final state pairs. Second, the OPE is singular where the exchange particle goes on its mass-shell, which is encoded in the pole. The remaining $z_{pk}$ behavior is analytic in the physical region we consider.

The $\Bc$ coefficients are constructed to be regular functions of $z_{pk}$ on the interval $-1 \le z_{pk} \le +1$, thus for a fixed $\lambda'$ and $\lambda$ we can freely expand it into Legendre polynomials as
\begin{align}
	\label{eq:Bcoef_expand}
	\Bc_{\lambda'\lambda}^{(\ell'\ell)}(p,k,z_{pk}) = \sum_{j = 0}^{\infty}(2j + 1) \, \Bc_{j,\lambda'\lambda}^{(\ell'\ell)}(p,k) \, P_{j}(z_{pk}) \, ,
\end{align}
where by the orthogonality of the Legendre polynomials we can obtain the projected coefficients
\begin{align}
	\label{eq:Bcoef_proj}
	\Bc_{j,\lambda'\lambda}^{(\ell'\ell)}(p,k) = \frac{1}{2}\int_{-1}^{+1}\!\diff z_{pk} \, P_{j}(z_{pk}) \, \Bc_{\lambda'\lambda}^{(\ell'\ell)}(p,k,z_{pk}) \, .
\end{align}
We stress here that although Eq.~\eqref{eq:Bcoef_expand} is an expansion involving an infinite number of terms, for a fixed $\ell'$ and $\ell$ only a finite number of projected coefficients Eq.~\eqref{eq:Bcoef_proj} will exist for some target total angular momentum $J$. In Sec.~\ref{sec:cases}, we give explicit examples of these coefficients for $\ell',\ell = \{0, 1\}$.

Since the OPE is a known function, we can easily tabulate the $\Ac$ and $\Bc$ coefficients for the particular scattering channels of interest by the procedure outlined above. While at first, this may seem like extra computational steps given that $\Gc$ has a known form, we find that this decomposition allows us to write down a generic analytic representation for the definite-parity partial wave amplitudes of the OPE. In doing so, we arrive at an exact result that isolates all the known singular structures of the OPE in the $(\sigma_p,s,\sigma_k)$ variables, and a set of coefficients which are determined from the identified $\Ac$ and $\Bc$ coefficients.

\subsection{Partial Wave Projected Exchange Propagator}
\label{sec:exchange_projected}

The OPE as given in Eq.~\eqref{eq:ope_manip} allows for explicit analytic partial wave projection. First we project Eq.~\eqref{eq:ope_manip} into the helicity basis using Eq.~\eqref{eq:helicity_pw_proj},
\begin{align}
	\label{eq:ope_proj}
	\Gc_{\ell'\lambda',\ell\lambda}^{J}(p,k) & \equiv \frac{1}{2} \int_{-1}^{+1} \!\diff\cos\theta_{pk} \, d_{\lambda\lambda'}^{(J)}(\theta_{pk}) \, \Gc_{\ell'\lambda',\ell\lambda}(\p,\k) \, , \nn \\[5pt]
    & = \frac{1}{4pk} \, \Ac_{\lambda'\lambda}^{(\ell'\ell)}(p,k)\,\int_{-1}^{+1}\!\diff z_{pk} \, \frac{\xi_{\lambda\lambda'}(z_{pk})}{\zeta_{pk} - z_{pk} } \, d_{\lambda\lambda'}^{(J)}(z_{pk}) \,  \nn \\[5pt]
	& \qquad + \frac{1}{4pk} \, \int_{-1}^{+1}\!\diff z_{pk} \, \xi_{\lambda\lambda'}(z_{pk}) \, d_{\lambda\lambda'}^{(J)}(z_{pk}) \, \Bc_{\lambda'\lambda}^{(\ell'\ell)}(p, k, z_{pk}) \, .
\end{align}
The integral in the first term is entirely in terms of known functions independent of spectator momenta, whereas the second integral involves the $\Bc$ coefficient which depends on the momenta chosen pairs. By using the expansion~\eqref{eq:Bcoef_expand}, we can remove the momentum dependence leaving an integral over functions which depend solely on $z_{pk}$ for the second term of Eq.~\eqref{eq:ope_proj},
\begin{align}
    \label{eq:ope_proj_2nd_term}
	(\mathrm{2\ts{nd}\,\,term}) = \frac{1}{4pk}\sum_{j} (2j+1) \Bc_{j,\lambda'\lambda}^{(\ell'\ell)}(p,k) \, \int_{-1}^{+1}\! \diff z_{pk} \, \xi_{\lambda\lambda'}(z_{pk}) \, d_{\lambda\lambda'}^{(J)}(z_{pk}) \, P_{j}(z_{pk}) \, . 
\end{align}

Both of the resulting integrals can be computed analytically by recognizing that the product the product $\xi_{\lambda\lambda'}(z_{pk}) \, d_{\lambda\lambda'}^{(J)}(z_{pk})$ is a regular function of $z_{pk}$ for any $J$, $\lambda'$, and $\lambda$ since  $d_{\lambda\lambda'}^{(J)}(z_{pk}) \propto \xi_{\lambda\lambda'}(z_{pk}) \, P^{\lvert \lambda - \lambda'\rvert, \lvert \lambda+\lambda' \rvert}_{J - J_{\min}}(z_{pk})$ and the singular behavior of the half-angle factor is removed since it is squared.
\footnote{An alternative approach to evaluating the first integral is to use the rotational $e_{\lambda'\lambda}^{(J)}$ functions as described in Ref.~\cite{andrews_gunson}. However, we find the approach presented in this manuscript is `easier' on the reader as we use the more commonly known Legendre function $Q_J$.}
This allows us to perform the following expansion,
\begin{align}
	\xi_{\lambda\lambda'}(z_{pk}) \, d_{\lambda\lambda'}^{(J)}(z_{pk}) = \sum_{j=0}^{\infty} (2j+1) \, \Cc_{j,\lambda'\lambda}^{J} \, P_{j}(z_{pk}) \, ,
\end{align}
where the expansion coefficients $\Cc_{j,\lambda'\lambda}^{(J)}$ are determined by 
\begin{align}
	\label{eq:C_integral}
	\Cc_{j,\lambda'\lambda}^{J} = \frac{1}{2} \int_{-1}^{+1}\! \diff z_{pk} \, \xi_{\lambda\lambda'}(z_{pk}) \, d_{\lambda\lambda'}^{(J)}(z_{pk}) \, P_{j}(z_{pk}) \, .
\end{align}
We note here that this integral is precisely what appears in Eq.~\eqref{eq:ope_proj_2nd_term}. So, both the first and second terms of Eq.~\eqref{eq:ope_proj} are related to the $\Cc$ coefficient Eq.~\eqref{eq:C_integral}. Equation~\eqref{eq:C_integral} can be evaluated in closed form, with the result being
\begin{align}
    \label{eq:C_result}
	\Cc_{j,\lambda'\lambda}^{J} & = \frac{(-1)^{\eta}}{2J_{\min} + 1} \, \sqrt{\frac{\lvert \lambda - \lambda' \rvert! \,  \lvert \lambda + \lambda' \rvert!}{(2J_{\min})!} } \,  \braket{J_{\min}\lambda' | J\lambda' , j 0} \braket{J_{\min} \lambda | J \lambda, j 0} \, ,
\end{align}
where $\eta = 0$ if $ \lambda' \ge \lambda$ and $\eta = \lambda' - \lambda$ if $\lambda' < \lambda$, and we recall that $J_{\min} = \max(\lvert \lambda \rvert, \lvert \lambda' \rvert)$. For convenience, we provide a derivation of this result in App.~\ref{sec:app.C_integral_eval}. 

Using Eq.~\eqref{eq:C_integral}, we write the helicity partial wave projection of the OPE as
\begin{align}
    \label{eq:G_hel_not_Q}
	\Gc_{\ell'\lambda',\ell\lambda}^{J}(p,k) & = \frac{1}{4pk} \, \Ac_{\lambda'\lambda}^{(\ell'\ell)}(p,k)\, \sum_{j} (2j+1) \, \Cc_{j,\lambda'\lambda}^{J} \, \int_{-1}^{+1}\!\diff z_{pk} \, \frac{P_j(z_{pk})}{\zeta_{pk} - z_{pk} } \,  \nn \\[5pt]
	& \qquad + \frac{1}{2pk}\sum_{j} (2j+1) \, \Cc_{j,\lambda'\lambda}^{J} \, \Bc_{j,\lambda'\lambda}^{(\ell'\ell)}(p,k)  \, .
\end{align}
The final integral can be expressed in terms of the well-known Legendre functions of the 2\ts{nd} kind
\begin{align}
    \label{eq:Q_fcn}
	Q_j(\zeta_{pk}) & = \frac{1}{2} \int_{-1}^{+1} \! \diff z_{pk} \, \frac{P_j(z_{pk})}{\zeta_{pk} - z_{pk}} \, ,
\end{align}
where the analytic structure is fixed by $\zeta_{pk} \to \zeta_{pk} + i\epsilon$.
Properties and examples of the Legendre functions are given in App.~\ref{sec:app.angular_momentum}. The Legendre $Q$ functions contain branch cuts when $\zeta_{pk} = \pm 1$, which originate from the on-shell exchange $u_{pk} = m_e^2$. Since the $\Ac$ and $\Bc$ coefficients are regular in the energy variables, the $Q$ functions contain the entire non-analytic structure of the partial wave projected OPE, which results in a branch cut in the complex $s$-plane for fixed $\sigma_k$ and $\sigma_p$. The on-shell constraint $\zeta_{pk} = \pm 1$ leads to an expression for the physical boundary region of real-particle exchanges~\cite{byckling1973particle}, given by $\Phi(p,k) \ge 1$ where
\begin{align}
    \label{eq:Kibble_exch}
	\Phi(p,k) & = \sigma_k \sigma_p (s + m_k^2 + m_p^2 + m_e^2 - \sigma_k - \sigma_p) \, \nn \\[5pt]
	& \quad - \sigma_k(s - m_p^2)(m_k^2 - m_e^2) - \sigma_p(s - m_k^2)(m_p^2 - m_e^2) \, \nn \\[5pt]
	& \quad - (s m_e^2 - m_p^2 m_k^2)(s + m_e^2 - m_p^2 - m_k^2 ) \, .
\end{align}

Combining Eqs.~\eqref{eq:G_hel_not_Q}, \eqref{eq:C_integral}, and \eqref{eq:Q_fcn}, we find a compact expression for the helicity partial wave projection of the OPE as
\begin{align}
    \label{eq:OPE_helicity_pw}
	\Gc_{\ell' \lambda', \ell \lambda}^{J}(p,k)  & = \frac{1}{2pk} \, \sum_{j}(2j+1)\, \Cc_{j,\lambda'\lambda}^{J} \, \left[ \, \Ac_{\lambda'\lambda}^{(\ell'\ell)}(p,k) \, Q_j(\zeta_{pk}) + \Bc_{j,\lambda'\lambda}^{(\ell'\ell)}(p,k) \right] \, ,
\end{align}
where $\zeta_{pk} = \zeta_{pk}(p,k)$ is defined in Eq.~\eqref{eq:ope_arg}. Note that the sum is finite since the $\Cc$ coefficients are zero for $j$ outside the range $\lvert J-J_{\min} \rvert \le j \le J+J_{\min}$. The pole term multiplying the $\Ac$ coefficients results in singular behavior in the energy variables, while the $\Bc$ coefficients are regular functions of energies, thereby giving additional short-distance physics to the ones already contained in the three-body $K$ matrix of the on-shell representations as discussed in Ref.~\cite{Jackura:2022gib}.

Having this analytic representation for the helicity partial wave projection of $\Gc$, we form the appropriate linear combinations to arrive at an expression in the spin-orbit basis using Eqs.~\eqref{eq:LS_amp_recouple} and \eqref{eq:spin_orbit_P},  
\begin{align}
	\Gc_{L'S',LS}^{(\ell'\ell),J}(p,k) = \sum_{\lambda',\lambda} \Pc_{\lambda'}^{(\ell')}(^{2S'+1}L'_J) \, \Gc_{\ell' \lambda', \ell \lambda}^{J}(p,k) \, \Pc_{\lambda}^{(\ell)}(^{2S+1}L_J) \,.
\end{align}
Inserting Eq.~\eqref{eq:OPE_helicity_pw}, we find the definite parity partial wave projections of the OPE for the $\3\to\3$ scattering of external spinless particles takes the form, as a matrix in $LS$-space,
\begin{align}
	\label{eq:ope_JP_result}
	\Gc^{(\ell'\ell),J}_{L'S',LS}(p,k) = \wt{\Kc}_{\Gc,L'S',LS}^{(\ell'\ell),J}(p,k) + \sum_{j} \wt{\Tc}_{j,L'S',LS}^{(\ell'\ell),J}(p,k) \, Q_j(\zeta_{pk}) \, ,
\end{align}
where $\wt{\Kc}_{\Gc}$ is a known short-distance contribution given in terms of the $\Bc$ coefficients, whereas $\wt{\Tc}_j$ are computed from a given set of $\Ac$ coefficients. The matrix elements of $\wt{\Kc}_\Gc$ are 
\begin{align}
    \label{eq:Ktilde_coef}
	\wt{\Kc}_{\Gc,L'S',LS}^{(\ell'\ell),J}(p,k) & \equiv  \frac{1}{2pk} \, \sum_{j}(2j+1)\, \nn \\[5pt]
	& \qquad\qquad  \times \sum_{\lambda',\lambda} \Pc_{\lambda'}^{(\ell')}(^{2S'+1}L'_J) \, \Cc_{j,\lambda'\lambda}^{J} \, \Bc_{j,\lambda'\lambda}^{(\ell'\ell) }(p,k) \, \Pc_{\lambda}^{(\ell)}(^{2S+1}L_J) \, ,
\end{align}
while the matrix elements of $\wt{\Tc}_j$ are
\begin{align}
    \label{eq:Ttilde_coef}
	\wt{\Tc}_{j, L'S',LS}^{(\ell'\ell),J}(p,k) \equiv \frac{1}{2pk} \, (2j+1)\, \sum_{\lambda',\lambda} \Pc_{\lambda'}^{(\ell')}(^{2S'+1}L'_J) \, \Cc_{j,\lambda'\lambda}^{J} \, \Ac_{\lambda'\lambda}^{(\ell'\ell)}(p,k) \, \Pc_{\lambda}^{(\ell)}(^{2S+1}L_J) \, .
\end{align}
We stress here that Eq.~\eqref{eq:ope_JP_result} is a generic result for an exchange of a spinless particle between pairs with any angular momentum that couple to some total $J^P$. The matrices $\wt{\Kc}_{\Gc}$ and $\wt{\Tc}_j$ are regular functions of the energies in the physical region. 

It is important to note that Eq.~\eqref{eq:ope_JP_result} is not a unique decomposition, as the Legendre functions of the $2\ts{nd}$-kind can be written, \cf Eqs.~\eqref{eq:app.Ql_explicit} and \eqref{eq:app.Wnm1} in App.~\ref{sec:app.angular_momentum}, as
\begin{align}
    Q_j(z) = P_j(z) \, Q_0(z) - W_{j-1}(z) \, ,
\end{align}
where $P_j$ are the Legendre polynomials and $W_{j-1}(z)$ is polynomial in $z$, defined for $j>0$ by
\begin{align}
    W_{j-1}(z) = \sum_{n=1}^{j}\frac{1}{n} \, P_{n-1}(z) \, P_{j-n}(z) \, ,
\end{align}
whereas for $j=0$, $W_{-1} = 0$. This relation allows one to shift the definitions of $\wt{\Kc}_\Gc$ and $\wt{\Tc}_j$ to absorb/remove terms regular in the kinematic variables, which can further simplify expressions for the partial wave projected OPE. Exploiting this relation, we reduce our result for the partial wave OPE to one involving only $Q_0(\zeta_{pk})$,
\begin{align}
    \label{eq:ope_JP_final_result}
    \Gc_{L'S',LS}^{(\ell'\ell),J}(p,k)  = \Kc_{\Gc,L'S',LS}^{(\ell'\ell),J}(p,k) + \Tc_{L'S',LS}^{(\ell'\ell),J}(p,k) \, Q_0(\zeta_{pk})
\end{align}
which is the expression claimed in Eq.~\eqref{eq:main_eq} as our primary result. The matrix elements of $\Kc_{\Gc}$ are given by 
\begin{align}
    \label{eq:K_coef}
    \Kc_{\Gc,L'S',LS}^{(\ell'\ell),J}(p,k) & = \wt{\Kc}_{\Gc,L'S',LS}^{(\ell'\ell),J}(p,k) - \sum_{j} \wt{\Tc}_{j,L'S',LS}^{(\ell'\ell),J}(p,k) \, W_{j-1}(\zeta_{pk}) \, , \nn \\[5pt]
    & =  \frac{1}{2pk} \, \sum_{j}(2j+1)\,\sum_{\lambda',\lambda} \Pc_{\lambda'}^{(\ell')}(^{2S'+1}L'_J) \, \Cc_{j,\lambda'\lambda}^{J} \,  \nn \\[5pt]
	& \qquad\qquad  \times \left[ \, \Bc_{j,\lambda'\lambda}^{(\ell'\ell) }(p,k) - W_{j-1}\, \Ac_{j,\lambda'\lambda}^{(\ell'\ell) }(p,k) \, \right] \, \Pc_{\lambda}^{(\ell)}(^{2S+1}L_J) \, ,
\end{align}
and the $\Tc$ coefficient is the sum over the $\wt{\Tc}_j$ coefficients weighted by Legendre polynomials,
\begin{align}
    \label{eq:T_coef}
    \Tc_{L'S',LS}^{(\ell'\ell),J}(p,k) & \equiv \sum_{j} \wt{\Tc}_{j,L'S',LS}^{(\ell'\ell),J}(p,k) \, P_j(\zeta_{pk}) \, , \nn \\[5pt]
    & =  \frac{1}{2pk} \, \sum_{j}(2j+1) P_{j}(\zeta_{pk}) \, \nn \\[5pt]
	& \qquad\qquad  \times  \sum_{\lambda',\lambda} \Pc_{\lambda'}^{(\ell')}(^{2S'+1}L'_J) \, \Cc_{j,\lambda'\lambda}^{J} \, \Ac_{\lambda'\lambda}^{(\ell'\ell)}(p,k) \, \Pc_{\lambda}^{(\ell)}(^{2S+1}L_J)
\end{align}

Equation~\eqref{eq:ope_JP_final_result}, along with \eqref{eq:K_coef} and \eqref{eq:T_coef}, are the main result of this work. The only remaining task for the user is to construct the $\Kc_\Gc$ and $\Tc$ matrices by identifying the appropriate $\Ac$ and $\Bc$ coefficients. In the next section, we illustrate the procedure for some low-spin cases.

\section{Case Studies}
\label{sec:cases}

In this section we examine the consequence of our main result, Eq.~\eqref{eq:ope_JP_result}, for three cases of low-spin systems: pairs with $\ell' = \ell = 0$, $\ell' = \ell = 1$, and the transition process $\ell' = 1, \ell = 0$. Any higher spin system can be found by following the procedure outlined in this section. This procedure is easily amenable to symbolic computation with software such as \texttt{Mathematica}, and such a notebook with examples is supplied in the Supplemental Material.

The main tasks are to identify the $\Ac$ and $\Bc$ coefficients as defined in Eqs.~\eqref{eq:Acoef} and \eqref{eq:Bcoef}, respectively, given for each case. Once these coefficients are determined, we compute the $\wt{\Kc}_\Gc$ and $\wt{\Tc}_j$ functions, Eqs.~\eqref{eq:Ktilde_coef} and \eqref{eq:Ttilde_coef}, respectively. These matrices are then fed into Eqs.~\eqref{eq:K_coef} and \eqref{eq:T_coef} for $\Kc_\Gc$ and $\Tc$, respectively, giving the analytic projection shown in Eq.~\eqref{eq:ope_JP_final_result}.

Throughout this section, we denote $\eta = \eta_k\eta_p\eta_e$ as the product of intrinsic parities of the initial and final spectators, and the exchange particle. As our focus is on hadronic processes, we also assume parity-conserving reactions with $P = \eta(-1)^{L+S} = \eta (-1)^{L'+S'}$. Additionally, since $S'=\ell'$ and $S=\ell$ always in this work, we introduce a convenient notation for the partial wave OPE,
\begin{align}
    \Gc(^{2S'+1}L'_{J}|^{2S+1}L_J) \equiv \Gc_{L'S',LS}^{(S'S),J}(p,k) \, , \nn 
\end{align}
where the dependencies on kinematic variables are left implicit.

\subsection{Pairs with \mathbshead{\ell' = \ell = 0}{ellpell0}}
\label{sec:pair_00}
The simplest case is when both the incoming and outgoing pairs are in relative $S$ waves, so that $\ell' = \ell = 0$. The complications with the spin-helicity function are eliminated since $\lambda' = \lambda = 0$, \ie $\Hc_{\lambda'\lambda}^{(00)} = \delta_{\lambda'0}\delta_{\lambda 0}$. This in turn indicates that the coefficients are $\Ac_{\lambda'\lambda}^{(00)} = \delta_{\lambda'0}\delta_{\lambda 0}$ and $\Bc_{\lambda'\lambda}^{(00)} = 0$. Therefore, the exchange propagator~\eqref{eq:ope_manip} reduces to the simple form
\begin{align}
	\Gc_{0\lambda',0\lambda} = \frac{1}{2pk} \, \frac{\delta_{\lambda'0}\delta_{\lambda 0}}{\zeta_{pk} - z_{pk} }  \, .
\end{align}
Recoupling to spin-orbit amplitudes is also trivial as only $S = \ell = 0$ and $S' = \ell' = 0$, restricting the allowed orbital angular momentum to be $J = L = L'$. The parity of the system is then $P = \eta (-1)^{J}$.

From Eq.~\eqref{eq:Ktilde_coef}, we see that $\wt{\Kc}_{\Gc}(^1J_J|^1J_J) = 0$ because $\Bc_{\lambda'\lambda}^{(00)} = 0$ for this case. Using Eq.~\eqref{eq:Ttilde_coef} and the identities for $\Pc_{\lambda}^{(0)}(^{1}J_J) = \delta_{\lambda 0}$ and $ \, \Cc_{j,\lambda'\lambda}^{J} = \delta_{Jj} \delta_{\lambda' 0} \delta_{\lambda 0}/ (2J+1)$ found from Eqs.~\eqref{eq:spin_orbit_P} and \eqref{eq:C_result}, respectively, one finds that the partial wave OPE for a target $J^{P}$ simplifies to 
\begin{align}
\label{eq:GJ_00}
	\Gc(^1J_J | ^1J_J) = \frac{1}{2pk} \, Q_{J}(\zeta_{pk}) \, .
\end{align}
As detailed in Eqs.~\eqref{eq:ope_JP_final_result}, \eqref{eq:K_coef}, and \eqref{eq:T_coef}, we can express this amplitude in terms of only $Q_0$. Explicitly, the $J=0$, and $1$ partial wave OPE amplitudes are given by
\begin{align}
    \label{eq:1S0to1S0}
    \Gc(^1S_0 | ^1S_0) & = \frac{1}{2pk} \, Q_0(\zeta_{pk}) \, ,  \\[5pt]
    \label{eq:1P1to1P1}
    \Gc(^1P_1 | ^1P_1) & = -\frac{1}{2pk} + \frac{\zeta_{pk}}{2pk} \, Q_0(\zeta_{pk}) \, .  
\end{align}
The $J=0$ OPE agrees with the well-known result found in many works on three-body scattering processes, \eg Refs.~\cite{Jackura:2018xnx,Dawid:2020uhn,Jackura:2020bsk,Dawid:2023jrj}.

We perform a simple check of Eq.~\eqref{eq:GJ_00} by verifying it satisfies the expected behavior near threshold. As discussed in Sec.~\ref{sec:helicity}, we can recognize that $\Gc$ can be thought of as an effective two-body amplitude by interpreting $\sqrt{\sigma_k}$ and $\sqrt{\sigma_p}$ as effective masses of the external states. As a result, one would expect that such amplitude $\Gc(^1J_J|^1J_J)$ scales as $(pk)^{J}$ in the vicinity of the nearest pair-spectator threshold $\sqrt{s} \sim \sqrt{\sigma_p} + m_p \sim \sqrt{\sigma_k} + m_k$, \cf Sec.~\ref{sec:spin_orbit}.

To reproduce this behavior, we note that near this threshold both $p$ and $k$ are small, $p,k \to 0$. Fixing $\sigma_k  > \sigma_k^{(\mathrm{thr}.)} $ and $\sigma_p > \sigma_p^{(\mathrm{thr}.)}$, we see from the definition of $\zeta_{pk}$, Eq.~\eqref{eq:ope_arg}, that near threshold $\zeta_{pk}$ diverges as $1/pk$,
\begin{align}
    \zeta_{pk} = \frac{\Nc}{2pk} + \Oc\!\left( pk^{-1}, p^{-1}k \right) \, , \nn 
\end{align}
where $\Nc = (\sqrt{\sigma_k} - m_p)^2 - m_e^2$ is a positive constant. From the behavior of Legendre function for large arguments, Eq.~\ref{eq:app.legQ_asymp}, we see that near threshold $Q_{J}(\zeta_{pk}) \to (pk)^{J + 1}$. Therefore, near threshold Eq.~\eqref{eq:GJ_00} satisfies the expected behavior of $\Gc(^1J_J|^1J_J)\sim (pk)^{J}$. Explicitly, we write the threshold expansions of Eqs.~\eqref{eq:1S0to1S0} and \eqref{eq:1P1to1P1} by using the asymptotic expansion of $Q_0(\zeta_{pk})$ for $\zeta_{pk} \to \infty$ as given in Eq.~\eqref{eq:app.legQ0_asymp},
\begin{align}
    \label{eq:LegQ0_asymp}
    Q_0(\zeta_{pk}) & = \frac{1}{\zeta_{pk}} + \frac{1}{3\zeta_{pk}^3} + \Oc\!\left( \zeta_{pk}^{\,-5} \right) \, .  
\end{align}
Since $\zeta_{pk}$ appears as a reciprocal in this expansion, we can write the Taylor series for $1/\zeta_{pk}$ for small $p$ and $k$ as
\begin{align}
    \frac{1}{\zeta_{pk}} = \frac{2pk}{\Nc} + \Oc\!\left( p^2k, pk^2 \right) \, . \nn
\end{align}
Then, the explicit threshold behavior for the $^1S_0 \to ^1S_0$ and $^1P_1 \to ^1P_1$ amplitudes is
\begin{align}
    \Gc(^1S_0 | ^1S_0) &  = \frac{1}{\Nc} + \Oc\!\left(p,k\right) \, , \nn \\[5pt]
    \Gc(^1P_1 | ^1P_1) & =  \frac{2}{3\Nc^2} \, pk + \Oc\!\left(p^2k,pk^2\right) \, , \nn 
\end{align}
which is consistent with the expected behavior.

\subsection{Pairs with \mathbshead{\ell' = \ell = 1}{ellpell1}}
\label{sec:pair_11}

We turn to the case where the pairs carry non-zero angular momenta $\ell' = \ell = 1$. In this case, the spin-helicity matrix has a non-trivial structure, and we must work out the $\Ac$ and $\Bc$ coefficients. Recall that given the elements of the spin-helicity matrix $\Hc$, we can solve for the $\Ac$ and $\Bc$ coefficients using Eqs.~\eqref{eq:Acoef} and \eqref{eq:Bcoef}, respectively. Most readily, the $\Ac$ coefficient is given by first isolating $\Ac(z_{pk})$ from Eq.~\eqref{eq:spin_hel_singular},
\begin{align}
    \label{eq:A_coef_11}
	\overbar{\Ac}_{\lambda'\lambda}^{(11)}(p,k,z_{pk})  = \frac{1}{\xi_{\lambda\lambda'}(z_{pk})}  \, \overbar{\Hc}_{\lambda'\lambda}^{(11)}(\p,\k) \,  ,
\end{align}
and then setting $z_{pk} = \zeta_{pk}$. For convenience, we introduced the $\ell'=\ell = 1$ spin-helicity function $\overbar{\Hc}$ and $\overbar{\Ac}$ coefficient which has a common factor removed,
\begin{align}
	\overbar{\Hc}_{\lambda'\lambda}^{(11)} & \equiv \frac{q_p^{\star}q_k^{\star}}{3pk} \, \Hc_{\lambda'\lambda}^{(11)} \, , \nn \\[5pt]
    \overbar{\Ac}_{\lambda'\lambda}^{(11)} & \equiv \frac{q_p^{\star}q_k^{\star}}{3pk} \, \Ac_{\lambda'\lambda}^{(11)} \, . \nn 
\end{align}

When both initial and final state pairs are in relative $P$ wave, the spin structure is such that there are only five independent functions which results from the reflection property Eq.~\eqref{eq:spin_helicity_reflection}. There is an accidental symmetry relating the $\lambda', \lambda = +1,+1$ and $+1,-1$ components, yielding only four independent functions. In terms of the polar angles previously defined, we obtain for the spin-helicity function
\begin{subequations}
    \label{eq:H_11}
	\begin{align}
		\overbar{\Hc}_{\pm 1\pm 1}^{(11)} = -\overbar{\Hc}_{\pm 1\mp 1}^{(11)} & = - \frac{1}{2} \frac{k_p^{\star} p_k^{\star}}{pk} \, \sin\chi_p^{\star}\sin\chi_k^{\star} \, , \\[5pt]
		\overbar{\Hc}_{\pm 10}^{(11)} & = \mp \frac{1}{\sqrt{2}} \frac{k_p^{\star} p_k^{\star}}{pk} \, \sin\chi_p^{\star}\cos\chi_k^{\star} \, , \\[5pt]
		\overbar{\Hc}_{0 \pm 1}^{(11)} & = \pm \frac{1}{\sqrt{2}} \frac{k_p^{\star} p_k^{\star}}{pk} \, \cos\chi_p^{\star}\sin\chi_k^{\star} \, ,\\[5pt]
		\overbar{\Hc}_{00}^{(11)} & = \frac{k_p^{\star} p_k^{\star}}{pk} \, \cos\chi_p^{\star}\cos\chi_k^{\star} \, .
	\end{align}
\end{subequations}
The Lorentz transformations Eqs.~\eqref{eq:k_boost} and \eqref{eq:p_boost} relate the polar angles $\chi_k^{\star}$ and $\chi_p^{\star}$ to the total CM frame polar angle $\theta_{pk}$. Note that if we compare Eq.~\eqref{eq:H_11} to expressions in Ref.~\cite{Kamal:1965vto}, we find disagreement with respect to overall phase factors. This is due to the azimuthal angle of $\pi$ for one of the momenta as detailed in Eq.~\eqref{eq:spin_helicity_pair_rest_frames}, which was neglected by the author of Ref.~\cite{Kamal:1965vto}. One can convince themselves that this phase factor is necessary and consistent with an alternative model for the OPE which replaces the spin-helicity matrix as given in Eq.~\eqref{eq:spin_helicity} with polarization Lorentz tensors contracted with momenta, \eg $\Hc_{\lambda'\lambda}^{(11)} \propto (\varepsilon^*(P_p,\lambda') \cdot k)(\varepsilon(P_k,\lambda) \cdot p)$, which can be seen by considering an effective Lagrangian of vector field $V_\mu$ coupling to two scalars, \eg $\Lc \supset - ig_p V_p^\mu \, \varphi_e \partial_{\mu} \varphi_k - ig_k V_k^{\mu} \, \varphi_e \partial_\mu \varphi_p$ where $g_p$, $g_k$ are effective couplings to the scalar $\varphi$ fields.

Recalling that $z_{pk} \equiv \cos\theta_{pk}$ and $\sin\theta_{pk} = \sqrt{1-z_{pk}^2}$, the half-angle factor for each helicity combination is given by
\begin{align}
	\xi_{\pm1\pm1}(z_{pk}) & = \frac{1+z_{pk}}{2} \, , \nn \\[5pt]
	\xi_{\pm1\mp1}(z_{pk}) & = \frac{1-z_{pk}}{2} \, , \nn \\[5pt]
	\xi_{\pm10}(z_{pk})  = \xi_{0\pm1}(z_{pk})  & = \frac{\sqrt{1-z_{pk}^2}}{2} \, , \nn \\[5pt]
	\xi_{00}(z_{pk})  & = 1 \, . \nn
\end{align}
Combining Eqs.~\eqref{eq:A_coef_11} and \eqref{eq:H_11} and using the boost relations Eqs.~\eqref{eq:p_boost} and \eqref{eq:k_boost} gives for the $\overbar{\Ac}(p,k,z_{pk})$ coefficients
\begin{subequations}
    \label{eq:A_coef_bar}
	\begin{align}
		\overbar{\Ac}_{\pm 1 \pm 1}^{(11)} & = -1 + z_{pk} \, ,  \\[5pt]
        \overbar{\Ac}_{\pm 1 \mp 1}^{(11)} & = 1 + z_{pk} \, ,  \\[5pt]
		\overbar{\Ac}_{\pm 1 0}^{(11)} & = 
  \pm \sqrt{2} \, \gamma_k  \, \left(\frac{\beta_k \omega_p}{p} + z_{pk} \right) \, , \\[5pt]
		\overbar{\Ac}_{0\pm 1}^{(11)} & = \mp 	
  \sqrt{2}\, \gamma_p \, \left(\frac{\beta_p \omega_k}{k} + z_{pk} \right) \, ,  \\[5pt]
		\overbar{\Ac}_{00}^{(11)} & = \gamma_p\gamma_k \left(\frac{\beta_p \omega_k}{k} + z_{pk} \right)\left(\frac{\beta_k \omega_p}{p} + z_{pk} \right)  \, . 
	\end{align} 
\end{subequations}
The on-shell coefficients $\overbar{\Ac}(p,k) \equiv \overbar{\Ac}(p,k,\zeta_{pk})$ are given by Eq.~\eqref{eq:A_coef_bar} with the substitution $z_{pk} \to \zeta_{pk}$. We again define the $\overbar{\Bc}$ coefficients in terms of $\Bc$ coefficients as 
\begin{align}
    \overbar{\Bc}_{\lambda'\lambda}^{(11)} & \equiv \frac{q_p^{\star}q_k^{\star}}{3pk} \, \Bc_{\lambda'\lambda}^{(11)}\, , \nn 
\end{align}
which are related to $\overbar{\Ac}$ through Eq.~\eqref{eq:Bcoef}, 
\begin{align}
	\overbar{\Bc}_{\lambda'\lambda}^{(11)}(p,k,z_{pk}) = \frac{1}{\zeta_{pk} - z_{pk}} \, \left[\, \overbar{\Ac}_{\lambda'\lambda}^{(11)}(p,k,z_{pk}) - \overbar{\Ac}_{\lambda'\lambda}^{(11)}(p,k)   \,\right] \, , \nn 
\end{align}
Evaluating the difference and dividing by the pole gives $\overbar{\Bc}(p,k,z_{pk})$
\begin{subequations}
    \label{eq:B_coef_bar}
	\begin{align}
        \overbar{\Bc}_{\pm 1 \pm 1}^{(11)} & = -1 \, , \\[5pt]
		\overbar{\Bc}_{\pm 1 \mp 1}^{(11)}  & = -1  \, ,  \\[5pt]
		\overbar{\Bc}_{\pm 10}^{(11)} & = \mp \sqrt{2}\, \gamma_k  \, ,  \\[5pt]
		\overbar{\Bc}_{0\pm 1}^{(11)} & = \pm \sqrt{2} \, \gamma_p  \, ,  \\[5pt]
		\overbar{\Bc}_{00}^{(11)} & = - \gamma_p\gamma_k \left(\frac{\beta_p \omega_k}{k} + \frac{\beta_k \omega_p}{p} + z_{pk} + \zeta_{pk} \right)  \, .  
	\end{align}
\end{subequations} 
Finally, we require the projected $\Bc$ coefficients defined by Eq.~\eqref{eq:Bcoef_proj}, repeated here for convenience,
\begin{align}
	\overbar{\Bc}_{j,\lambda'\lambda}^{(11)}(p,k) = \frac{1}{2}\int_{-1}^{+1}\!\diff z_{pk} \, P_{j}(z_{pk}) \, \overbar{\Bc}_{\lambda'\lambda}^{(11)}(p,k,z_{pk}) \, . \nn
\end{align}
Upon substituting Eq.~\eqref{eq:B_coef_bar} we find only two non-zero terms in the expansion 
\begin{subequations}
    \begin{align}
    	\overbar{\Bc}_{0,\lambda'\lambda}^{(11)} & = -|\lambda\,\lambda'| \, 
        -\sqrt{2} \gamma_k \, \lambda'\, \delta_{\lambda ,0} 
        +\sqrt{2} \gamma_p \,\lambda\, \delta_{\lambda' ,0}   \,  \nn \\[5pt]
    	&\qquad - \gamma_p\gamma_k \left(\frac{\beta_p \omega_k}{k} + \frac{\beta_k \omega_p}{p} +  \zeta_{pk} \right)  \, \delta_{\lambda',0}\delta_{\lambda ,0} \, , \\[5pt]
    	\overbar{\Bc}_{1,\lambda'\lambda}^{(11)} &= -\frac{1}{3} \, \gamma_p\gamma_k \, \delta_{\lambda',0}\delta_{\lambda ,0}  \, , \\[5pt]
    	\overbar{\Bc}_{j,\lambda'\lambda}^{(11)} &= 0 \, , \quad \mathrm{for} \,\, j > 1 \, . 
    \end{align}
\end{subequations}

Having found the $\Ac$ and $\Bc$ coefficients for $\ell'=\ell=1$, we now construct $\Kc_\Gc$ and $\Tc$ matrices for some target $J^P$. Trivially $S' = S = 1$, therefore if $J = 0$ only $L'=L=1$ contributes, while for $J > 0$ the allowed orbital angular momenta are $L = J-1, J, J+1$ and $L' = J-1, J, J+1$ for the initial and final states, respectively.

\subsubsection{Total \mathbshead{J=0}{J0}}

Let us first consider a target $J = 0$, where only $L' = L = 1$ is allowed and the corresponding parity is $\eta$. Thus, we need only compute a single $^3P_0$ amplitude with the spin-orbit recoupling from Eq.~\eqref{eq:spin_orbit_P} being $\Pc_\lambda^{(1)}(^3P_0) = - \delta_{\lambda,0}$. Evaluating the expression for $\wt{\Kc}_\Gc$ and $\wt{\Tc}_j$ using  Eqs.~\eqref{eq:Ktilde_coef} and \eqref{eq:Ttilde_coef} respectively, we find
\begin{align}
    \wt{\Kc}_{\Gc}(^3P_0 | ^3P_0) &= -\frac{3\gamma_k \gamma_p}{2 q_k^{\star} q_p^{\star}} \left( \frac{\beta_p \omega_k}{k}   + \frac{\beta_k \omega_p}{p}  + \zeta_{pk} \right) \, , \nn \\[5pt]
    \wt{\Tc}_{j}(^3P_0 | ^3P_0) &=\delta_{j0} \, \frac{3\gamma_k \gamma_p}{2  q_k^{\star} q_p^{\star}} \left(\frac{\beta_p \omega_k}{k} + \zeta_{pk} \right)\left(\frac{\beta_k \omega_p}{p} + \zeta_{pk} \right) \, . \nn 
\end{align}
Since only $j=0$ contributes for this case, feeding these matrices into Eqs.~\eqref{eq:K_coef} and \eqref{eq:T_coef} gives trivially $\Kc_\Gc = \wt{\Kc}_\Gc$ and $\Tc = \wt{\Tc}_0$. Adding these contributions together as dictated by Eq.~\eqref{eq:ope_JP_final_result}, one finds 
\begin{align}
\label{eq:G3P0}
	\Gc(^3P_0 | ^3P_0) &= 
 -\frac{3\gamma_k \gamma_p}{2 q_k^{\star} q_p^{\star}} \left( \frac{\beta_p \omega_k}{k}   + \frac{\beta_k \omega_p}{p}  + \zeta_{pk} \right)
 \nn\\[5pt]
 &\qquad
 + 
 \frac{3\gamma_k \gamma_p}{2  q_k^{\star} q_p^{\star}} \left(\frac{\beta_p \omega_k}{k} + \zeta_{pk} \right)\left(\frac{\beta_k \omega_p}{p} + \zeta_{pk} \right)
 Q_0(\zeta_{pk}).
\end{align}

As before, we check the threshold behavior of this amplitude by fixing $\sigma_k$ and $\sigma_p$ and expanding for small $p$ and $k$. Just as in Sec.~\ref{sec:pair_00}, $\zeta_{pk}$ diverges as $\zeta_{pk} = \Nc / 2pk$ as $p,k \to 0$, thus $Q_0(\zeta_{pk})$ admits an expansion as Eq.~\eqref{eq:LegQ0_asymp}. Moreover, $\gamma_p = 1 + \Oc(p^2)$, $\omega_p = m_p + \Oc(p^2)$, and $E_p = \sqrt{\sigma_p} + \Oc(p^2)$ as $p \to 0$, with similar expansions for variables of the $k$ spectator as $k\to 0$. Near the pair-spectator thresholds, one finds that the expansion of Eq.~\eqref{eq:G3P0} for $p,k \to 0$ is given by
\begin{align}
    \Gc(^3P_0|^3P_0) = \frac{1}{\Nc q_p^\star q_k^\star}\left(1 + \frac{3m_pm_k}{\sqrt{\sigma_p}\sqrt{\sigma_k}} \right) \, pk + \Oc\!\left(p^2k,pk^2\right) \, , \nn 
\end{align}
where the relative momenta $q_p^\star$ and $q_k^\star$ are finite positive constants since $\sigma_p$ and $\sigma_k$ are fixed above their respective thresholds. Therefore, as expected $\Gc(^{3}P_0|^{3}P_0) \sim pk$ near threshold.

\subsubsection{Total \mathbshead{J=1}{J1}, Parity \mathbshead{\eta}{eta}}

Next let us consider a target $J = 1$ with a parity $\eta$, which enforces $L' = L = 1$, \ie $^3P_1 \to ^3P_1$. From Eq.~\eqref{eq:spin_orbit_P} the spin-orbit coupling is  $\Pc_\lambda^{(1)}(^3P_1) = - \lambda / \sqrt{2}$. From Eq.~\eqref{eq:Ktilde_coef} we find that
$\wt{\Kc}_{\Gc}(^3P_1|^3P_1) = 0$, while from Eq.~\eqref{eq:Ttilde_coef} the $\wt{\Tc}_j$ factors are
\begin{align}
    \wt{\Tc}_j(^3P_1|^3P_1) = -\frac{1}{2 q_p^{\star}q_k^{\star}} \left( \, 2\delta_{j0} - 3\zeta_{pk}\,\delta_{j1} + \delta_{j2} \, \right) \, , \nn 
\end{align}
with all coefficients $j>2$ being zero. Thus, the partial wave OPE Eq.~\eqref{eq:ope_JP_result} is
\begin{align}
    \label{eq:ope_11_J1_v1}
    \Gc(^3P_1|^3P_1) = -\frac{1}{q_p^\star q_k^\star}Q_0(\zeta_{pk}) + \frac{3\zeta_{pk}}{2q_p^\star q_k^\star}Q_1(\zeta_{pk}) - \frac{1}{2q_p^\star q_k^\star}Q_2(\zeta_{pk}) \, .
\end{align}

Before applying the simplifications of Eqs.~\eqref{eq:K_coef} and \eqref{eq:T_coef}, we first perform an intermediary manipulation of Eq.~\eqref{eq:ope_11_J1_v1} by using the Bonnet recursion relation for the Legendre function $Q_1$, see Eq.~\ref{eq:app.legQ.recur}, to simplify the expression to
\begin{align}
    \label{eq:ope_11_J1}
    \Gc(^3P_1|^3P_1) = -\frac{1}{2q_p^{\star}q_k^{\star}} \Big[\, Q_0(\zeta_{pk}) - Q_2(\zeta_{pk}) \, \Big] \, .
\end{align}
We note that this expressions agrees up to an overall sign with the result of Ref.~\cite{Kamal:1965vto}, in the context of studying the binding of the $\omega$ meson via $\pi$-exchange between the $\pi$ and the resonating $\pi\pi \to \rho$ subsystems. The difference in overall sign is due to the error in Ref.~\cite{Kamal:1965vto} from not considering the correct azimuthal angle of one of the momenta as discussed in the beginning of Sec.~\ref{sec:pair_11}. Finally, we use $Q_2(\zeta_{pk}) = P_2(\zeta_{pk}) Q_0(\zeta_{pk}) - 3\zeta_{pk} / 2$ to express Eq.~\eqref{eq:ope_11_J1} in the form of Eq.~\eqref{eq:ope_JP_final_result},
\begin{align}
    \Gc(^3P_1 | ^3P_1) = -\frac{3}{4q_p^\star q_k^\star} \zeta_{pk} + \frac{3}{4q_p^\star q_k^\star} \left(\, \zeta_{pk}^2 - 1 \,\right) \, Q_0(\zeta_{pk}) \, .
\end{align}

Near threshold we expect $\Gc(^3P_1|^3P_1) \sim pk$, which is verified by following the same procedure as in the previous cases, and finding
\begin{align}
    \Gc(^3P_1 | ^3P_1) = - \frac{1}{\Nc q_p^\star q_k^\star} \, pk + \Oc\left(p^2k, pk^2\right) \, , \nn 
\end{align}
which agrees with the expected behavior.

\subsubsection{Total \mathbshead{J=1}{J1}, Parity \mathbshead{-\eta}{meta}}
\label{sec:pair11_J1P_meta}

Our final example for this case is $J=1$ with parity $-\eta$. Here we encounter a coupled channel system in $S$ and $D$ waves since $L',L = J \pm 1$. The spin-orbit factors are given by
\begin{align}
	\Pc_{\lambda}^{(1)}(^{3}S_1) &= \sqrt{\dfrac{1}{3}}  \, , \nn \\[5pt]
    \Pc_{\lambda}^{(1)}(^{3}D_1) &= \sqrt{\dfrac{1}{6}} \, \lvert \lambda \rvert - \sqrt{\dfrac{2}{3}} \, \delta_{\lambda,0} \, . \nn
\end{align}
Feeding this, the $A$ and $B$ coefficients, and other building blocks into Eqs.~\eqref{eq:Ktilde_coef} and \eqref{eq:Ttilde_coef}, then through Eqs.~\eqref{eq:K_coef} and \eqref{eq:T_coef}, gives the following expressions for the $^3S_1 \to {^3S_1}$, $^3S_1 \to {^3D_1}$, $^3D_1 \to {^3S_1}$, and $^3D_1\to{^3D_1}$ OPE amplitudes: 
\begin{align}
    \Gc(^3L'_1|^3L_1) = \Kc_\Gc(^3L'_1|^3L_1) + \Tc(^3L'_1|^3L_1) \, Q_0(\zeta_{pk})
\end{align}
with the following $\Kc_\Gc$ and $\Tc$ matrices: 
\begin{subequations}
    \begin{align}
        \Kc_{\Gc}(^3S_1 | ^3S_1) & = \frac{1}{2q_p^\star q_k^\star} \Bigg\{ \gamma_p\left[ \zeta_{pk}\left( \frac{\beta_p \omega_k}{k} + \zeta_{pk} \right)   - \frac{2}{3} \right] +  \gamma_k \left[ \zeta_{pk}\left( \frac{\beta_k \omega_p}{p} + \zeta_{pk} \right)  - \frac{2}{3} \right] \nn \\[5pt]
        &  \quad - \gamma_p \gamma_k \left[ \zeta_{pk}\left(  \frac{\beta_k \omega_p}{p} + \frac{\beta_p \omega_k}{k} + \zeta_{pk} \right)  + \frac{\beta_p \beta_k \omega_p \omega_k}{pk} + \frac{1}{3} \right] - \zeta_{pk}^2 + \frac{2}{3} \Bigg\} \, , \\[5pt]
        \Kc_{\Gc}(^3D_1 | ^3S_1) & = \frac{1}{2\sqrt{2}q_p^\star q_k^\star} \Bigg\{  -2\gamma_p\left[ \zeta_{pk}\left( \frac{\beta_p \omega_k}{k} + \zeta_{pk} \right)   - \frac{2}{3} \right] + \gamma_k \left[ \zeta_{pk}\left( \frac{\beta_k \omega_p}{p} + \zeta_{pk} \right)   - \frac{2}{3} \right]  \nn \\[5pt]
        &  \quad + 2\gamma_p \gamma_k \left[ \zeta_{pk}\left(  \frac{\beta_k \omega_p}{p} + \frac{\beta_p \omega_k}{k} + \zeta_{pk} \right)  + \frac{\beta_p \beta_k \omega_p \omega_k}{pk} + \frac{1}{3} \right] - \zeta_{pk}^2 + \frac{2}{3} \Bigg\} \, , \\[5pt]
        \Kc_{\Gc}(^3D_1 | ^3D_1) & = \frac{1}{4q_p^\star q_k^\star} \Bigg\{  -2\gamma_p \left[ \zeta_{pk}\left( \frac{\beta_p\omega_k}{k} + \zeta_{pk}\right) - \frac{2}{3}\right] - 2\gamma_k \left[ \zeta_{pk}\left( \frac{\beta_k\omega_p}{p} + \zeta_{pk}\right) - \frac{2}{3}\right]   \nn \\[5pt]
        & \quad - 4\gamma_p \gamma_k \left[  \zeta_{pk} \left(  \frac{\beta_k \omega_p}{p} + \frac{\beta_p \omega_k}{k} + \zeta_{pk} \right)  + \frac{\beta_p \beta_k \omega_p \omega_k}{pk} + \frac{1}{3}  \right] - \zeta_{pk}^2 + \frac{2}{3}\Bigg\} \, ,
    \end{align}
\end{subequations}
for $\Kc_\Gc$, and
\begin{subequations}
    \begin{align}
        \Tc(^3S_1 | ^3S_1) & = \frac{1}{2q_p^\star q_k^\star} \Bigg\{ \gamma_p(1 - \zeta_{pk}^2 ) \left( \frac{\beta_p \omega_k}{k} + \zeta_{pk} \right)  +  \gamma_k(1 - \zeta_{pk}^2 ) \left( \frac{\beta_k \omega_p}{p} + \zeta_{pk} \right)   \nn \\[5pt]
        &  \quad + \gamma_p \gamma_k \,\zeta_{pk} \left[ \zeta_{pk}\left(  \frac{\beta_k \omega_p}{p} + \frac{\beta_p \omega_k}{k} + \zeta_{pk} \right)  + \frac{\beta_p \beta_k \omega_p \omega_k}{pk}  \right] + \zeta_{pk}^3 - \zeta_{pk} \Bigg\} \, , \\[5pt]
        \Tc(^3D_1 | ^3S_1) & = \frac{1}{2\sqrt{2}q_p^\star q_k^\star} \Bigg\{ -2\gamma_p(1 - \zeta_{pk}^2) \left( \frac{\beta_p \omega_k}{k} + \zeta_{pk} \right)  + \gamma_k(1 - \zeta_{pk}^2 ) \left( \frac{\beta_k \omega_p}{p} + \zeta_{pk} \right)   \nn \\[5pt]
        &  \quad - 2\gamma_p \gamma_k \,\zeta_{pk} \left[ \zeta_{pk}\left(  \frac{\beta_k \omega_p}{p} + \frac{\beta_p \omega_k}{k} + \zeta_{pk} \right)  + \frac{\beta_p \beta_k \omega_p \omega_k}{pk}  \right] + \zeta_{pk}^3 - \zeta_{pk} \Bigg\} \, , \\[5pt]
        \Tc(^3D_1 | ^3D_1) & = \frac{1}{4q_p^\star q_k^\star} \Bigg\{ -2\gamma_p(1 - \zeta_{pk}^2) \left( \frac{\beta_p \omega_k}{k} + \zeta_{pk} \right)  - 2\gamma_k(1 - \zeta_{pk}^2) \left( \frac{\beta_k \omega_p}{p} + \zeta_{pk} \right)   \nn \\[5pt]
        &  \quad + 4\gamma_p \gamma_k \,\zeta_{pk} \left[ \zeta_{pk}\left(  \frac{\beta_k \omega_p}{p} + \frac{\beta_p \omega_k}{k} + \zeta_{pk} \right)  + \frac{\beta_p \beta_k \omega_p \omega_k}{pk}  \right] + \zeta_{pk}^3 - \zeta_{pk} \Bigg\} \, .
    \end{align}
\end{subequations}
for the $\Tc$ matrices. The $\Kc_\Gc$ and $\Tc$ coefficients for the $^3D_1 \to {^3S_1}$ process are found by interchanging $k \leftrightarrow p$ in the $^3S_1 \to {^3D_1}$ coefficients, noting the symmetry $\zeta_{pk} = \zeta_{kp}$ which can be seen from Eq.~\eqref{eq:ope_arg}.

Examining the threshold behavior as in previous cases, we find that the $^3S_1 \to {^3S_1}$ OPE has the following expansion near threshold
\begin{align}
    \Gc(^3S_1 | ^3S_1) = \frac{1}{\Nc q_p^\star q_k^\star} \left( \frac{m_p}{\sqrt{\sigma_k}} \, k^2 + \frac{m_k}{\sqrt{\sigma_p}} \, p^2 \right)  + \Oc\!\left(p^3, k^3, p^2k, pk^2\right) \, . \nn 
\end{align}
If we fix $\sigma_p < \sigma_k$, then the threshold we approach first is $\sqrt{s} \sim \sqrt{\sigma_k} + m_k$. As we approach this threshold, then the amplitude does approach a constant, as $p$ will be finite this threshold. However, if both $k$ and $p$ approach threshold simultaneous, \eg in the case where $m_p = m_k$ and $\sigma_p = \sigma_k$, then the amplitude scales as $k^2$, which is faster than the requisite constant scaling we expect for $S$ waves. Although this behavior may be surprising, it is not inconsistent with the requirement that the amplitude is equal to a finite constant at threshold. 

Repeating this exercise for $^3S_1 \to {^3D_1}$ and $^3D_1 \to {^3D_1}$ waves, we find the following expansions
\begin{align}
    \Gc(^3D_1 | ^3S_1) & = -\frac{\sqrt{2}m_k}{\Nc q_p^\star q_k^\star \sqrt{\sigma_p}} \, p^2 + \Oc\!\left( p^3, p^2k \right) \, , \nn \\[5pt]
    \Gc(^3D_1 | ^3D_1) & = \frac{2}{15\Nc^2 q_p^\star q_k^\star} \left(1 + \frac{10 m_p m_k}{\sqrt{\sigma_p}\sqrt{\sigma_k}} \right) \, p^2 k^2 + \Oc\!\left( p^3k^2, p^2k^3 \right) \, , \nn 
\end{align}
where the threshold expansion of the $^3D_1 \to {^3S_1}$ amplitude is found by interchanging $k\leftrightarrow p$ on $\Gc(^3D_1 | ^3S_1)$. Both of these follow the expected threshold behavior. This completes our set of examples for $\ell' = \ell = 1$. Next, we will look at examples for transitions between $\ell = 0$ and $\ell' = 1$.

\subsection{Pairs with \mathbshead{\ell' = 1}{ellp1} and \mathbshead{\ell = 0}{ell0}}
\label{sec:pair_10}

Here we consider an initial pair with spin $\ell = 0$, and a final pair with spin $\ell' = 1$. Such transitions are allowed in general, and observed in nature, \eg in $\sigma\pi\to \rho\pi$ in the $I(J^{P}) = 1(1^+)$ channel of $3\pi$ scattering. Repeating the same strategy as in the previous cases, the spin-helicity matrix is given by
\begin{align}
    \Hc_{\lambda'\lambda}^{(10)} = \delta_{\lambda 0} \, \sqrt{4\pi}\, \left( \frac{ k_p^\star }{q_p^\star} \right) \, Y_{1\lambda'}^*(\bh{k}_p^\star)
\end{align}
which corresponds to an $\Ac$ coefficient
\begin{subequations}
    \begin{align}
        \Ac^{(10)}_{\pm 1 0} & = \mp \frac{\sqrt{6}k}{q_p^{\star}} \, , \\[5pt]
        \Ac^{(10)}_{0 0} & = -\frac{\sqrt{3} \gamma_p k}{q_p^{\star}} \, \left( \frac{\beta_p \omega_k}{k} + \zeta_{pk} \right) \, ,
    \end{align}
\end{subequations}
and a $B$ coefficient
\begin{align}
    \Bc_{\lambda'\lambda}^{(10)} & = \delta_{\lambda'0} \delta_{\lambda 0} \, \frac{\sqrt{3} \gamma_p k}{q_p^{\star}} \,.
\end{align}
Therefore, there is only one contribution to the $\Bc_j$ amplitudes, $\Bc_{j, \lambda'\lambda}^{(10)} = \delta_{j0} \, \Bc_{\lambda'\lambda}^{(10)}$. The coefficients are then fed into the expressions for the $\Kc_\Gc$ and $\Tc$ matrices. Since total angular momentum and parity are conserved, an initial state with $S=0$, $L=J$, and $P = \eta(-1)^{J}$ restricts the final state, with $S'=1$, to have an $L'$ quantum number be $L' = 1$ for $J=0$, and $L' = J \pm 1$ for $J > 0$.

\subsubsection{Total \mathbshead{J=0}{J0}}

Following the same procedure as in the previous cases, we have for $J=0$, in which the system parity is $\eta$, the $^1S_0 \to {^3}P_0$ OPE is given by
\begin{align}
    \Gc(^3P_0|^1S_0) & = -\frac{\sqrt{3}\gamma_p}{2pq_p^\star} + \frac{\sqrt{3}\gamma_p}{2pq_p^\star} \left( \frac{\beta_p \omega_k}{k} + \zeta_{pk} \right)  Q_0(\zeta_{pk}) \, ,
\end{align}
which has a threshold expansion
\begin{align}
    \Gc(^3P_0|^1S_0) & = \frac{\sqrt{3} m_k}{\Nc q_p^\star \sqrt{\sigma_p}} \, p + \Oc\!\left(p^2, pk \right) \, , \nn
\end{align}
which agrees with the expected behavior.

\subsubsection{Total \mathbshead{J=1}{J1}}

Our final example is for $J = 1$, which must be in a $-\eta$ parity states due to the initial $^1P_1$ state. There are two options for the transition, either $^1P_1 \to {^3S_1}$ or $^1P_1\to{^3D_1}$. The partial wave OPE amplitudes for these transitions are
\begin{align}
    \Gc(^3S_1|^1P_1) & = \frac{1}{2pq_p^\star} \left( \gamma_p \left( \frac{\beta_p \omega_k}{k} + \zeta_{pk} \right) - \zeta_{pk} \right) \, \nn \\[5pt]
    & \qquad - \frac{1}{2p q_p^\star} \left[ \gamma_p \zeta_{pk} \left(\frac{\beta_p \omega_k}{k} + \zeta_{pk} \right)   - \zeta_{pk}^2 + 1 \right] Q_0(\zeta_{pk}) \, , \\[5pt]
    \Gc(^3D_1|^1P_1) & = -\frac{1}{2\sqrt{2}pq_p^\star} \left( 2\gamma_p \left(\frac{\beta_p \omega_k}{k} + \zeta_{pk} \right) + \zeta_{pk} \right) \, \nn \\[5pt]
    & \qquad + \frac{1}{2\sqrt{2}pq_p^\star}\left[ 2\gamma_p \zeta_{pk} \left( \frac{\beta_p \omega_k}{k} + \zeta_{pk} \right) + \zeta_{pk}^2 - 1 \right] Q_0(\zeta_{pk}) \, .
\end{align}

We note that the threshold behavior for fixed $\sigma_k$, $\sigma_p$ is
\begin{align}
    \Gc(^3S_1 | ^1P_1) & = - \frac{1}{\Nc q_p^\star} \, k + \Oc\!\left(k^2, pk\right) \, , \nn \\[5pt]
    \Gc(^3D_1 | ^1P_1) & = \frac{2\sqrt{2} m_k}{3\Nc^2 q_p^\star \sqrt{\sigma_p}} \, p^2 k + \Oc\!\left(p^3 k, p^2 k^2\right) \, , \nn 
\end{align}
as expected. 

Transitions from $\ell = 1 $ and $\ell' = 0$ states can be obtained by interchanging the initial and final state, $k \leftrightarrow p$ in the above expressions. Any higher angular momentum state can be found by the same procedure outlined here. In the next section, we examine some applications of the above results for the scattering of three pions.

\section{Application -- \mathbshead{3\pi \to 3\pi}{threepi}}
\label{sec:application}

As a final illustration, we apply the results in Sec.~\ref{sec:cases} for $3\pi \to 3\pi$ scattering, and plot the partial wave OPEs for some selected allowed quantum numbers of three pions in various kinematic regions. We limit the total energy of the three pion system such that inelastic processes are forbidden, \ie $3m_{\pi} \le \sqrt{s} < 5m_{\pi}$ where $m_{\pi}$ is the pion mass. Therefore, the exchange amplitude consists only of pion interactions, $m_k = m_p = m_e = m_\pi$. Furthermore, we will only consider physical scattering kinematics, so that the physical boundary is set by Eq.~\eqref{eq:Kibble_exch} where all masses are set to the pion mass, \ie $\Phi(p,k) \ge 0$ with $\Phi(p,k) = \sigma_k \sigma_p (s + 3m_\pi^2 - \sigma_k - \sigma_p) - m_\pi^2(s - m_\pi^2)^2$.

We also assume the isospin limit for pions, that is the pions have a flavor symmetry characterized by their isospin $I_\pi = 1$ and $G$ parity $G_\pi = -1$. Isospin symmetry restricts the allowed partial wave contributions for the three-pion system. Two pions are in either $I = 0$, $1$, or $2$ states with positive $G$ parity. Bose symmetry restricts even partial waves, \eg $S$ and $D$ waves, to be in either an $I=0$ or $2$ state, whereas odd waves, \eg $P$ waves, can only be in the $I=1$ state. For three pions systems, which have negative $G$ parity, the allowed total isospin representations are $I_{3\pi} = 0, 1, 2, 3$. There are multiple contributing three-pion partial waves per target $J^P$, we summarize the lowest allowed three-pion waves in Tab.~\ref{tab:3piQN} for each total isospin $I_{3\pi}$, total angular momentum $J \le 1$, and up through two-pions in relative $P$ wave. We label a three-pion partial wave with $([\pi\pi]_{\ell}^{I}\pi)_L$, where the two-pion system is in a relative $\ell$ wave and isospin $I$, and the pair-spectator pion is in an orbital angular momentum $L$, \eg $([\pi\pi]_P^1\pi)_S$ describes a three pion system where two of the pions are in an isovector $P$ wave and the recoiling pion is in a relative $S$ wave with the pair.
\begin{table}[t]
	\centering
	\caption{Contributions of $([\pi\pi]_\ell^I\pi)_L$ partial waves to $\pi\pi\pi$ in total isospin $I_{3\pi}$ and for total angular momentum $J\le 1$. Lowest angular momenta are considered, where the two pion pairs are in $\ell \le P$ wave, and the orbital angular momentum between the pair and the spectators is $L \le D$ wave.} 
	\label{tab:3piQN}
	\begin{tabularx}{0.55\textwidth}{@{}s|s|Y@{}}
		\rule{0pt}{4ex}  
		$I_{3\pi}^{G}$ & $J^{PC}$ & $([\pi\pi]_\ell^I \pi)_{L}$ \\[5pt]
		\hline
		\hline
		%
		\multirow{3}{4mm}{$3^{-}$} & $0^{-+}$ & $([\pi\pi]_S^2 \pi)_{S}$  \\
		 & $1^{-+}$ & none \\
		 & $1^{++}$ & $([\pi\pi]_S^2 \pi)_{P}$ \\ \hline
		%
		\multirow{3}{4mm}{$2^{-}$} & $0^{--}$ & $([\pi\pi]_S^2 \pi)_{S}$,  $([\pi\pi]_P^1 \pi)_{P}$  \\
		 & $1^{--}$ & $([\pi\pi]_P^1 \pi)_{P}$ \\
		 & $1^{+-}$ & $([\pi\pi]_S^2 \pi)_{P}$, $([\pi\pi]_P^1 \pi)_{S}$, $([\pi\pi]_P^1 \pi)_{D}$  \\ \hline
		%
		\multirow{3}{4mm}{$1^{-}$} & $0^{-+}$ & $([\pi\pi]_S^{0,2} \pi)_{S}$,  $([\pi\pi]_P^1 \pi)_{P}$ \\
		 & $1^{-+}$ & $([\pi\pi]_P^1 \pi)_{P}$ \\
		 & $1^{++}$ & $([\pi\pi]_S^{0,2} \pi)_{P}$, $([\pi\pi]_P^1 \pi)_{S}$, $([\pi\pi]_P^1 \pi)_{D}$ \\ \hline
		%
		\multirow{3}{4mm}{$0^{-}$} & $0^{--}$ & $([\pi\pi]_P^1 \pi)_{P}$  \\
		 & $1^{--}$ & $([\pi\pi]_P^1 \pi)_{P}$ \\
		 & $1^{+-}$ & $([\pi\pi]_P^1 \pi)_{S}$, $([\pi\pi]_P^1 \pi)_{D}$ \\ \hline
		\hline
	\end{tabularx}
\end{table}

%
\begin{figure*}[t]
	\centering
	\includegraphics[width=0.85\textwidth]{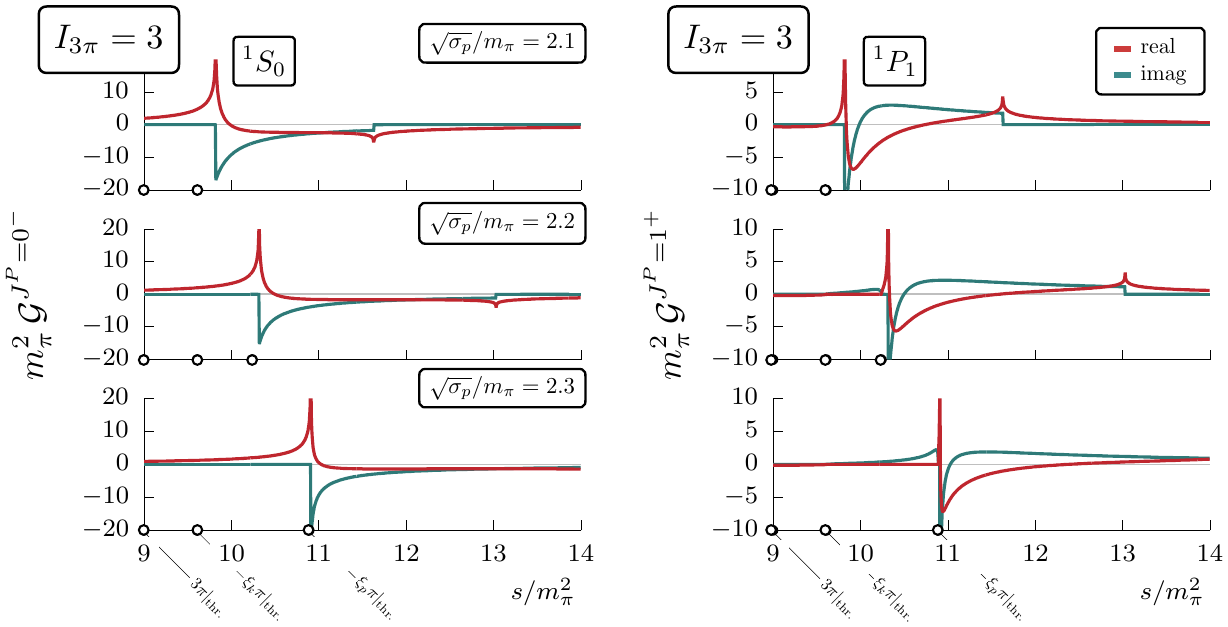}
	\caption{Real and imaginary parts of the OPE for $I_{3\pi} = 3$ and total spin-parity $J^P = 0^-$ and $1^+$ for $\pi\pi$ sub-channels in an isotensor $S$ wave state. Each panel is plotted for a fixed initial $\sqrt{\sigma_k}  = 2.1 m_\pi$. Each row is plotted at a different $\sigma_p$: $\sqrt{\sigma_p} = 2.1 m_\pi$ for the top row, $\sqrt{\sigma_p} = 2.2 m_\pi$ for the middle row, and $\sqrt{\sigma_p} = 2.3 m_\pi$ for the bottom row. The spectroscopic label $^{2S+1} L_J$ indicates the partial wave OPE contribution to the $([\pi\pi]_S^{2}\pi)_L$ amplitude. Thresholds are indicated for the $3\pi$ production at $s = (3m_\pi)^2$, the effective initial pair-spectator $\xi_k\pi$ at $s = (\sqrt{\sigma_k} + m_\pi)^2 = (3.1m_\pi)^2$, and the effective initial pair-spectator $\xi_p\pi$ at $s = (\sqrt{\sigma_p} + m_\pi)^2$, which is located at $s = (3.1m_\pi)^2$, $(3.2m_\pi)^2$, and $(3.3 m_\pi)^2$ for each $\sigma_p$ shown.}
	\label{fig:isospin3}
\end{figure*}

%
\begin{figure*}[t]
	\centering
	\includegraphics[width=1\textwidth]{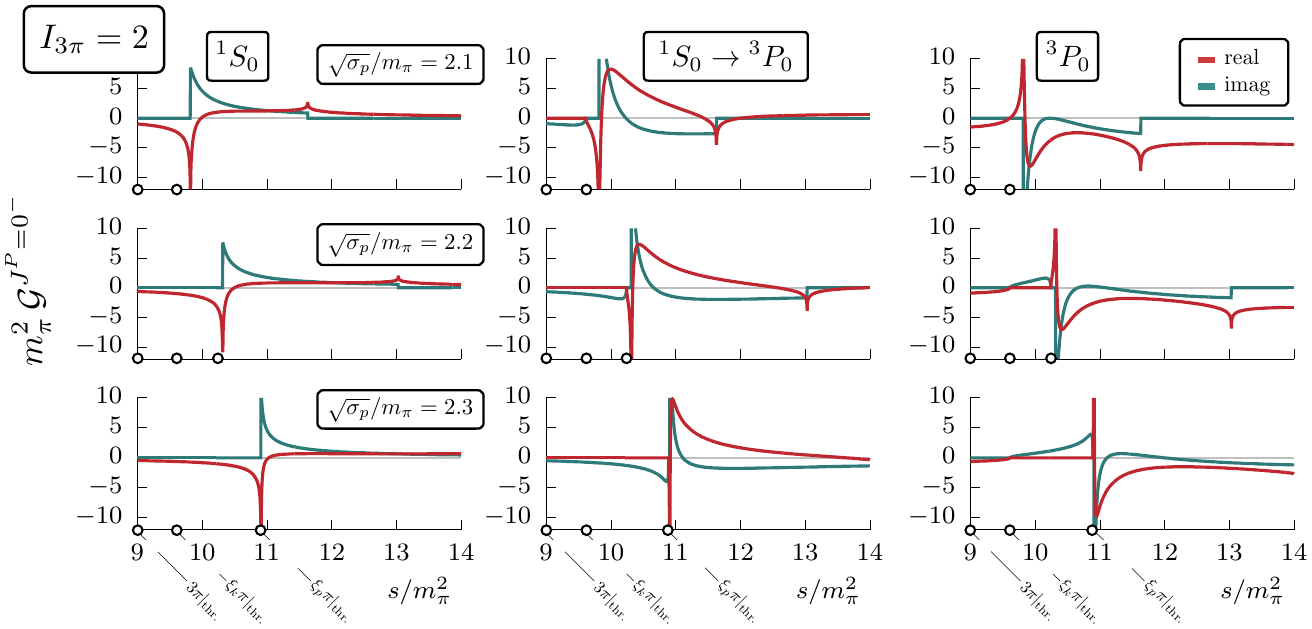}
	\caption{Same as Fig.~\ref{fig:isospin3} for $I_{3\pi} = 2$ and $J^P = 0^-$. Here there are two kinds of $\pi\pi$ pairs, one which is in an isotensor $S$ wave, with a pair-spectator system in $^1S_0$, and another pair in an isovector $P$ wave in which the pair-spectator system is $^3P_0$. These waves are allowed to mix through total angular momentum and isospin conservation, leading to a non-zero $^1S_0 \to {^3P_0}$ amplitude.}
	\label{fig:isospin2}
\end{figure*}

%
\begin{figure*}[t]
	\centering
	\includegraphics[width=0.98\textwidth]{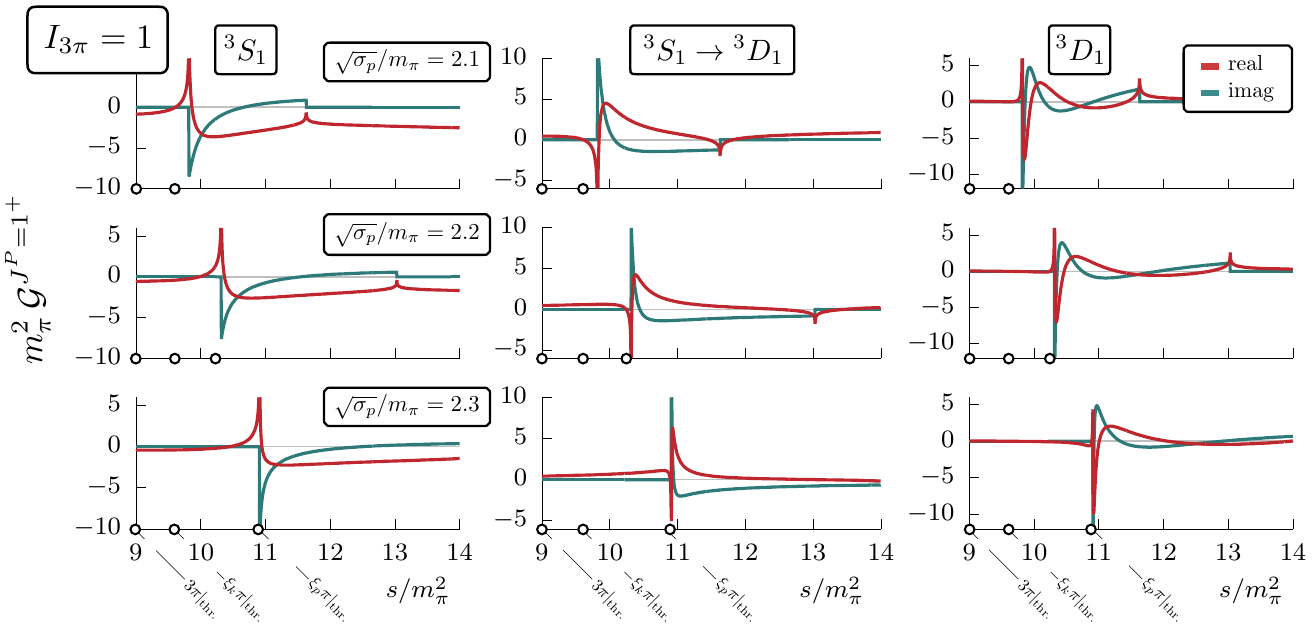}
	\caption{Same as Fig.~\ref{fig:isospin3} for $I_{3\pi} = 1$ and $J^P = 1^+$. Shown are the contributions from isovector $P$ wave $\pi\pi$ pairs.}
	\label{fig:isospin1pt1}
\end{figure*}

%
\begin{figure*}[t]
	\centering
	\includegraphics[width=0.98\textwidth]{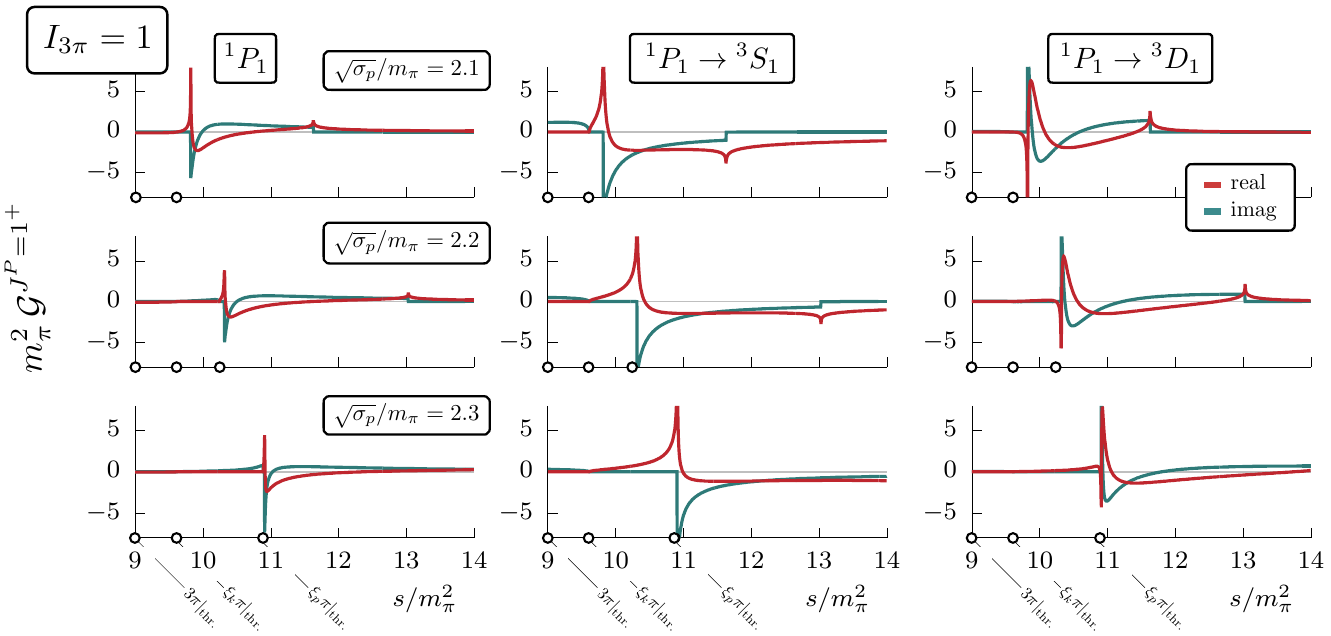}
	\caption{Same as Fig.~\ref{fig:isospin3} for $I_{3\pi} = 1$ and $J^P = 1^+$, where now the contributions from isoscalar $S$ wave $\pi\pi$ pairs as well as the mixing to isovector $P$ wave pairs are shown.}
	\label{fig:isospin1pt2}
\end{figure*}

%
\begin{figure}[t]
	\centering
	\includegraphics[width=0.48\textwidth]{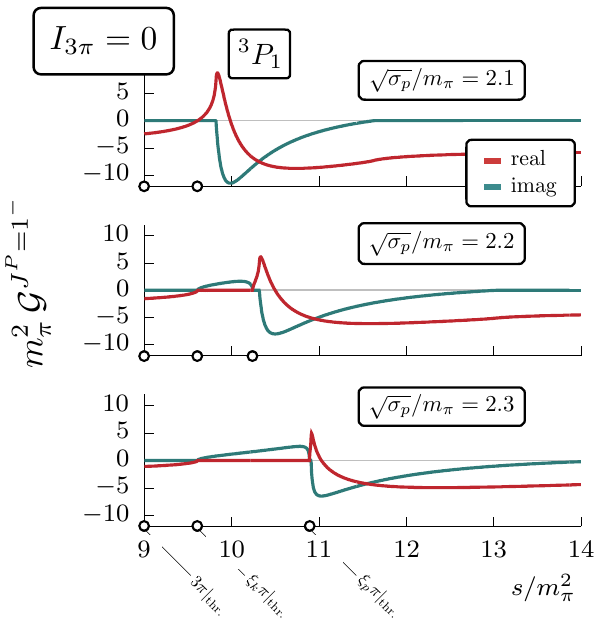}
	\caption{Same as Fig.~\ref{fig:isospin3} for $I_{3\pi} = 0$ and $J^P = 1^-$. Only isovector $P$ wave $\pi\pi$ pairs are allowed to couple to $^3P_1$ amplitudes in this channel.}
	\label{fig:isospin0}
\end{figure}

Our results in Sec.~\ref{sec:cases} can be applied immediately to these partial waves, with the exception of the inclusion of appropriate isospin recoupling coefficients. These can be included in a straightforward manner as detailed in Refs.~\cite{Ascoli:1975mn,Hansen:2020zhy}. The result is the OPE in Eq.~\eqref{eq:ope_JP_final_result} contains three additional quantum numbers,
\begin{align}
    \left[ \Gc^{(\ell'\ell),J}_{L'S',LS} \, \right]_{I',I}^{I_{3\pi}} & = \Gc^{(\ell'\ell),J}_{L'S',LS} \, \braket{I' ,I_{3\pi} |I ,I_{3\pi}}
\end{align}
where $I$ is the initial pair isospin, $I'$ is the final pair isospin, and $I_{3\pi}$ is the total isospin of the three pion system. The mutiplicative factor $\braket{I', I_{3\pi} | I, I_{3\pi}}$ is the three pion isospin recoupling coefficient, which is defined in terms of the Wigner 6-j symbol as~\cite{VMK}
\begin{align}
    \label{eq:isospin_recoupling}
    \braket{I',I_{3\pi} | I ,I_{3\pi}} & = \sqrt{(2I'+1)(2I+1)} \,  
    \begin{Bmatrix} 
    1 & 1 & I \\
    1 & I_{3\pi} & I'
    \end{Bmatrix} \, .
\end{align}
Explicit values can be found in Refs.~\cite{Ascoli:1975mn,Hansen:2020zhy}, or by direct computation via Eq.~\eqref{eq:isospin_recoupling}. The isospin recoupling coefficients introduce a weight factor for the particular isospin channel.

We plot a representative partial wave OPE of each isospin channel to show the generic behavior. For each channel, we plot the OPE as a function of $s$ in the range $9 \le s/m_{\pi}^2 \le 14$ at fixed $\sqrt{\sigma_k} / m_{\pi} = 2.1$ and for three values of $\sigma_p$, $\sqrt{\sigma_p} / m_{\pi} = \{ 2.1, 2.2, 2.3 \}$. In each plot, we highlight the $3\pi$ threshold at $s = (3m_\pi)^2$, the initial pair-spectator threshold which we indicate by $\xi_k\pi$ at $s = (\sqrt{\sigma_k} + m_\pi)^2$, where we remind the reader that $\xi_k$ represents a quasi-particle of mass $\sqrt{\sigma_k}$, and the final pair-spectator threshold $\xi_p\pi$ at $s = (\sqrt{\sigma_p} + m_\pi)^2$. Note that when both $\sigma_k = \sigma_p$, then the initial and final pair-spectator thresholds overlap. 

In our numerical evaluation of the partial wave OPE, we ensure that we approach the real energy axes by introducing an artificial imaginary shift. To control the limit, we introduce for $\sigma_k$, and $\sigma_p$ a shift $\sigma \to \sigma +i\epsilon_\sigma$, and for $s$ we introduce a shift $s \to s+i\epsilon_s$, with the restriction that $\epsilon_s > \epsilon_\sigma$, meaning we assume that we approach the real $\sigma$-axes first before approach the real $s$-axis. This limiting procedure then gives a positive imaginary part to the spectator momentum for unphysical energies. To ensure the correct behavior required by $S$ matrix unitarity, we set $\zeta_{pk} \to \zeta_{pk} + i\epsilon$ appearing in the argument of the $Q$ functions as discussed with Eq.~\eqref{eq:Q_fcn}. To ensure the proper behavior for the imaginary part of $\Gc$ required by $S$ matrix unitarity, we restrict $\epsilon > \epsilon_s > \epsilon_\sigma$.

First, we consider the $I_{3\pi} = 3$ channel, which from Tab.~\ref{tab:3piQN} the lowest waves include $J^P = 0^-$ and $1^+$. The real and imaginary parts of both the $\Gc(^1S_0 | ^1S_0)$ and $\Gc(^1P_1 | ^1P_1)$ amplitudes are shown in Fig.~\ref{fig:isospin3}. There is a clear movable singularity in both amplitudes which arise from when the exchanged pion goes on-mass-shell, that is when $\zeta_{pk} = \pm 1$. The analytic structure of the OPE has been well studied in the literature, see for example Ref.~\cite{Jackura:2018xnx}, so we only highlight a few important features. In the physical kinematic region, the imaginary part of each partial wave OPE is constrained by $S$ matrix unitarity, through the imaginary part of $Q_0$, giving
\begin{align}
    \label{eq:imag_unitarity}
    \im \Gc^{J^P} = - \frac{\pi}{2} \, \Tc^{J^P} \, \Theta(\Phi(p,k)) \Theta(p) \Theta(k) \, ,
\end{align}
where $\Theta$ is the Heaviside step function. Including isospin, we multiply Eq.~\eqref{eq:imag_unitarity} by the recoupling coefficient~\eqref{eq:isospin_recoupling}. For fixed $\sigma_p$ and $\sigma_k$, we can solve $\zeta_{pk} = \pm 1$ for the branch points in $s$ which are given by
\begin{align}
    \label{eq:branch_points}
    s^{(\pm)} & = \frac{1}{2m_\pi^2} \Bigg[ \, (\sigma_k - m_\pi^2)(\sigma_p - m_\pi^2) + m_\pi^2(\sigma_p + \sigma_k + m_\pi^2) \, \nn\\[5pt]
    & \qquad\qquad  \pm \lambda^{1/2}(m_\pi^2,m_\pi^2,\sigma_k)\lambda^{1/2}(m_\pi^2,m_\pi^2,\sigma_p) \, \Bigg] \, .
\end{align}
These movable branch points correspond to the non-zero imaginary part of the OPE above the highest pair-spectator threshold in Fig.~\ref{fig:isospin3}. The existence of these branch points is independent of the partial wave of the OPE, as seen in the $^1S_0$ and $^1P_1$ amplitudes of Fig.~\ref{fig:isospin3}.

Next we examine three pions in $I_{3\pi} = 2$ in the $J^P = 0^-$ channel, which is shown in Fig.~\ref{fig:isospin2}. According to Tab.~\ref{tab:3piQN}, two types of pairs contributes to this partial wave, $([\pi\pi]_S^2\pi)_S$ and $([\pi\pi]_P^1\pi)_P$. Therefore, we have three contributing OPE amplitudes, $^1S_0$ for isotensor pairs, $^3P_0$ for isovector pairs, and a mixing amplitude between isotensor and isovector pairs in $^1S_0 \to ^3P_0$. As in the $I_{3\pi} = 3$ case, each amplitude has a non-analytic structure arising from on-shell pion exchange with branch points given by Eq.~\eqref{eq:branch_points}.

Figures~\ref{fig:isospin1pt1} and \ref{fig:isospin1pt2} show partial wave OPE amplitudes which contribute to the $J^P = 1^+$ channel of $I_{3\pi} = 1$. Figure~\ref{fig:isospin1pt1} show the contributions coming from isovector $P$ wave $\pi\pi$ pairs, while Fig.~\ref{fig:isospin1pt2} shows contributions from the isoscalar $S$ wave pairs and its mixing with isovector $P$ wave pion pairs. Note that we do not plot contributions from isotensor pion pairs which are also part of this wave as shown in Tab.~\ref{tab:3piQN}. Physically, this case is most relevant for $3\pi$ scattering in the isovector $a_1$ channel, which allows for dynamical mixing between resonating $\rho\pi \leftrightarrow \sigma\pi$ systems, \cf the $a_1$ listing in Ref.~\cite{ParticleDataGroup:2022pth} and references therein. Notice that for the $^3S_1$ amplitude in Fig.~\ref{fig:isospin1pt1} that in the top panel with $\sigma_k = \sigma_p$, the scaling behavior at the pair-spectator threshold does grow from zero as $p^2$ as indicated in our threshold expansion discussed in Sec.~\ref{sec:pair11_J1P_meta}. However, when $\sigma_p \ne \sigma_k$ as in the middle and bottom panel, then the threshold behavior does approach a constant at the $\xi_p \pi$ threshold.

Finally, we show in Fig.~\ref{fig:isospin0} the $I_{3\pi} = 0$, $J^P = 1^-$ channel. Table~\ref{tab:3piQN} list one entry for this channel, $([\pi\pi]_P^1)_P$, therefore only the $^3P_1$ partial wave OPE contributes. A physical application of for this amplitude is in the isoscalar $\omega$ meson, which couples strongly to the $\rho\pi$ channel in $P$ wave.

\section{Summary}
\label{sec:summary}

We have shown a generic procedure to project relativistic scattering amplitudes of three spinless particles to definite $J^P$ partial waves, with focus on it application to the kinematic singularity arising from on-shell particle exchanges between two-body sub-channel scattering processes. The procedure as presented in Sec.~\ref{sec:ope}, specifically in the final projection results shown in Eqs.~\eqref{eq:ope_JP_final_result}, \eqref{eq:K_coef}, and \eqref{eq:T_coef} allow one to systematically compute the contribution from the one-particle exchange, which was illustrated in Sec.~\ref{sec:cases} for some low-lying spins of immediate interest, \eg in the scattering of three pions as discussed in Sec.~\ref{sec:application}. These results can then be supplied into the corresponding integral equations~\cite{Hansen:2015zga,Jackura:2022gib}, along with some parameterized and constrained three-body $K$ matrix, \eg ones constrained from lattice QCD calculations with finite-volume quantization conditions, to reconstruct the complete on-shell $\3\to\3$ hadronic scattering amplitude.

Our resulting analytic representation for the one-particle exchange of definite $J^P$ allows one to avoid performing a numerical integration over a singular function, which is generally slowly convergent, and allows the practitioner to have full control over the analytic behavior in the complex $s$-, $\sigma_p$-, and $\sigma_k$-planes. Controlling the analytic structure of aspects of three-body amplitudes has been shown to be important for the analytic continuation of the $\3\to\3$ amplitude, \eg for searching for resonant structures in hadron spectroscopy~\cite{Dawid:2023jrj, Dawid:2023kxu}. Looking forward, our results can be immediately used in the community in further theoretical and phenomenological studies of three-hadron resonance production. Furthermore, they can be extended to accomadate a more general class of reactions, such as those with external particles with arbitrary spin.

\section*{Acknowledgements}

The authors would like to thank S. Dawid, J. Dudek, M. Hansen, V. Mathieu, F. Romero-L\'opez, S. Sharpe, and A. Szczepaniak for useful comments and discussions. AWJ acknowledges the support of the USDOE ExoHad Topical Collaboration, contract DE-SC0023598. RAB acknowledges the support of the USDOE Early Career award, contract DE-SC0019229.

\appendix

\section{Recapitulation of Angular Momentum Functions}
\label{sec:app.angular_momentum}

This appendix is devoted to collecting useful identities and properties of the Legendre functions and Wigner rotation matrix elements which we use throughout this work. While these relations can be found in the literature (which we refer to as appropriate), we feel that this summary serves to assist the reader in understanding the technical aspects of our work.

\subsection{Legendre Functions of the \mathbshead{1\ts{st}}{1st}-kind}
The Legendre functions of the 1\ts{st} kind, $P_{\ell}(z)$, are the regular solutions of Legendre's differential equation~\cite{abramowitz+stegun}, which can be expressed explicitly for $\ell \in \Nbb_0$ and $-1 \le z \le +1$ by the Rodrigues' formula,
\begin{align}
    P_\ell(z) = \frac{1}{2^\ell \, \ell!} \, \frac{\diff^\ell}{\diff z^\ell}(z^2 - 1)^\ell \, .
\end{align}
We consider only $\ell \in \Nbb_0$, therefore the functions are analytic in $z \in \Cbb$ for each $\ell$. The Legendre functions form an orthogonal set of functions over the interval $-1\le z \le +1$, 
\begin{align}
	\label{eq:app.leg.orth}
	\int_{-1}^{+1} \diff z \, P_{\ell'} (z) P_{\ell}(z) = \frac{2}{2\ell + 1} \, \delta_{\ell'\ell}  \, .
\end{align}
The first few Legendre functions are
\begin{align}
	P_0(z) = 1 \, , \quad 
	P_1(z) = z \, , \quad 
	P_2(z) = \frac{1}{2}(3z^2 - 1) \, . \nn 
\end{align}
Given $P_0$ and $P_1$, all remaining $P_\ell$ can be generated through the Bonnet recursion relation for $\ell > 1$,
\begin{align}
	\label{eq:app.leg.recur}
	\ell P_\ell(z) = z(2\ell-1) \, P_{\ell - 1}(z) - (\ell - 1) \, P_{\ell - 2}(z) \, .
\end{align}
There are many additional properties and identities which can be found in Ref.~\cite{abramowitz+stegun}. Here we state one integral relation,
\begin{align}
    \label{eq:leg_P_int_id}
    \int_{-1}^{+1}\!\diff z \, z^\ell P_\ell(z) = \frac{2^{\ell+1}(\ell!)^2}{(2\ell+1)!} \, ,
\end{align}
which is useful in the asymptotic expansion of the Legendre functions of the 2\ts{nd} kind as discussed next in section.

\subsection{Legendre Functions of the \mathbshead{2\ts{nd}}{2nd}-kind}
A second class of solutions of Legendre's differential equation are the Legendre functions of the 2\ts{nd} kind, $Q_\ell(z)$. For every $\ell \in \Nbb_0$, the $Q_\ell$ functions are related to the $P_\ell$ functions through the Neumann relation~\cite{abramowitz+stegun},
\begin{align}
    \label{eq:Q_neumann}
	Q_\ell(z) = \frac{1}{2} \int_{-1}^{+1} \! \diff z' \, \frac{P_\ell(z')}{z - z'} \, .
\end{align}
The integral has endpoint singularities, leading to branch points in $Q_\ell$ in the complex $z$-plane at  $z = \pm 1$ for each $\ell$. We choose to orient the branch cut such that the function is analytic on $z \in \Cbb \, / \, \{z \, |\, -1 \le z \le +1 \}$, which is equivalent to choosing $z \to z + i\epsilon$ in Eq.~\eqref{eq:Q_neumann} and taking the limit $\epsilon \to 0^+$ after integration. The Neumann relation~\eqref{eq:Q_neumann} allows us to easily identify the discontinuity of $Q_{\ell}$ across the branch cut,
\begin{align}
	\mathrm{Disc} \, Q_\ell(z) = -i\pi P_{\ell}(z) \, \Theta(1 - \lvert z \rvert^2) \, ,
\end{align} 
where $\Theta$ is the Heaviside step function.

The first few Legendre functions of the 2\ts{nd} kind are given by
\begin{align}
    \label{eq:app.explicit_Q}
	Q_0(z) = \frac{1}{2} \log\left( \frac{z + 1}{z - 1} \right) \, , \quad 
	Q_1(z) = P_1(z) Q_0(z) - 1 \, , \quad 
	Q_2(z) = P_2(z) Q_0(z) - \frac{3z}{2} \, .  
\end{align}
Combining the Neumann relation~\eqref{eq:Q_neumann} with the Bonnet recursion relation for the $P_\ell$ functions, Eq.~\eqref{eq:app.leg.recur}, yields a recursion relation for integer $\ell > 1$ given $Q_0$ and $Q_1$,
\begin{align}
	\label{eq:app.legQ.recur}
	\ell Q_\ell(z) = z(2\ell-1) \, Q_{\ell - 1}(z) - (\ell - 1) \, Q_{\ell - 2}(z) \, .
\end{align}
One can then construct an explicit expression for $Q_\ell$ for any $\ell \in \Nbb_0$ and $z$ on the cut plane,
\begin{align}
    \label{eq:app.Ql_explicit}
    Q_\ell(z) = P_\ell(z) \, Q_0(z) - W_{\ell - 1}(z) \, ,
\end{align}
where $W_{\ell - 1}$ is defined for $\ell > 0$ as
\begin{align}
    \label{eq:app.Wnm1}
    W_{\ell - 1}(z) = \sum_{n=1}^{\ell} \frac{1}{n} P_{n-1}(z) \, P_{\ell - n}(z) \, ,
\end{align}
with the $\ell = 0$ case defined as $W_{-1} = 0$~\cite{abramowitz+stegun}.

The behavior of $Q_\ell(z)$ as $z\to\infty$ can be found by expanding the Neumann relation, Eq.~\eqref{eq:Q_neumann}, for large $z$,
\begin{align}
    \label{eq:app.legQ_expand}
    Q_{\ell}(z) = \frac{1}{2} \sum_{n = 0}^{\infty} \frac{1}{z^{n+1}} \int_{-1}^{+1}\!\diff z' \, (z')^n\,P_{\ell}(z') \, .
\end{align}
The integral is identically zero for $n < \ell$, thus the leading asymptotic behavior is given when $n = \ell$, which from Eq.~\eqref{eq:leg_P_int_id} gives
\begin{align}
    \label{eq:app.legQ_asymp}
    Q_\ell(z) \xrightarrow[z \to \infty]{} \frac{2^\ell (\ell!)^2}{(2\ell + 1)!} \, \frac{1}{z^{\ell + 1}} \, .
\end{align}
By direct evaluation of Eq.~\eqref{eq:app.legQ_expand}, the explicit asymptotic expansion for the $\ell = 0$ function is given by
\begin{align}
    \label{eq:app.legQ0_asymp}
    Q_0(z) = \sum_{n=0}^{\infty} \frac{z^{-2n-1}}{2n+1}   = \frac{1}{z} + \frac{1}{3z^3} + \Oc\!\left(\frac{1}{z^5}\right)\, .
\end{align}
%

\subsection{Spherical Harmonics}
For spinless particles, orbital angular momentum states are represented by the spherical harmonics~\cite{VMK} (with the Condon–Shortley phase convention)
\begin{align}
	Y_{\ell m}(\theta,\varphi) & = \braket{\theta,\varphi|\ell m} \, , \nn \\[5pt]
	& = (-1)^{\ell}\sqrt{\frac{(2\ell+1)(\ell - m)!}{4\pi(\ell + m)!}} \, P_{\ell}^m(\cos\theta) \, e^{im \varphi} \, ,
\end{align}
where $P_{\ell}^m$ are the associated Legendre functions,
\begin{align}
    P_\ell^m(z) = (1-z^2)^{m/2} \,\frac{\diff^m}{\diff z^m}P_\ell(z) \, .
\end{align}
See Ref.~\cite{abramowitz+stegun} for properties of $P_\ell^m$. The spherical harmonics for $m \ge 0$ and $\ell \le 2$ are explicitly
\begin{align}
	&Y_{00} = \phantom{-}\sqrt{\frac{1}{4\pi}} \, , & \nn \\[5pt]
	&Y_{10} = \phantom{-}\sqrt{\frac{3}{4\pi}}  \,  \cos\theta \, ,& 
	&Y_{11} = -\sqrt{\frac{3}{8\pi}} \, e^{i\varphi} \, \sin\theta \, , &   \nn \\[5pt]
	&Y_{20} = \phantom{-}\sqrt{\frac{5}{16\pi}} \,  (3\cos^2\theta - 1) \, , &  
	&Y_{21} = -\sqrt{\frac{15}{8\pi}} \,  e^{i\varphi} \, \sin\theta\cos\theta  \, , & 
	&Y_{22} = \phantom{-}\sqrt{\frac{15}{32\pi}} \,  e^{2i\varphi} \, \sin^2 \theta \, , & \nn 
\end{align}
where $m < 0$ components are given by the reflection property
\begin{align}
	Y_{\ell m}^{*}(\theta,\varphi) = (-1)^{m} \, Y_{\ell \, -m}(\theta,\varphi) \, .
\end{align}
The spherical harmonics are orthonormal over the entire solid angle,
\begin{align}
	\int_0^{2\pi}\diff \varphi \int_{-1}^{+1} \diff \cos\theta \, Y_{\ell' m'}^{*}(\theta,\varphi)Y_{\ell m}(\theta,\varphi) = \delta_{\ell'\ell} \, \delta_{m'm} \, ,
\end{align}
and satisfy the spherical addition theorem
\begin{align}
	P_{\ell}(z) = \frac{4\pi}{2\ell + 1} \sum_{m = -\ell}^{\ell} \, Y_{\ell m}^{*}(\theta',\varphi')Y_{\ell m}(\theta,\varphi) \, ,
\end{align}
where $z = \cos\theta \cos\theta' + \sin\theta\sin\theta' \cos(\varphi' - \varphi)$ and $P_{\ell}$ are the Legendre functions of the 1\ts{st}-kind. Note that if $m = 0$ for all $\ell \in \Nbb_0$, then the spherical harmonics are related to the Legendre functions as
\begin{align}
	Y_{\ell 0}(\theta,\varphi) = \sqrt{\frac{2\ell + 1}{4\pi}} \, P_{\ell}(z) \, . \nn 
\end{align}
%

\subsection{Wigner Rotation Matrix Elements}
Here we summarize some useful properties of Wigner rotation matrix elements. A detailed review can be found in Ref.~\cite{VMK}. A rotation of $\theta$ about an axis $\bh{n}$ is given by the unitary operator $\Rc_{\bh{n}}(\theta) = e^{-i \theta\, \mathbf{J}\cdot\bh{n}}$ where $\mathbf{J}$ is the angular momentum operator. For a generic rotation about the Euler angles $\alpha,\beta,\gamma$ defined by 
\begin{align}
	\Rc(\alpha,\beta,\gamma) \equiv \Rc_{\bh{z}}(\alpha) \cdot \Rc_{\bh{y}}(\beta) \cdot \Rc_{\bh{z}}(\gamma) \, . \nn 
\end{align}
Wigner $D$ matrix elements of $\Rc$ in a basis $\ket{jm}$, with representation $j$ and projection $m$ spanning $-j \le m \le j$, are defined as
\begin{align}
	D_{mm'}^{(j)}(\alpha,\beta,\gamma) & \equiv \bra{jm} \Rc(\alpha,\beta,\gamma) \ket{jm'} \, , \nn \\[5pt]
	& = e^{-im\alpha} \, d_{mm'}^{(j)}(\beta) \, e^{-im' \gamma} \, ,
\end{align}
where $d_{mm'}^{(j)}$ are the Wigner ``little'' $d$ matrix elements. The $d$ matrix elements can be expressed in terms of the Jacobi polynomials~\cite{VMK}
\begin{align}
	\label{eq:wigner_d_jacobi}
	d_{mm'}^{(j)}(\beta) = (-1)^{\eta} \sqrt{\frac{\sigma!(\sigma+\mu+\nu)!}{(\sigma+\mu)!(\sigma+\nu)!}} \, \xi_{mm'}(\cos\beta) \, P_\sigma^{(\mu,\nu)}(\cos\beta) \, ,
\end{align}
where $\mu = \lvert m - m' \rvert$, $\nu = \lvert m + m' \rvert$, $\sigma = j - (\mu + \nu)/2$, and the phase power $\eta = 0$ if $ m' \ge m$ and $\eta = m' - m$ if $m' < m$.  The function $\xi_{mm'}$ is known as the \emph{half-angle factor}~\cite{Collins:1971ff} and is defined as
\begin{align}
	\xi_{mm'}(z) = \left( \frac{1 - z}{2} \right)^{\lvert m - m' \rvert / 2} \left( \frac{1 + z}{2} \right)^{\lvert m + m' \rvert / 2} \, ,
\end{align}
which is singular at $z = \pm 1$ and symmetric under the interchange $m \leftrightarrow m'$, 
\begin{align}
	\xi_{mm'}(z) = \xi_{m'm}(z) \, . \nn 
\end{align}
The Jacobi polynomials $P_\sigma^{(\mu,\nu)}(z)$ are regular functions of real $z$,
\begin{align}
    P_\sigma^{(\mu,\nu)}(z) = \sum_{n=0}^\sigma \frac{(\sigma+\mu)! (\sigma+\nu)!}{n!(\sigma+\mu-n)!(\nu+n)!(\sigma-n)!} 
    \, \left( \frac{z-1}{2} \right)^{\sigma - n} \, \left( \frac{z+1}{2} \right)^{n} \, ,
\end{align}
where we require $\sigma$, $\sigma + \mu$, $\sigma + \nu$, and $\sigma + \mu + \nu$ are non-negative integers~\cite{abramowitz+stegun}.

The Wigner $d$ matrix elements themselves have numerous symmetry relations, most importantly for this work
\begin{align}
    d_{mm'}^{(j)} = (-1)^{m'-m}d_{m'm}^{(j)} = d_{-m'-m}^{(j)} \, , \nn 
\end{align}
and
\begin{align}
    \label{eq:app.d_sym}
    d_{m m'}^{(j)}(\beta) = d_{m'm}^{(j)}(-\beta) ,\ ,
\end{align}
which are discussed in Ref.~\cite{VMK}.

Like the spherical harmonics, the Wigner $d$ matrix elements respect an orthogonality condition over the interval $-1 \le \cos\beta \le +1$,
\begin{align}
	\int_{-1}^{+1}\diff \cos\beta \, d_{mm'}^{(j)}(\beta) \, d_{\bar{m}\bar{m}'}^{(\bar{j})}(\beta) = \frac{2}{2j+1} \, \delta_{j\bar{j}} \, \delta_{m\bar{m}} \, \delta_{m'\bar{m}'} \, 
\end{align}
which can be seen by the definition~\eqref{eq:wigner_d_jacobi} and using the orthogonality condition of the Jacobi polynomials over the same domain
\begin{align}
    \label{eq:Jacobi_orthog}
	\int_{-1}^{+1}\!\diff z \, & (1-z)^{\mu} (1+z)^{\nu} \,  P_{\sigma}^{(\mu,\nu)}(z) P_{\rho}^{(\mu,\nu)}(z) \, \nn \\[5pt]
	& = \frac{2^{\mu+\nu+1}}{2\sigma + \mu + \nu + 1} \, \frac{(\sigma + \mu)! (\sigma + \nu )!}{\sigma! \, (\sigma + \mu + \nu)!} \, \delta_{\sigma \rho} \, ,
\end{align}
for $\mu,\nu > -1$~\cite{abramowitz+stegun}. The addition theorem for Wigner $d$ matrix elements~\cite{VMK} is given by
\begin{align}
	\sum_{m''} d_{m m''}^{(j)}(\beta_1) \, d_{m''m'}^{(j)}(\beta_2) \, e^{-im'' \varphi} = D_{mm'}^{(j)}(\alpha,\beta,\gamma) \, ,
\end{align}
where the Euler angles $\alpha$, $\beta$, and $\gamma$ are given by the following relations~\cite{VMK}
\begin{subequations}
	\begin{align}
		\cot\alpha & = \cos\beta_1 \cot\varphi + \cot\beta_2 \,\frac{\sin\beta_1}{\sin\varphi} \, , \\[5pt]
		\cos\beta & = \cos\beta_1\cos\beta_2 - \sin\beta_1\sin\beta_2 \cos\varphi \, , \\[5pt]
		\cot\gamma & = \cos\beta_2 \cot\varphi + \cot\beta_1 \,\frac{\sin\beta_2}{\sin\varphi}  \, .
	\end{align}
\end{subequations}
The sign of $\alpha$ and $\gamma$ can be fixed by the relation
\begin{align}
	\frac{\sin\alpha}{\sin\beta_1} = \frac{\sin\gamma}{\sin\beta_2} = \frac{\sin\varphi}{\sin\beta} \, .
\end{align}
Finally, the Wigner $D$ matrix elements are related to the spherical harmonics as
\begin{align}
	\label{eq:Ylm_to_Wigner}
	Y_{\ell m}(\theta,\varphi) = \sqrt{\frac{2\ell + 1}{4\pi}} \, D_{m0}^{(\ell)\,*}(\varphi,\theta,0) \, .
\end{align}
%

\section{Evaluation of the \mathbshead{\Cc}{Ccoef} integral \texorpdfstring{Eq.~\eqref{eq:C_integral}}{Eq}}
\label{sec:app.C_integral_eval}

Here we provide a derivation of the closed-form solution of the $\Cc$ integral Eq.~\eqref{eq:C_integral}, repeated here for convenience
\begin{align}
	\Cc_{j,\lambda'\lambda}^{J} = \frac{1}{2} \int_{-1}^{+1}\! \diff z \, \xi_{\lambda\lambda'}(z) \, d_{\lambda\lambda'}^{(J)}(z) \, P_{j}(z) \, . \nn
\end{align}
To evaluate this integral, we first recognize that $P_j(z) = d_{00}^{(j)}(z)$ for any $j$ and $z$. Therefore, we can use the Clebsch-Gordan expansion~\cite{VMK} to reduce the product of Wigner $d$ functions to a single element
\begin{align}
	d_{\lambda\lambda'}^{(J)}(z) \, P_j(z) & = d_{\lambda\lambda'}^{(J)}(z) \, d_{00}^{(j)}(z) \, , \nn \\[5pt]
	& = \sum_{n = \lvert J - j \rvert}^{J+j} \braket{n\lambda' | J\lambda' , j 0} \braket{n \lambda | J \lambda, j 0} \, d_{\lambda\lambda'}^{(n)} (z) \, ,
\end{align}
which reduces the coefficient to
\begin{align}
	\Cc_{j,\lambda'\lambda}^{J} = \frac{1}{2} \sum_{n = \lvert J - j \rvert}^{J+j} \braket{n\lambda' | J\lambda' , j 0} \braket{n \lambda | J \lambda, j 0} \int_{-1}^{+1}\! \diff z \, \xi_{\lambda\lambda'}(z) \, d_{\lambda \lambda'}^{(n)} (z) \, . 
\end{align}

Next we write the Wigner $d_{\lambda\lambda'}^{(n)}$ matrix element in terms of the Jacobi polynomials $P_{\sigma}^{(\mu,\nu)}$, as given in App.~\ref{sec:app.angular_momentum}. Using the expression Eq.~\eqref{eq:wigner_d_jacobi}, the integral takes the form 
\begin{align}
	\label{eq:C_integral_jacobi}
	\Cc_{j,\lambda'\lambda}^{J} & = \frac{1}{2} \sum_{n = \lvert J - j \rvert}^{J+j} \braket{n\lambda' | J\lambda' , j 0} \braket{n \lambda | J \lambda, j 0} \, \nn \\[5pt]
	& \qquad \times (-1)^{\eta} \sqrt{\frac{\sigma!(\sigma+\mu+\nu)!}{(\sigma+\mu)!(\sigma+\nu)!}} \int_{-1}^{+1}\! \diff z \, \xi_{\lambda\lambda'}^2(z) \, P_{\sigma}^{(\mu,\nu)} (z) \, ,
\end{align}
where $\mu = \lvert \lambda - \lambda'\rvert$, $\nu = \lvert\lambda + \lambda' \rvert$, and $\sigma = n - (\mu + \nu)/2$. The phase is such that $\eta = 0$ if $ \lambda' \ge \lambda$ and $\eta = \lambda' - \lambda$ if $\lambda' < \lambda$. The advantage here is that the Jacobi polynomials $P_{\sigma}^{(\mu,\nu)}(z)$ are orthogonal over the interval $z \in [-1,1]$ with the weight $(1-z)^{\mu}(1+z)^{\nu}$. This is precisely the form of the integral in Eq.~\eqref{eq:C_integral_jacobi} since $\xi_{\lambda\lambda'}^2(z) \propto (1-z)^{\mu}(1+z)^{\nu}$ which removes the square root singular behavior, and for any $\mu$, $\nu$. We also note that for any $\mu$ and $\nu$ with $\sigma = 0$~\cite{abramowitz+stegun}, 
\begin{align}
	P_0^{(\mu,\nu)}(z) = 1 \, .
\end{align}
Therefore, applying the orthogonality condition Eq.~\eqref{eq:Jacobi_orthog} yields the relation
\begin{align}
	\int_{-1}^{+1} \! \diff z \, & \xi_{\lambda\lambda'}^{2}(z) \, P_{\sigma}^{(\mu,\nu)}(z) \, \nn \\[5pt]
	& = \frac{1}{2^{\mu+\nu}} \int_{-1}^{+1}\!\diff z \, (1-z)^{\mu} (1+z)^{\nu} \,  P_{\sigma}^{(\mu,\nu)}(z) P_{0}^{(\mu,\nu)}(z) \, , \nn \\[5pt]
	& = \frac{2}{2\sigma + \mu + \nu + 1} \, \frac{(\sigma + \mu)! (\sigma + \nu )!}{\sigma! \, (\sigma + \mu + \nu)!} \, \delta_{\sigma 0} \, .
\end{align}
Inserting this result into Eq.~\eqref{eq:C_integral_jacobi} gives
\begin{align}
	\Cc_{j,\lambda'\lambda}^{J} & = \sum_{n = \lvert J - j \rvert}^{J+j} \braket{n\lambda' | J\lambda' , j 0} \braket{n \lambda | J \lambda, j 0} \, \nn \\[5pt]
	& \qquad \times  \frac{(-1)^{\eta}}{2\sigma + \mu + \nu + 1} \, \sqrt{\frac{(\sigma + \mu)! (\sigma + \nu )!}{\sigma! \, (\sigma + \mu + \nu)!} } \, \delta_{\sigma 0}\, .
\end{align}
The Kronecker delta enforces $\sigma = 0$, which fixes $n = (\mu + \nu)/2$. Thus the sum in Eq.~\eqref{eq:C_integral_jacobi} has only a single non-zero term, giving
\begin{align}
	\Cc_{j,\lambda'\lambda}^{J} & = \braket{J_{\min}\lambda' | J\lambda' , j 0} \braket{J_{\min} \lambda | J \lambda, j 0}   \frac{(-1)^{\eta}}{\mu + \nu + 1} \, \sqrt{\frac{\mu! \, \nu!}{(\mu + \nu)!} } \, .
\end{align}
Finally, we note the relation between $J_{\min}$ and $\mu$, $\nu$,
\begin{align}
	J_{\min} = \max(\lvert \lambda' \rvert, \lvert \lambda \rvert) = \frac{1}{2}\left( \lvert \lambda' - \lambda \rvert + \lvert \lambda' + \lambda \rvert \right) = \frac{\mu + \nu}{2} \, , \nn 
\end{align}
giving our final result for the $\Cc$ coefficient Eq.~\eqref{eq:C_result}, repeated here for convenience
\begin{align}
	\Cc_{j,\lambda'\lambda}^{J} & = \frac{(-1)^{\eta}}{2J_{\min} + 1} \, \sqrt{\frac{\lvert \lambda' - \lambda \rvert! \,  \lvert \lambda' + \lambda \rvert!}{(2J_{\min})!} } \,  \braket{J_{\min}\lambda' | J\lambda' , j 0} \braket{J_{\min} \lambda | J \lambda, j 0}  \, . \nn 
\end{align}
%

\section{Partial wave projection in generic reference frames}
\label{sec:app.generic_frames}

In the main body of our work, Sec.~\ref{sec:ref_frames} and \ref{sec:helicity}, we stated that without loss of generality, that we can orient a coordinate system $XYZ$ so that the initial pair momentum in the CM frame was aligned with the $Z$-axis, $\P_k = \bh{Z}$, and the final pair momentum lies in the $XZ$-plane at a polar angle $\theta_p$ from the $Z$-axis, so that the CM frame scattering angle $\theta_p = \theta_{pk}$. In this Appendix, we show that one can construct a partial wave expansion with respect to a generic space-fixed coordinate system. One practical reason for considering expansions in a generic coordinate system is in implementing the results of this work to $\3\to\3$ amplitudes which are constructed by summing over all pair-spectator combinations, which is what is proposed in the formulation of the scattering formalism~\cite{Hansen:2015zga,Jackura:2022gib}. As different pair-spectator systems require their own coordinate system to define momenta and angles, imposing a global external coordinate system with which all pair-spectator systems can be related allows one to define an expansion for the complete amplitude. This application is outside the scope of this work, thus we did not discuss details, but point the reader to Refs.~\cite{Ascoli:1975mn,Mikhasenko:2019vhk,JPAC:2019ufm} which discuss aspects of this procedure. However, we intend that this appendix be useful for future study on that application as well as extensions to analyses of higher few-body systems.

\begin{figure}[t]
	\centering
	\includegraphics[width=0.48\textwidth]{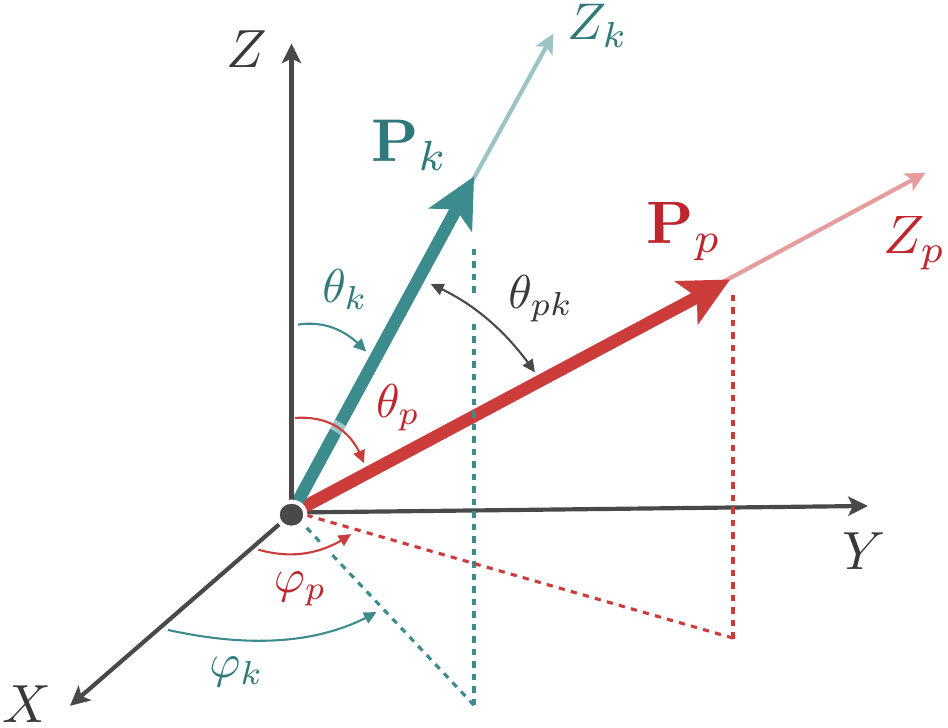}
	\caption{Orientations of the momenta $\P_k$ and $\P_p$ with respect to the generic space-fixed coordinate system $(X,Y,Z)$. The $Z_k$ and $Z_p$ axes are associated with body-fixed coordinate systems which are defined fixed to the momenta $\P_k$ and $\P_p$, respectively.}
	\label{fig:pair_momenta_cm_angles}
\end{figure}
Let $\P_k$ and $\P_p$ be the initial and final pair momenta in the total CM frame defined with respect to some generic space-fixed coordinate system $XYZ$. Then, the polar and azimuthal angles of $\P_k$ are $\theta_k$ and $\varphi_k$, respectively, while the polar and azimuthal angles of $\P_p$ are $\theta_p$ and $\varphi_p$, respectively. These orientations are depicted in Fig.~\ref{fig:pair_momenta_cm_angles}, where we also introduce the CM frame effective scattering angle $\theta_{pk}$ defined with the usual addition of spherical angles
\begin{align}
    \label{eq:app.cos_theta_pk}
    \cos\theta_{pk} & = \bh{P}_p \cdot \bh{P}_k \, , \nn \\[5pt]
    & = \cos\theta_p \cos\theta_k + \sin\theta_p \sin\theta_k \cos(\varphi_k - \varphi_p) \, , 
\end{align}
which follows from decomposing $\bh{P}_p$ and $\bh{P}_k$ into Cartesian components with respect to the space-fixed coordinate system. The reaction plane in the CM frame is now defined with a unit normal vector 
\begin{align}
    \bh{n} = \frac{\P_k \times \P_p }{\lvert \P_k \times \P_p  \rvert} \, , \nn 
\end{align}
and is still used to define the azimuthal angles of the initial and final state pair rest frames as discussed in Sec.~\ref{sec:ref_frames}. Note that when we orient $\bh{P}_k = \bh{Z}$, we recover that $\theta_{p} = \theta_{pk}$ and $\bh{n} = \bh{Y}$, as is chosen in the main text.

The partial wave expansion proceeds as in Sec.~\ref{sec:helicity}, up through Eq.~\eqref{eq:helicity_expand_simplify}, which we repeat here for convenience,
\begin{align}
    \label{eq:app.hel_pw_expand}
	\Mc_{\ell'\lambda', \ell\lambda}(\p,\k) & = \sum_{J = J_{\min}}^{\infty} (2J+1) \,  \Mc_{\ell'\lambda' , \ell\lambda}^{J}(p,k) \, \sum_{m_J = -J}^{J} D_{m_{J}\lambda'}^{(J)\,*}(\bh{P}_p) \, D_{m_{J}\lambda}^{(J)}(\bh{P}_k) \, . \nn 
\end{align}
The task now is to simplify the sum on $m_J$. Similar to spherical harmonics, the Wigner $D$ matrix elements have an addition theorem which allows us to reduce the composition of two rotation functions to a single rotation, \cf App.~\ref{sec:app.angular_momentum}. For the rotation functions in Eq.~\eqref{eq:helicity_expand_simplify}, we find the following~\cite{wigner_euler}
\begin{align}
    \sum_{m_J} D_{m_J \lambda'}^{(J)\,*}(\bh{P}_p)D_{m_J \lambda}^{(J)}(\bh{P}_k) & = \sum_{m_J} D_{m_J \lambda'}^{(J)\,*}(\varphi_p,\theta_p,0)D_{m_J \lambda}^{(J)}(\varphi_k,\theta_k,0) \, , \nn \\[5pt]
    & = \sum_{m_J} d_{m_J\lambda'}^{(J)}(\theta_p) \, d_{m_J\lambda}^{(J)}(\theta_k) \, e^{-im_J(\varphi_{k} - \varphi_{p}) }\, , \nn \\[5pt]
    & = \sum_{m_J} d_{\lambda m_J}^{(J)}(-\theta_k) d_{m_J \lambda'}^{(J)}(\theta_p) \, e^{-im_J(\varphi_k - \varphi_p)} \, ,
\end{align}
where in the last line we used the symmetry identity Eq.~\eqref{eq:app.d_sym}. From the addition theorem, we find that we can express this sum as a single Wigner $D$ matrix element,
\begin{align}
    \label{eq:app.use_wigner_add}
    \sum_{m_J} D_{m_J \lambda'}^{(J)\,*}(\bh{P}_p)D_{m_J \lambda}^{(J)}(\bh{P}_k) & = \sum_{m_J} d_{\lambda m_J}^{(J)}(-\theta_k) d_{m_J \lambda'}^{(J)}(\theta_p) \, e^{-im_J(\varphi_k - \varphi_p)} \, \nn \\[5pt]
    &= e^{i\lambda\varphi_{pk}} \, d_{\lambda\lambda'}^{(J)} (\theta_{pk}) \, e^{i\lambda' \psi_{pk}} \, , \nn \\[5pt]
    & = D_{\lambda\lambda'}^{(J) \, * }(\varphi_{pk},\theta_{pk},\psi_{pk}) \, ,
\end{align}
where the resulting Euler angles are given by the usual addition of rotation matrices as summarized in App.~\ref{sec:app.angular_momentum}. Explicitly, the total CM frame scattering angle $\theta_{pk}$ is given by $\cos\theta_{pk} \equiv \bh{P}_p\cdot\bh{P}_k$ as is defined in Eq.~\eqref{eq:app.cos_theta_pk}, while the azimuthal angles $\varphi_{pk}$ and $\psi_{pk}$ are fully specified by the following relations
\begin{align}
	\label{eq:euler_combine}
	\begin{gathered}
		\cot\varphi_{pk} =  -\cos\theta_{k} \, \cot(\varphi_k - \varphi_p) + \cot\theta_p \, \frac{\sin\theta_k}{\sin(\varphi_k - \varphi_p)}  \, , \\[5pt]
		\cot\psi_{pk} = -\cos\theta_{p} \, \cot(\varphi_k - \varphi_p) + \cot\theta_k \,\frac{\sin\theta_p}{\sin(\varphi_k - \varphi_p)} \, , \\[5pt]
		-\frac{\sin\varphi_{pk}}{\sin\theta_{p}} = \frac{\sin\psi_{pk}}{\sin\theta_{k}} = \frac{\sin(\varphi_k - \varphi_p)}{\sin\theta_{pk}} \, .
	\end{gathered}	
\end{align}

After using the addition theorem on the Wigner $d$ matrix elements, Eq.~\eqref{eq:app.use_wigner_add}, we find that the helicity partial wave expansion Eq.~\eqref{eq:app.hel_pw_expand} reduces to
\begin{align}
    \label{eq:app.hel_pw_expand_reduce}
	\Mc_{\ell'\lambda', \ell\lambda}(\p,\k) = \sum_{J}(2J+1) \, \Mc_{\ell'\lambda' , \ell \lambda}^J(p,k) \, D_{\lambda\lambda'}^{(J)\,*}(\varphi_{pk},\theta_{pk},\psi_{pk}) \, ,
\end{align}
which can be inverted to project a helicity amplitude into its partial waves
\begin{align}
    \label{eq:app.hel_pw.proj}
	\Mc_{\ell'\lambda',\ell \lambda}^J(p,k) & = \frac{1}{8\pi^2} \int_{0}^{2\pi} \!\diff\psi_{pk} \int_{0}^{2\pi} \!\diff \varphi_{pk} \, \nn \\[5pt]
	& \qquad\qquad  \times \int_{-1}^{+1} \!\diff\cos\theta_{pk} \, D_{\lambda\lambda'}^{(J)}(\varphi_{pk},\theta_{pk},\psi_{pk}) \, \Mc_{\ell'\lambda',\ell \lambda}(\p,\k) \, .
\end{align}
Comparing to the kinematics outlined in Sec.~\ref{sec:amplitudes}, it seems that there are additional independent variables in the form of the azimuthal dependencies $\varphi_{pk}$ and $\psi_{pk}$. However, these azimuthal angles are non-dynamical in the sense that they orient the reaction plane with respect to our arbitrary external coordinate system. To see this, first consider the simple limit where $\bh{\P}_k = \bh{Z}$. Therefore, $\theta_k = \varphi_k = 0$. From Eq.~\eqref{eq:euler_combine} we find that as $\theta_k \to 0$ and $\varphi_k \to 0$, that the Euler angles $\varphi_{pk} \to \varphi_{p}$, $\theta_{pk} \to \theta_{p}$, and $\psi_{pk} \to 0$. Thus, we have removed one of the azimuthal angles, and the angular momentum composition rule Eq.~\eqref{eq:app.use_wigner_add} gives
\begin{align}
	\sum_{m_J} D_{m_J\lambda'}^{(J)\,*}(\bh{P}_p) D_{m_J\lambda}^{(J)}(\bh{P}_k) & = D_{\lambda\lambda'}^{(J)\,*}(\varphi_{p},\theta_{p},0) \nn  \, .
\end{align}
The remaining angle $\varphi_{pk} = \varphi_p$ is not dynamical, as it only orients the reaction plane with respect to the external coordinate system, $\cos\varphi_p = \bh{P}_p \cdot \bh{X}$.  Therefore, we can rotate the system about the helicity quantization axis by and angle of $-\varphi_{p}$, which preserves the helicity~\cite{Jacob:1959at}, to eliminate this redundant angle and arrive at our result in Eq.~\eqref{eq:Wigner_D_addition}. 

In a similar manner, we can simultaneously rotate away both azimuthal angles in Eq.~\eqref{eq:app.hel_pw_expand_reduce},
\begin{align}
    \label{eq:app.rotate_helicity}
	\sum_{\lambda',\lambda}D_{\bar{\lambda}'\lambda'}^{(J)} (0,0,\varphi_{pk}) \, \Mc_{\ell' \lambda', \ell \lambda}(\p,\k) \, D_{\lambda \bar{\lambda}}^{(J)}(\psi_{pk},0,0) = \Mc_{\ell' \bar{\lambda}', \ell \bar{\lambda}}(\p,\k) \, \Big\rvert_{\varphi_{pk} = \psi_{pk} = 0} \, .
\end{align}
This transformation leaves the magnitude of the pair momenta invariant as they are aligned with their respective quantization axes. We conclude that the angles $\psi_{pk}$ and $\varphi_{pk}$ are non-dynamical variables and are arbitrary rotations which arise from the definitions of the three-particle states with respect to the spatial coordinate system. Therefore, if one rotates the system as in Eq.~\eqref{eq:app.rotate_helicity} to remove the $\varphi_{pk}$ and $\psi_{pk}$ angles, we arrive exactly at the expansion as originally presented in Eq.~\eqref{eq:helicity_pw_expand} of Sec.~\ref{sec:helicity}.

Finally, we consider in detail the consequence of a generic reference frame to the OPE. In particular, we show that the dependence on the non-dynamical angles explicitly cancels in performing the partial-wave projection of the OPE. We begin by rewriting Eq.~\eqref{eq:spin_helicity} here for convenience,
\begin{align}
	\Hc_{\lambda'\lambda}^{(\ell'\ell)}(\p,\k) \equiv \left(\frac{k_p^{\star}}{q_p^{\star}}\right)^{\ell'} 4\pi \, Y_{\ell'\lambda'}^{*}(\bh{\k}_p^{\star}) Y_{\ell\lambda}(\bh{\p}_k^{\star}) \left(\frac{p_k^{\star}}{q_k^{\star}}\right)^{\ell} \, . \nn 
\end{align}
We need to decompose the vectors $\k_p^\star$ and $\p_k^\star$ with respect to the general coordinate system, as well as describe their Lorentz transformations. In the total CM frame, the spectator momenta are anti-parallel to the pair momenta, \eg for the final spectator $\p = -\P_p$, thus its polar and azimuthal angles with respect to the space-fixed coordinate system are $\pi - \theta_p$ and $\varphi_p + \pi$, respectively. Let us focus on obtaining the angles $\p_k^\star$ in terms of the various coordinate system. First, Let us define the Lorentz transformations between the total CM frame and the pair rest frame for spectator $k$,
\begin{subequations}
    \begin{align}
        \label{eq:app.Lorentz_par}
        (\p_k^\star)_\parallel & = \gamma_k(\p_\parallel - \bs{\beta}_k \omega_p) \, , \nn \\[5pt]
        & = \gamma_k\left( \frac{\p\cdot \bs{\beta}_k}{\beta_k^2} -\omega_p \right)  \, \bs{\beta}_k \, , \\[5pt]
        \label{eq:app.Lorentz_per}
        (\p_k^\star)_\perp & = \p_\perp \equiv \p - \p_\parallel \, , \nn \\[5pt]
        & = \p - \left(\frac{\p\cdot \bs{\beta}_k}{\beta_k^2} \right) \, \bs{\beta}_k \, ,
    \end{align}
\end{subequations}
where we recall that $\bs{\beta}_k = \P_k / E_k$ and $\gamma_k = 1/\sqrt{1 - \beta_k^2}$.

Let us first define the coordinates $(x_k,y_k,z_k)$ which are fixed to the reaction plane for this system. Similar to Sec.~\ref{sec:ref_frames}, the $z_k$-axis is defined as $\bh{z}_k = \P_k / \lvert \P_k \rvert$, and the $y_k$-axis is given by $\bh{y}_k = \P_k \times \P_p / \lvert \P_k \times \P_p \rvert$, which is the unit normal to the reaction plane. Therefore, the $x_k$-axis is defined by the unit vector $\bh{x}_k = \bh{y}_k \times \bh{z}_k$. Note that $\lvert \P_k \times \P_p \rvert = pk\sin\theta_{pk} = pk\sqrt{1-\cos^2\theta_{pk}}$. Thus we find that the Lorentz transformation Eq.~\eqref{eq:app.Lorentz_par} is given by
\begin{align}
    \label{eq:pkstar_parallel}
    (\p_k^\star)_\parallel \cdot \bh{z}_k & = p_k^\star\, \cos(\chi_k^\star) \, , \nn \\[5pt]
    & = \gamma_k(- p\cos\theta_{pk} -\omega_p \beta_k ) \, ,
\end{align}
where $\bh{z}_k = \bh{\bs{\beta}}_k$, and from the perpendicular component Eq.~\eqref{eq:app.Lorentz_per},
\begin{align}
    \label{eq:pkstar_perp}
    (\p_k^\star)_{\perp} \cdot \bh{x}_k & = - p_k^\star\sin\chi_k^\star \, , \nn \\[5pt]
    & = - p \sin\theta_{pk} \, .
\end{align}
Note that we have recovered the Lorentz transformations as detailed in Sec.~\ref{sec:exchange}.

\begin{figure*}[t]
	\centering
	\includegraphics[width=0.98\textwidth]{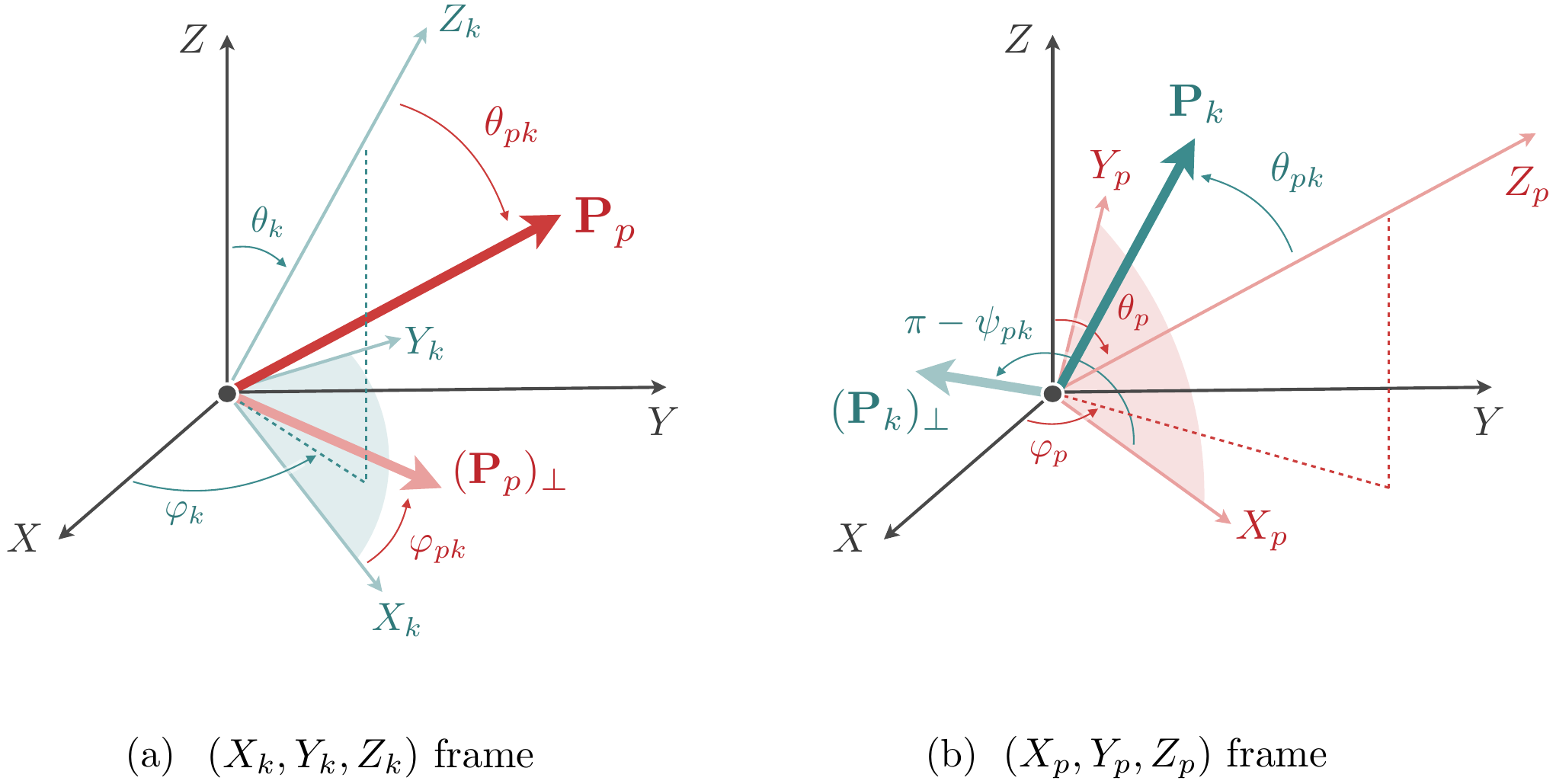}
	\caption{ (a) Polar and azimuthal angles of $\P_p$ with respect to the body-fixed $(X_k,Y_k,Z_k)$ frame. The polar angle is $\theta_{pk}$ which is defined with respect to the $Z_k$-axis. The perpendicular component $\p_\perp$ lies in the $X_kY_k$-plane, which is shaded blue, with an azimuthal angle defined about the $Z_k$-axis with respect to $X_k$. (b) Polar and azimuthal angles of $\P_k$ with respect to the body-fixed $(X_p,Y_p,Z_p)$ system. The polar angle is defined as $\theta_{pk}$, while the azimuthal angle is defined in the $X_pY_p$-plane, shaded red, with respect to the $X_p$ axis to be $\pi - \psi_{pk}$.}
	\label{fig:euler_body_fixed_frames}
\end{figure*}
Next we define \emph{body-fixed} coordinates $(X_k,Y_k,Z_k)$ which are defined fixed to the pair momentum $\P_k$. Specifically, the $Z_k$-axis is defined as $\bh{Z}_k = \bh{P}_k$, and the $Y_k$-axis is perpendicular to the plane formed by the vector $Z_k$ and the space-fixed $Z$-axis as $\bh{Y}_k = \bh{Z} \times \bh{Z}_k / \lvert \bh{Z} \times \bh{Z}_k \rvert$, where we note that $\lvert \bh{Z} \times \bh{Z}_k \rvert = \sin\theta_k$. Then, the $X_k$-axis is given by $\bh{X}_k = \bh{Y}_k \times \bh{Z}_k$. Note that if we align $\P_k$ with the $Z$-axis, then the body-fixed coordinate system is identical to the space-fixed coordinate system $(X,Y,Z)$. 

To compute the angles of the momentum $\p_k^\star$ with respect to this coordinate system, we notice that $\bh{Z}_k = \bh{z}_k$, thus the parallel component of $\p_k^\star$ has the same form as the Lorentz transformations with respect to the reaction plane coordinate system, Eq.~\eqref{eq:pkstar_parallel}.
Therefore, we only need to work with the perpendicular component to determine the azimuthal angles, \ie $\bh{X}_k \cdot \p_k^\star = \bh{X}_k \cdot (\p_k^\star)_\perp$ and $\bh{Y}_k \cdot \p_k^\star = \bh{Y}_k \cdot (\p_k^\star)_\perp$. Moreover, since the perpendicular component is unchanged under Lorentz transformations, Eq.~\eqref{eq:app.Lorentz_per}, and $\bh{\bs{\beta}}_k = \bh{Z}_k$ with $\bh{X}_k\cdot \bh{Z}_k = \bh{Y}_k \cdot \bh{Z}_k = 0$, then the azimuthal angles of $(\p_k^\star)_\perp$ are completely determined by $\p$. Thus, we evaluate the following scalar products, $\bh{X}_k \cdot \p_k^\star = \bh{X}_k \cdot \p$ and $\bh{Y}_k \cdot \p_k^\star = \bh{Y}_k \cdot \p$. Finally, recall that $\p = -\P_p$, so decomposing $\P_p$ with respect to the $(X_k,Y_k,Z_k)$ body fixed coordinates immediately yields the angles of $\p$.

\begin{figure*}[t]
	\centering
	\includegraphics[width=0.98\textwidth]{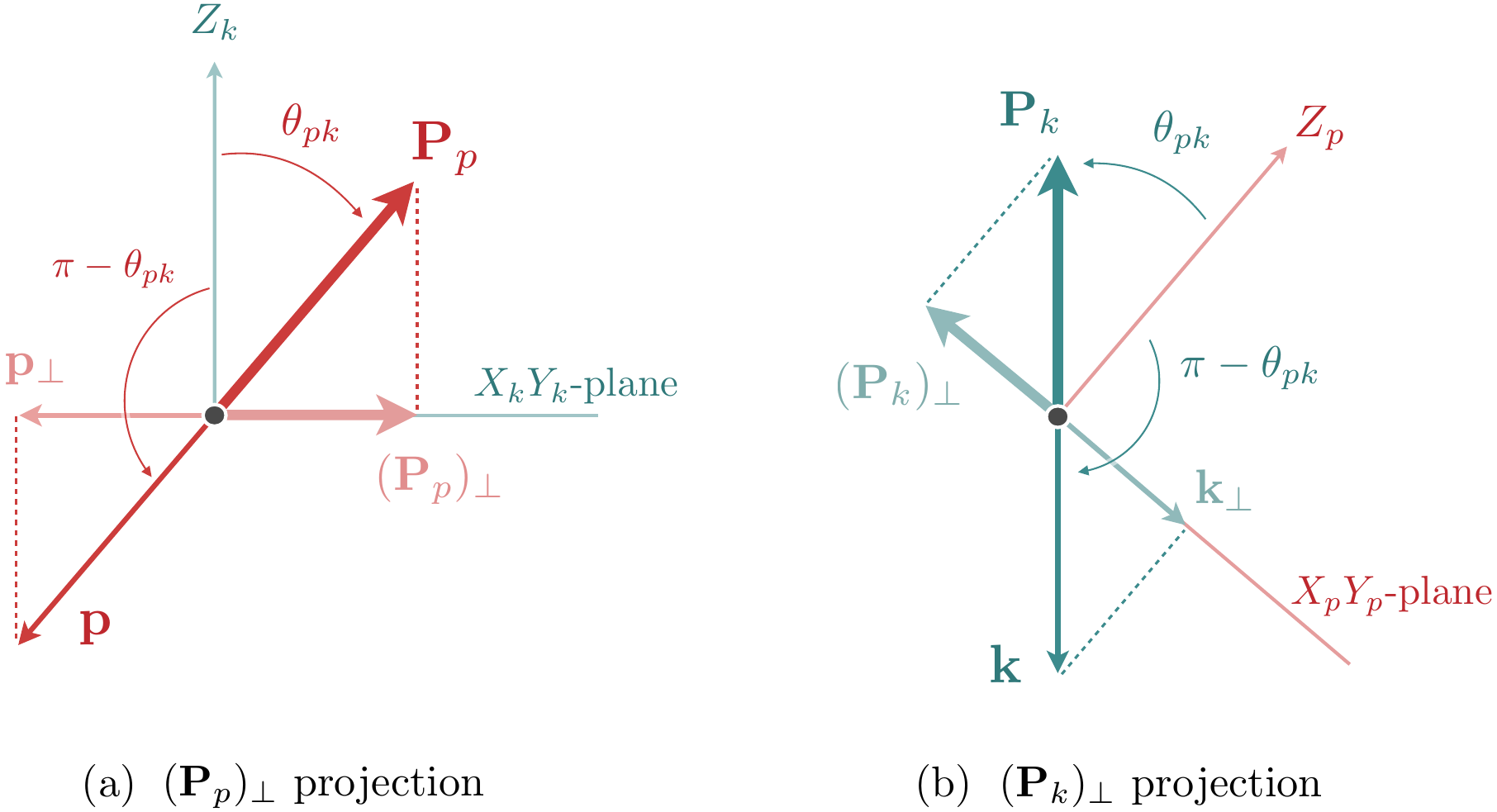}
	\caption{Projection of the perpendicular components of (a) $\P_p$ and $\p$ to the $X_kY_k$-plane, and (b) $\P_k$ and $\k$ to the $X_pY_p$-plane.}
	\label{fig:euler_projection}
\end{figure*}

In what we follows, we prove that azimuthal angle of $\p$ is $\varphi_{pk} + \pi$. We do this by first assuming this relation, which can be qualitatively argued from the addition of Wigner $D$ matrices shown Eq.~\eqref{eq:app.use_wigner_add}. Then, we relate the angles in the space-fixed to those in \emph{body-fixed} coordinates, illustrated in Fig.~\ref{fig:euler_body_fixed_frames} and defined below. Finally, we show that the resulting relations are equivalent to Eq.~\eqref{eq:euler_combine}. Once we establish this relation, we turn our attention to prove that the azimuthal angle of $\k$ is $-\psi_{pk}$, following the same procedure.

\begin{figure*}[t]
	\centering
	\includegraphics[width=0.98\textwidth]{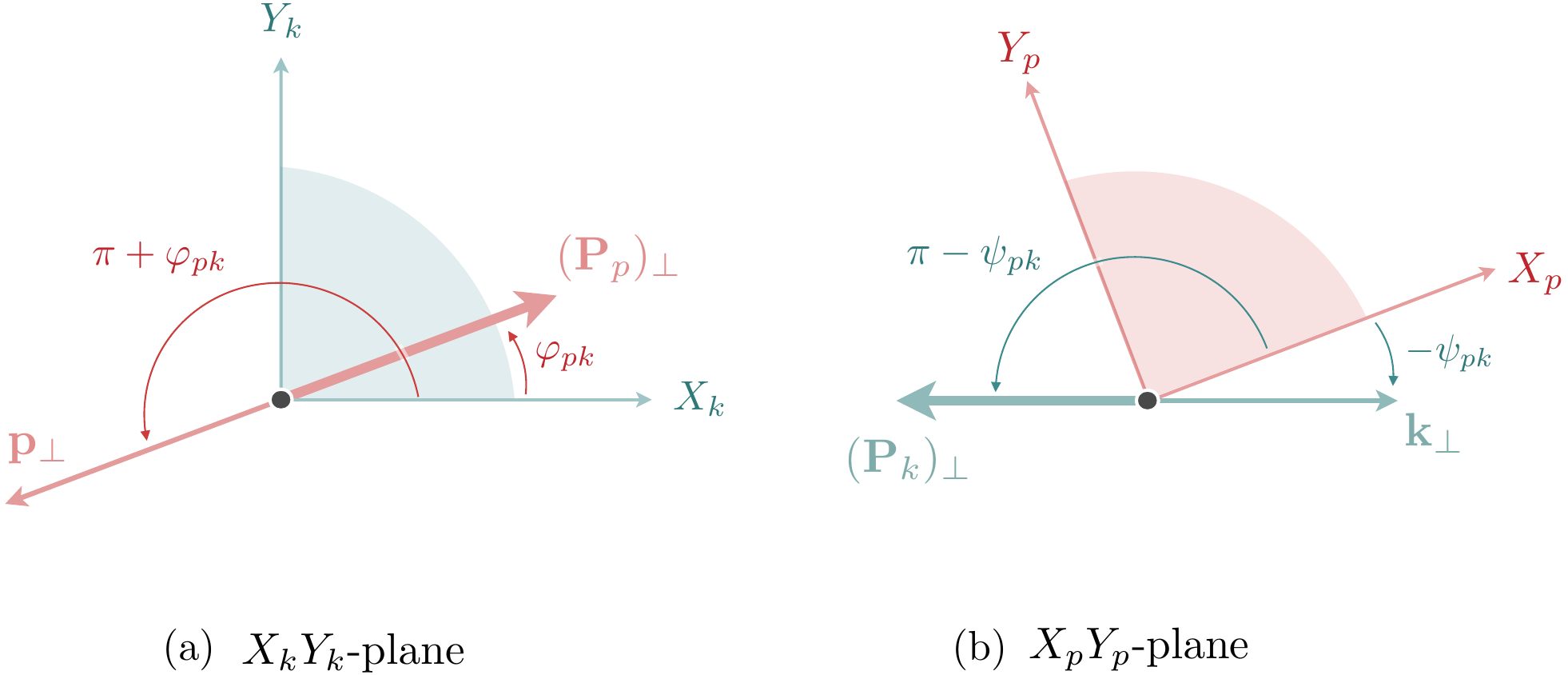}
	\caption{(a) Azimuthal angles of $\P_p$ and $\p$ defined in the $X_kY_k$-plane. (b) Azimuthal angles of $\P_k$ and $\k$ defined in the $X_pY_p$-plane.}
	\label{fig:euler_XY_plane}
\end{figure*}

We begin by first relating the vector $\P_p$ to the body-fixed coordinate system $(X_k,Y_k,Z_k)$, as illustrated in Fig.~\ref{fig:euler_body_fixed_frames} (a). Resolving $\P_p$ into a Cartesian coordinate system reveal that geometrically, $\bh{Y}_k \cdot \P_p = p\, \sin\theta_{pk} \, \sin\varphi_{pk}$ and $\bh{X}_k \cdot \P_p = p\,\sin\theta_{pk} \, \cos\varphi_{pk}$. This is further depicted in Figs.~\ref{fig:euler_projection} (a) and \ref{fig:euler_XY_plane} (a), which shows the decomposition in the $X_kZ_k$- and $X_kY_k$-planes, respectively. Here $\theta_{pk}$ is the angle of $\P_p$ with respect to the $Z_k$-axis (which is the definition of the effective CM frame scattering angle), and we have defined $\varphi_{pk}$ as the azimuthal angle of $\P_p$ with respect to the $X_k$-axis. Therefore, the polar and azimuthal angles of $\p$ in this frame are $\pi - \theta_{pk}$ and $\pi + \varphi_{pk}$, respectively.

To verify that $\varphi_{pk}$ is identical to the results from the addition theorem in Eq.~\eqref{eq:euler_combine}, we compute the scalar products with the relations between the unit vectors of the body-fixed coordinate system to those of the space-fixed system. By direct evaluation, we find
\begin{subequations}
   \begin{align}
        \label{eq:app.Yk_component}
        \bh{Y}_k \cdot \p_k^\star & = \bh{Y}_k \cdot (\p_k^\star)_\perp = \bh{Y}_k \cdot \p \, , \nn\\[5pt]
        & = - p\, \sin\theta_{pk} \, \sin\varphi_{pk} \, , \nn \\[5pt]
        & = p\, \sin\theta_p \, \sin(\varphi_k - \varphi_p) \, , \\[5pt]
        \label{eq:app.Xk_component}
        \bh{X}_k \cdot \p_k^\star & = \bh{X}_k \cdot (\p_k^\star)_\perp = \bh{X}_k \cdot \p \, , \nn\\[5pt]
        & = - p\,\sin\theta_{pk} \, \cos\varphi_{pk} \, , \nn \\[5pt]
        & = p \, \left(\cos\theta_p \, \sin\theta_k - \cos\theta_k \, \sin\theta_p \, \cos(\varphi_k - \varphi_p) \right) \, ,
    \end{align} 
\end{subequations}
where the second line in each of these equalities comes from the geometric decomposition shown in Figs.~\ref{fig:euler_projection} (a) and \ref{fig:euler_XY_plane} (a), while the third line is the result where we use the definitions of the unit vectors of the $(X_k, Y_k, Z_k)$ system with respect to the $(X,Y,Z)$ system. We emphasize that the negative signs on the second line of each equation is because $\p = -\P_p$.

The cotangent of the azimuthal angle of $\p_k^\star$ is given by taking the ratio of the $X_k$-component, Eq.~\eqref{eq:app.Xk_component}, to the $Y_k$-component, Eq.~\eqref{eq:app.Yk_component},
\begin{align}
    \frac{\bh{X}_k \cdot \p_k^\star}{\bh{Y}_k \cdot \p_k^\star} & = \cot\varphi_{pk}  \, , \nn \\[5pt]
    & = - \cos\theta_k \, \cot(\varphi_k - \varphi_p) + \cot\theta_p \, \frac{\sin\theta_k}{\sin(\varphi_k - \varphi_p)}  \, ,
\end{align}
which is identical to the one found in Eq.~\eqref{eq:euler_combine}. The sign is fixed from the relationship between $\sin\theta_{pk}$ and $\sin\theta_{p}$ from the $Y_k$-component in Eq.~\eqref{eq:app.Yk_component},
\begin{align}
    \sin\varphi_{pk} = - \frac{\sin\theta_p}{\sin\theta_{pk}} \, \sin(\varphi_k - \varphi_p) \, , \nn 
\end{align}
where the negative sign fixes the relative orientation to the coordinate system. Therefore, we conclude that the azimuthal angle of $\p_k^\star$ is $\varphi_{pk} + \pi$, where $\varphi_{pk}$ is given by Eq.~\eqref{eq:euler_combine}. Since $Y_{\ell \lambda}(\theta,\varphi) \propto P_\ell^\lambda(\cos\theta) \, e^{i\lambda \varphi}$, we find that the angular dependence is given by 
\begin{align}
    Y_{\ell \lambda}(\bh{p}_k^\star) \propto P_\ell^\lambda(\cos\chi_k^\star) \, e^{i\lambda\pi}\,e^{i\lambda \varphi_{pk}} \, . \nn 
\end{align}
Note that if we align $\P_k$ with the $Z$-axis, then $\varphi_{pk} = 0$, and we recover the $\p_k^\star$ piece contributing to Eq.~\eqref{eq:spin_helicity_pair_rest_frames}, namely that there is a phase $e^{i\lambda\pi} = (-1)^\lambda$.

We now show that the azimuthal angle of $\k$ is $-\psi_{pk}$. We follow the same procedure as before, first assuming that $-\psi_{pk}$ is the azimuth of $\k$, then relating the angles of $\k$ in the space-fixed system to the \emph{body-fixed} coordinates, \cf Fig.~\ref{fig:euler_body_fixed_frames}. We then recover the expressions shown from the addition theorem in Eq.~\eqref{eq:euler_combine}. We repeat the above analysis for the vector $\k_p^\star$ with respect to a body-fixed coordinate system $(X_p,Y_p,Z_p)$, where the $Z_p$-axis is $\bh{Z}_p = \bh{P}_p$, $\bh{Y}_p = \bh{Z} \times \bh{Z}_p / \lvert \bh{Z} \times \bh{Z}_p \rvert$, and $\bh{X}_p = \bh{Y}_p \times \bh{Z}_p$. These coordinates are fixed to the momentum $\P_p$. The Lorentz transformations are along the $Z_p$-axis,
\begin{subequations}
    \begin{align}
        (\k_p^\star)_\parallel & = \gamma_p(\k_\parallel - \bs{\beta}_p \omega_k) \, , \nn \\[5pt]
        & = \gamma_p\left( \frac{\k\cdot \bs{\beta}_p}{\beta_p^2} -\omega_k \right)  \, \bs{\beta}_p \, , \\[5pt]
        (\k_p^\star)_\perp & = \k_\perp \equiv \k - \k_\parallel \, , \nn \\[5pt]
        & = \k - \left(\frac{\k\cdot \bs{\beta}_p}{\beta_p^2} \right) \, \bs{\beta}_p \, ,
    \end{align}
\end{subequations}
with $\bs{\beta}_p = \P_p / E_p$ and $\gamma_p = 1/\sqrt{1 - \beta_p^2}$. Along the $Z_p$-axis, the Lorentz transformations give the known relation, Eq.~\eqref{eq:p_boost}, between $\chi_p^\star$ and $\theta_{pk}$, similar to the previous case.
\begin{align}
    (\k_p^\star)_\parallel \cdot \bh{Z}_p & = k_p^\star\, \cos(\chi_p^\star) \, , \nn \\[5pt]
    & = \gamma_p(- k\cos\theta_{pk} -\omega_k \beta_p ) \, ,
\end{align}
The azimuthal angles of $\k_p^\star$ are again found by considering its perpendicular component with respect to the $Z_p$-axis. For this case, we show that the azimuthal angle of $\k_p^\star$ is $-\psi_{pk}$, which is illustrated in Fig.~\ref{fig:euler_body_fixed_frames} (b). As with the previous case, the angles of $\k_p^\star$ in the $(X_p,Y_p,Z_p)$ body-fixed system are related to those of $\k = -\P_k$. Figures~\ref{fig:euler_projection} (b) and \ref{fig:euler_XY_plane} (b) show the decomposition of $\P_k$ with respect to the $(X_p,Y_p,Z_p)$ coordinate system. We find that $\bh{X}_p\cdot \P_k = k\sin\theta_{pk}\cos(\pi-\psi_{pk}) = -k\sin\theta_{pk}\cos\psi_{pk}$ and $\bh{Y}_p\cdot \P_k = k\sin\theta_{pk}\sin(\pi-\psi_{pk}) = k\sin\theta_{pk}\sin\psi_{pk}$, where $\theta_{pk}$ and $\pi - \psi_{pk}$ are the polar azimuthal angles of $\P_k$, respectively. The polar and azimuthal angles of $\k = -\P_k$ in this frame are therefore $\pi - \theta_{pk}$ and $-\psi_{pk}$, respectively, as shown in Figs.~\ref{fig:euler_projection} (b) and \ref{fig:euler_XY_plane} (b).

As before, we connect $\psi_{pk}$ to Eq.~\eqref{eq:euler_combine} by resolving $\k_p^\star$ into $X_k$- and $Y_k$-components and using the definitions of the body-fixed frame unit vectors in terms of the external coordinate system. We find
\begin{subequations}
   \begin{align}
        \label{eq:app.Yp_component}
        \bh{Y}_p \cdot \k_p^\star & = \bh{Y}_p \cdot (\k_p^\star)_\perp = \bh{Y}_p \cdot \k \, , \nn\\[5pt]
        & = - k\, \sin\theta_{pk} \, \sin\psi_{pk} \, \nn \\[5pt]
        & = - k\, \sin\theta_k \, \sin(\varphi_k - \varphi_p) \, , \\[5pt]
        \label{eq:app.Xp_component}
        \bh{X}_p \cdot \k_p^\star & = \bh{X}_p \cdot (\k_p^\star)_\perp = \bh{X}_p \cdot \k \, , \nn\\[5pt]
        & = k\, \sin\theta_{pk} \, \cos\psi_{pk} \, , \nn \\[5pt]
        & = k \, \left(\cos\theta_k \, \sin\theta_p - \cos\theta_p \, \sin\theta_k \, \cos(\varphi_k - \varphi_p) \right) \, ,
    \end{align} 
\end{subequations}
where the second line of each equality is due to the geometric decomposition of $\k$, noting that $\sin(\pi - \theta_{pk}) = \sin\theta_{pk}$, $\sin(-\psi_{pk}) = -\sin\psi_{pk}$, and $\cos(-\psi_{pk})$. The third line is due to the relation of the body-fixed unit vectors to the space-fixed coordinate system. 

We find that the cotangent of the azimuthal angle of $\k_p^\star$ is
\begin{align}
    \frac{\bh{X}_p \cdot \k_p^\star}{\bh{Y}_p \cdot \k_p^\star} & = -\cot\psi_{pk}  \, , \nn \\[5pt]
    & = \cos\theta_p \, \cot(\varphi_k - \varphi_p) - \cot\theta_k \, \frac{\sin\theta_p}{\sin(\varphi_k - \varphi_p)}\, ,
\end{align}
exactly as found by the addition theorem in Eq.~\eqref{eq:euler_combine}. The sign fixed from Eq.~\eqref{eq:app.Yp_component}, from which
\begin{align}
    \sin\psi_{pk} = \frac{\sin\theta_k}{\sin\theta_{pk}} \, \sin(\varphi_k - \varphi_p) \, . \nn 
\end{align}
We conclude that the azimuthal angle of $\k_p^\star$ is $-\psi_{pk}$ with $\psi_{pk}$ defined in Eq.~\eqref{eq:euler_combine}. Therefore, the spherical harmonic $Y_{\ell'\lambda'}^*(\bh{k}_p^\star)$ is
\begin{align}
    Y_{\ell' \lambda'}^*(\bh{k}_p^\star) \propto P_{\ell'}^{\lambda'}(\cos\chi_p^\star) \, e^{i\lambda' \psi_{pk}}\,, \nn
\end{align}
where we note that because of the complex conjugation, the argument of the exponential is $-i\lambda'(-\psi_{pk}) = i\lambda'\psi_{pk}$. We conclude that the spin-helicity matrix in a general coordinate system has the form
\begin{align}
    \Hc_{\lambda'\lambda}^{(\ell'\ell)}(\p,\k) \propto e^{i\lambda'\psi_{pk}} \, e^{i \lambda \varphi_{pk}} \, , \nn 
\end{align}
which is the exact structure we found by the generic partial wave expansion Eq.~\eqref{eq:app.hel_pw_expand_reduce}, and therefore from the projection Eq.~\eqref{eq:app.hel_pw.proj} we find that the azimuthal dependence completely disappears, leaving what we found Sec.~\ref{sec:ope}.

\section{Formulary for low-lying partial wave OPE amplitudes}
\label{sec:app.formulary}

The purpose of this appendix is for those interested in using the results presented for the low-spin cases in their analysis. We collect the primary results for the $^{2S+1}L_J \to {^{2S'+1}L'_J}$ partial wave OPE amplitudes presented in Sec.~\ref{sec:cases}, as well as a minimal set of kinematics needed. The OPE is a function of the total three-body CM energy $\sqrt{s}$, as well as either the initial and final spectator momenta $k$ and $p$, respectively, or the initial and final state pair invariant mass-squares $\sigma_k$ and $\sigma_p$, respectively. Equation~\eqref{eq:pk_cm} relates these two sets of variables.
\begin{figure}[t]
	\centering
	\includegraphics[width=0.7\textwidth]{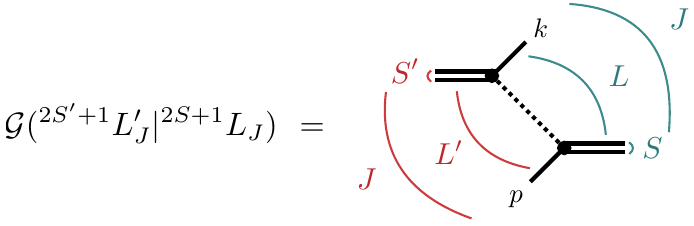}
	\caption{Angular momentum couplings of the OPE, as described in the text.}
	\label{fig:ope_pw}
\end{figure}

The initial and final spectators have masses $m_k$ and $m_p$, respectively, the exchange particle has mass $m_e$, and the product of their intrinsic parities is $\eta$, \eg if considering three pions, then $\eta = -1$. The initial and final pairs have spins $S$ and $S'$, respectively. The initial pair is coupled with its spectator into an orbital state $L$. Similarly, the final state pair is coupled to its spectator into an $L'$ orbital state. Finally, both the initial and final state spin and orbital angular momenta are, respectively, coupled to a total angular momentum $J$. These couplings are illustrated in Fig.~\ref{fig:ope_pw}. The parity of a particular state $J$ is given by $P = \eta(-1)^{L+S} = \eta(-1)^{L'+S'}$. 

From Eq.~\eqref{eq:ope_JP_final_result}, the partial wave OPE is then given by
\begin{align}
	\Gc(^{2S'+1}L'_J | ^{2S+1}L_J) & \equiv \Gc_{L'S',LS}^{(S'S),J}(p,k) \, , \nn \\[5pt]
	& = \Kc_{\Gc}(^{2S'+1}L'_J | ^{2S+1}L_J) + \Tc(^{2S'+1}L'_J | ^{2S+1}L_J) \, Q_0(\zeta_{pk})
\end{align}
where expressions for $\Kc_\Gc$ and $\Tc$ are given below for $J^P = \{0^\eta,1^\eta,1^{-\eta}\}$ and $S',S = \{0,1\}$. The function $Q_0$ is known and given by Eq.~\eqref{eq:app.explicit_Q}, where the argument $\zeta_{pk}$ is defined in terms of energies and momenta in Eq.~\eqref{eq:ope_arg}. For notational convenience, we introduce the three quantities which appear frequently,
\begin{subequations}
	\begin{align}
		f_{pk} & = \gamma_p \left( \, \frac{\beta_p \omega_k}{k} + \zeta_{pk} \, \right) \, , \\[5pt]
		g_{pk} & = \gamma_p \gamma_k \left( \frac{\beta_p \omega_k}{k}   + \frac{\beta_k \omega_p}{p}  + \zeta_{pk} \right) \, , \\[5pt]
		h_{pk} & = \frac{\gamma_p \gamma_k\beta_p \beta_k \omega_p \omega_k}{pk} \, ,
	\end{align}
\end{subequations}
where $\omega_k = \sqrt{m_k^2 + k^2}$, $\omega_p = \sqrt{m_p^2 + p^2}$, $\gamma_p = 1/\sqrt{1-\beta_p^2}$, $\gamma_k = 1/\sqrt{1-\beta_k^2}$ with $\beta_p$ and $\beta_k$ being the boost velocities as discussed in Eqs.~\eqref{eq:k_boost} and \eqref{eq:p_boost}. Note that $g_{pk}$ and $h_{pk}$ are symmetric under interchange of $p\leftrightarrow k$ since $\zeta_{pk}$ is symmetric. However, $f_{pk} \ne f_{kp}$. Finally, the expressions below contain $q_p^\star$ and $q_k^\star$, which are the relative momenta of the pairs in their respective CM frames as defined in Eq.~\eqref{eq:pair_q}.

\begin{enumerate}[label=\textbf{\arabic*}.]

	\litem{J^P = 0^\eta\,\,\mathrm{Amplitudes}}

	There are two possible partial waves, a singlet $^1S_0$ and triplet $^3P_0$, resulting in a $2\times 2$ matrix.
	\begin{enumerate}[label=\textbf{(\roman*)}]
		\litem{^1S_0 \to {^1S_0}}
		\begin{subequations}
		\begin{align}
			\Kc_\Gc(^1S_0 | ^1S_0) & = 0 \, , \\[5pt]
			\Tc(^1S_0 | ^1S_0) & = \frac{1}{2pk} \, .
		\end{align}
		\end{subequations}

		\litem{^1S_0 \to {^3P_0}}
		\begin{subequations}
		\begin{align}
			\label{eq:app.results.1S0_to_3P0.K}
		    \Kc_\Gc(^3P_0|^1S_0) & = -\frac{\sqrt{3}}{2pq_p^\star}\, \gamma_p \, , \\[5pt]
		    \label{eq:app.results.1S0_to_3P0.T}
		    \Tc(^3P_0|^1S_0) & = \frac{\sqrt{3}}{2pq_p^\star} \, f_{pk}  \,  .
		\end{align}
		\end{subequations}
		\litem{^3P_0 \to {^1S_0}}
		
		Given by Eqs.~\eqref{eq:app.results.1S0_to_3P0.K} and \eqref{eq:app.results.1S0_to_3P0.T} with $p \leftrightarrow k$ interchange.
		\litem{^3P_0 \to {^3P_0}}
		\begin{subequations}
		\begin{align}
			\label{eq:app.results.3P0_to_3P0.K}
		    \Kc_{\Gc}(^3P_0 | ^3P_0) &= -\frac{3}{2 q_k^{\star} q_p^{\star}} \, g_{pk} \, , \\[5pt]
		    \label{eq:app.results.3P0_to_3P0.T}
		    \Tc(^3P_0 | ^3P_0) &=  \frac{3}{2  q_k^{\star} q_p^{\star}} \, f_{pk} \, f_{kp} \, . 
		\end{align}
		\end{subequations}

		\end{enumerate}
		
		\litem{J^P = 1^{\eta}\,\,\mathrm{Amplitudes}}
		
		There is a single $^3P_1$ partial wave in this channel.
		\begin{enumerate}[label=\textbf{(\roman*)}]
		\litem{^3P_1 \to {^3P_1}}
		\begin{subequations}
		\begin{align}
			\label{eq:app.results.3P1_to_3P1.K}
		    \Kc(^3P_1 | ^3P_1) &= -\frac{3}{4q_p^\star q_k^\star} \, \zeta_{pk} \, , \\[5pt]
		    \label{eq:app.results.3P1_to_3P1.T}
		    \Tc(^3P_1 | ^3P_1) &= \frac{3}{4q_p^\star q_k^\star} \left(\, \zeta_{pk}^2 - 1 \,\right) \, .
		\end{align}
		\end{subequations}
		\end{enumerate}
		
		\litem{J^P = 1^{-\eta}\,\,\mathrm{Amplitudes}}
		
		There are three possible partial waves, one singlet $^1P_1$ and two triplets $^3S_1$ and $^3D_1$, resulting in a $3\times 3$ matrix.
		
		\begin{enumerate}[label=\textbf{(\roman*)}]
		\litem{^1P_1 \to {^1P_1}}
		\begin{subequations}
		\begin{align}
			\label{eq:app.results.1P1_to_1P1.K}
			\Kc_\Gc(^1P_1 | ^1P_1) &= -\frac{1}{2pk} \, , \\[5pt]
			\label{eq:app.results.1P1_to_1P1.T}
			\Tc(^1P_1 | ^1P_1) &= \frac{1}{2pk} \,\zeta_{pk} \, .
		\end{align}
		\end{subequations}

		\litem{^1P_1 \to {^3S_1}}
		\begin{subequations}
		\begin{align}
			\label{eq:app.results.1P1_to_3S1.K}
			\Kc_\Gc(^3S_1|^1P_1) & = \frac{1}{2pq_p^\star} \left( f_{pk} - \zeta_{pk} \right) \, \\[5pt]
		    \label{eq:app.results.1P1_to_3S1.T}
		    \Tc(^3S_1|^1P_1) & = - \frac{1}{2p q_p^\star} \left[ \zeta_{pk} f_{pk}  - \zeta_{pk}^2 + 1 \right] \, ,
		\end{align}
		\end{subequations}

		\litem{^1P_1 \to {^3D_1}}
		\begin{subequations}
		\begin{align}
			\label{eq:app.results.1P1_to_3D1.K}
		    \Kc_\Gc(^3D_1|^1P_1) & = -\frac{1}{2\sqrt{2}pq_p^\star} \left( 2f_{pk} + \zeta_{pk} \right) \, \\[5pt]
		    \label{eq:app.results.1P1_to_3D1.T}
		    \Tc(^3D_1|^1P_1) & = \frac{1}{2\sqrt{2}pq_p^\star}\left[ 2 \zeta_{pk} f_{pk} + \zeta_{pk}^2 - 1 \right] \, .
		\end{align}
		\end{subequations}

		\litem{^3S_1 \to {^1P_1}}
		
		Given by Eqs.~\eqref{eq:app.results.1P1_to_3S1.K} and \eqref{eq:app.results.1P1_to_3S1.T} with $p \leftrightarrow k$ interchange.
		
		\litem{^3S_1 \to {^3S_1}}
		\begin{subequations}
		    \begin{align}
				\label{eq:app.results.3S1_to_3S1.K}
		        \Kc_{\Gc}(^3S_1 | ^3S_1) & = \frac{1}{2q_p^\star q_k^\star} \Bigg[  \frac{2}{3}(1-\gamma_p)(1-\gamma_k) - \gamma_p\gamma_k  \nn \\[5pt]
		        &  \qquad\qquad\quad + \zeta_{pk} \left( f_{pk} +  f_{kp}  - \zeta_{pk}\right) - \left(\zeta_{pk}g_{pk} + h_{pk}\right)  \Bigg] \, , \\[5pt]
		        \label{eq:app.results.3S1_to_3S1.T}
		        \Tc(^3S_1 | ^3S_1) & = \frac{1}{2q_p^\star q_k^\star} \Bigg[ (1 - \zeta_{pk}^2 ) \left( f_{pk} + f_{kp} - \zeta_{pk} \right)  \nn\\[5pt]
		        & \qquad\qquad\quad +  \,\zeta_{pk} \left( \zeta_{pk}g_{pk} + h_{pk}  \right) \Bigg] \, ,
		    \end{align}
		\end{subequations}
		
		\litem{^3S_1 \to {^3D_1}}
		\begin{subequations}
		    \begin{align}
			\label{eq:app.results.3S1_to_3D1.K}
		        \Kc_{\Gc}(^3D_1 | ^3S_1) & = \frac{1}{2\sqrt{2}q_p^\star q_k^\star} \Bigg[  \frac{2}{3}(1+2\gamma_p)(1-\gamma_k) +2 \gamma_p\gamma_k   \nn \\[5pt]
		        &  \qquad\qquad\quad + \zeta_{pk} \left(-2f_{pk} + f_{kp} - \zeta_{pk}\right) + 2 \left( \zeta_{pk}g_{pk}  + h_{pk}  \right)  \Bigg] \, , \\[5pt]
		        \label{eq:app.results.3S1_to_3D1.T}
		        \Tc(^3D_1 | ^3S_1) & = \frac{1}{2\sqrt{2}q_p^\star q_k^\star} \Bigg[ (1 - \zeta_{pk}^2) \left( -2f_{pk}  +  f_{kp} - \zeta_{pk}\right)  \nn \\[5pt]
		        &  \qquad\qquad\quad - 2 \,\zeta_{pk} \left( \zeta_{pk}g_{pk}  + h_{pk}  \right)  \Bigg] \, ,
		    \end{align}
		\end{subequations}
		
		\litem{^3D_1 \to {^1P_1}}
		
		Given by Eqs.~\eqref{eq:app.results.1P1_to_3D1.K} and \eqref{eq:app.results.1P1_to_3D1.T} with $p \leftrightarrow k$ interchange.
		
		\litem{^3D_1 \to {^3S_1}}
		
		Given by Eqs.~\eqref{eq:app.results.3S1_to_3D1.K} and \eqref{eq:app.results.3S1_to_3D1.T} with $p \leftrightarrow k$ interchange.
		
		\litem{^3D_1 \to {^3D_1}}
		\begin{subequations}
			\begin{align}
				\label{eq:app.results.3D1_to_3D1.K}
				\Kc_{\Gc}(^3D_1 | ^3D_1) & = \frac{1}{4q_p^\star q_k^\star} \Bigg[ \frac{2}{3}(1+2\gamma_p)(1+2\gamma_k) - 4\gamma_p\gamma_k \nn \\[5pt]
		        & \qquad\qquad\quad  -\zeta_{pk}\left(2f_{pk} + 2f_{kp} + \zeta_{pk} \right)  - 4 \left(  \zeta_{pk} g_{pk}  + h_{pk} \right)  \Bigg] \, , \\[5pt]
		        \label{eq:app.results.3D1_to_3D1.T}
		        \Tc(^3D_1 | ^3D_1) & = \frac{1}{4q_p^\star q_k^\star} \Bigg[ -(1 - \zeta_{pk}^2) \left( 2 f_{pk} + 2 f_{kp} + \zeta_{pk} \right)  \nn \\[5pt]
		        & \qquad\qquad\quad + 4 \,\zeta_{pk} \left( \zeta_{pk}g_{pk}  + h_{pk}  \right)  \Bigg] \, .
		    \end{align}
		\end{subequations}
	\end{enumerate}
\end{enumerate}

\bibliography{bibi.bib}

\end{document}